\documentstyle[prd,aps,eqsecnum,epsf]{revtex}
 \def \tablerule{\noalign {\vskip3truept\hrule\vskip3truept}}
 \def\tablerule{\noalign {\hrule}}
 
 \def\gsim{\mathrel{
 \rlap{\raise 0.511ex \hbox{$>$}}{\lower 0.511ex
 \hbox{$\sim$}}}}
 \def\lsim{\mathrel{
 \rlap{\raise 0.511ex \hbox{$<$}}{\lower 0.511ex
 \hbox{$\sim$}}}}

\begin{document}

\title{Frequency-domain  P-approximant filters for time-truncated
inspiral gravitational wave  signals from compact binaries}

\author{Thibault Damour$^1$, Bala R. Iyer$^{2,3,4}$ and B.S. Sathyaprakash$^3$}

\address{$^1$ \it Institut des Hautes Etudes Scientifiques, 91440
Bures-sur-Yvette, France  }
\address{$^2$ \it Raman Research Institute, Bangalore 560 080, India}
\address{$^3$ \it Cardiff University, P.O. Box 913, Cardiff,
CF2 3YB, U.K. }
\address{$^4$ \it Albert Einstein Institute, D-14476, Golm, Germany}
\date{\today}
%\date{December 27, 1999}
\maketitle

\begin{abstract}

Frequency-domain filters for time-windowed
 gravitational waves from inspiralling
compact binaries are constructed which combine the excellent performance
of our previously developed time-domain P-approximants 
with the analytic convenience  of the stationary
phase approximation without a serious loss in event rate.
These Fourier-domain representations 
incorporate  the ``edge oscillations''
due to the (assumed) abrupt shut-off of the time-domain
signal caused by the relativistic plunge at the last stable orbit.
These new analytic approximations, the SPP-approximants, 
are  not only   {\it effectual} for detection
and {\it faithful} for parameter estimation,  
but are  also  
computationally inexpensive to generate (and are  {\it faster}
by factors up to 10, as compared to the corresponding time-domain templates).
The SPP approximants  should provide data 
analysts the Fourier-domain templates for massive black hole binaries
of total mass $m\lesssim 40 M_\odot$,
the most likely sources for LIGO and VIRGO.
\end{abstract}


\section {Introduction and Summary }\label{sec:INTRO}

The discovery of the first binary pulsar in 1974 \cite{ht75} has had a very
important impact on gravitational wave research. First, it proved the reality 
of gravitational radiation by measuring the orbital period decay \cite{ht94}
entailed by the propagation at the velocity of light of the gravitational
interaction between the two neutron stars making up the system \cite{td83}.
Second, it provided the first experimental evidence that General Relativity
correctly describes gravity in the strong-field regime \cite{twdw92}. Third,
 it led to a shift in perception regarding the most promising sources for
future gravitational wave (GW) detectors, away from the then assumed,
violent --- but less predictable --- gravitational collapse associated with supernovae,
to the more predictable, final inspiralling phase of 
compact binaries  of neutron stars and black holes driven by gravitational
radiation-reaction.
This also led to the thrust in the laser interferometric gravitational wave 
detectors
which are inherently broad-band rather than in the narrow-band bar detectors.

\subsection{Data analysis algorithms for inspiral wave searches}
\label{sec:IA}

Consider a compact binary system like the binary pulsar after it has been
inspiralling inwards for three hundred million years due to 
gravitational radiation-reaction.
The inspiral waveform enters the detector
bandwidth  during the last few minutes of   evolution of the binary.
Our ability, in principle, to compute 
the  waveform very accurately, 
allows us  to track the
 gravitational wave phase and  enhance the signal-to-noise ratio by
integrating the signal for the  interval
during which it lasts in the detector band. 
This, in turn, requires a template  with which  
 the detector output   may be filtered.
   Though template waveforms should, optimally, be exact  copies of the
      expected signal, in practice,
 they are constructed by some approximation
scheme and will differ from the 
actual signal in the detector output. Consequently, the 
overlap of template and signal waveforms will be less than 
if they had exactly matched, leading to a  loss 
of potential events.
Data analysis issues like these  for inspiralling compact binaries
of neutron stars and black holes have been formulated and addressed
for the last twelve years \cite{kt87,bs91},
even though interferometric gravitational wave  detectors like the GEO600 
\cite{geo}
or LIGO \cite{ligo} and VIRGO  \cite{virgo} are a year or three in the future.
 Much of the  work in this area has addressed practical issues of direct 
relevance to
data analysis strategies. These include:
construction of  templates for  detection\cite{sd91},  the number of 
templates, 
their placement, spacing, 
the required computing power and  the storage or memory requirement
\cite{bs94},
the order of post-Newtonian (PN) approximation
adequate for detection \cite{dp97,dis98,warn}, 
parameter estimation  by covariance
matrix \cite{cf94,pw95,kks95} and Monte Carlo simulations\cite{bsd96}, 
determination of cosmological parameters \cite{fc93}, tests of general relativity \cite{bs94a},
one step versus hierarchical searches \cite{md96}, effects of  
precession\cite{ta96a}
and of eccentricity \cite{kks95,mp99}.
  For the time-domain waveform,  all of these works use the 
restricted post-Newtonian  approximation to quasi-circular inspiral.
  This keeps the crucial phase information to the best order of approximation  
then available \cite{bdiww95},  but restricts the amplitude to be Newtonian
and  the  harmonic to  the second harmonic of the orbital frequency.
  Such an approximation should be adequate for the on-line search of
   gravitational wave signals \cite{3mn}. Evidently, it is assumed that the offline
   analysis of the data will use the best available (unrestricted 
   post-Newtonian) representation of the inspiral signals. 

\subsection{Modelling inspiral waveforms}
\label{sec:IB}

The post-Newtonian approximation   is basically a Taylor
 expansion (in powers of $v/c$) and all the above treatments use as
 building blocks the straightforward Taylor expansions
in $v/c$ of some intermediate
 quantities (orbital energy and gravitational-wave flux). We shall refer
 to the templates based on such straightforward PN expansions as 
 ``Taylor approximants'' (or simply T-approximants).
The very slow convergence and oscillatory behaviour
of the PN expansion, and therefore of the sequence of Taylor approximants,
 made imperative a search for
better   approximants for phasing. This prompted us \cite{dis98} (later 
referred to as DIS) to propose new approximants, with much improved 
convergence properties, for application to gravitational-wave data analysis
 problems.  

In  DIS \cite{dis98}, 
we showed how to construct a new type of {\it time-domain}
approximant, called ``P-approximants'', 
which not only  converged faster and more
monotonically, 
but  were also  more {\it effectual} (larger overlaps for detection) 
and {\it faithful}
(smaller biases for parameter estimation) 
than the standard T-approximants. 
Our construction was 
two-pronged: on the one hand, it introduced   new basic energy and flux
functions, and on the other hand, it made a  systematic use of Pad\'e 
techniques (a well-known convergence-acceleration technique) to construct
 successive approximants of our new basic energy and flux functions.
  These new functions form a  pivotal aspect of our construction and
   successfully  handle issues related to
appearance of non-rational functions in the energy function and logarithmic
terms in the flux function that for long proved to be hurdles to the 
application
of well-known Pad\'e techniques to this problem. 
For initial LIGO, the 2.5PN P-
approximants are likely to provide overlaps in excess of 96.5\% with exact 
waveforms\footnote{ 
This statement was proven in DIS by quantifying the convergence of
the sequence of P-approximants towards some `fiducial exact' waveform. 
In the test-particle case this waveform  used the known Schwarzschild's energy function $E(v)$
and Poisson's  numerically computed GW flux \cite{numflux}.
In the comparable mass case, 
it was constructed by modelling the
$\eta$- dependent higher post-Newtonian corrections 
to the best known analytical results.}
so that  more than 90\% of
the potential events can be detected. In contrast, 
the corresponding  2.5PN Taylor
approximants can only detect about 50\% of the potential events
for massive systems (at the price of large biases $\sim 15 \%$).
Later studies
have confirmed the performance of these P-approximants \cite{ep98} and
assessed \cite{bs98} their need in related contexts of space based interferometers
like LISA.

\subsection{Fourier representation of inspiral signals and validity of the 
stationary phase approximation (SPA)}
\label{sec:IC}

 Independent of the choice between T- and P-approximants,
another  desirable approximation in data analysis  for
inspiralling compact binaries is the stationary phase approximation (SPA),
which is a simple, explicit analytic approximation to the
Fourier transform of the time-domain chirp (see, e.g., \cite{SPA}).
In fact, most work on inspiral waveforms (except DIS) has used only
SPA approximants to the frequency-domain chirps.
In the course of our P-approximant work we noticed a progressive worsening of
 the overlap between the SPA and  the ``exact'' Fourier transform --- numerically
computed by a fast Fourier transform (FFT) of the time-domain signal ---
(see Table II of DIS) and commented on these `inaccuracies of the SPA'.
In the above by SPA one  means  not only the problem of the formal
accuracy of the stationary phase  estimate 
to the Fourier transform of an analytically extended, mathematical signal
 but also some issues linked to the physics,
and observability, of the real signal. In particular, in DIS, we were considering
templates which are  shut off, in the time-domain, at the last stable orbit (LSO).
The present paper will also consider such {\it time-truncated
inspiral signals}.
We shall discuss this point in more detail below, but the idea is that
the post-inspiral signal (plunge + merger) will have a 
frequency content very different from the inspiral one (probably pushed to
much higher frequencies). It should, therefore make sense to try to 
construct  filters that represent as best as possible an inspiral signal
which lasts only up to some maximum time (time-windowing).
For such signals, DIS noted a worsening of the usual (frequency-windowed)
SPA approximation, both as the total mass of the system increases and as
the post-Newtonian approximation order is increased, and mentioned that
this worsened performance was due to the fact that ``such systems emit
many less wave cycles in the effective detector bandwidth''
 centered (for
initial LIGO) near $f_{\rm det} = 167$ Hz.
In this paper  $f_{\rm det}$  denotes  
the frequency at which the  noise power spectrum per logarithmic
bin of the detector is the least 
(or equivalently the frequency at which the detector is
most sensitive to a broad-band burst). 
To avoid   irrelevant, uncontrolled sources of inaccuracy, DIS  used the FFT of 
the time-windowed chirp 
rather than its SPA to generate the frequency-domain waveform and make 
comparisons  between
the T- and P-approximants.

The use of FFT rather than SPA in DIS makes the P-approximant 
computationally expensive.
As will be discussed in detail in Sec.~\ref{sec:COMPU}, the use of SPA or
similar  frequency-domain representations is far less expensive.
The obvious need to incorporate this desirable feature  makes urgent
 and mandatory  a critical investigation of the possibility of
marrying together the excellent performance of the P-approximants to the
relative inexpensiveness of the SPA  without a serious  loss in event rate.

Recently, some issues related to the accuracy of the SPA have been 
investigated.
 For general chirps, Chassande-Mottin and Flandrin \cite{cf98} have studied  
  whether
the usual conditions assumed for the validity of the SPA
 are necessary and sufficient and attempted a quantitative control
  of the approximation.
 Droz {\it et al} \cite{dkpo99} have examined  other issues related
to the accuracy of the SPA of particular relevance to gravitational wave 
data analysis. 
Unlike DIS,  by SPA, Droz {\it et al} imply only the
stationary phase estimate of the Fourier transform and discuss separately the issue of
windowing --- the fact that the signal in the time-domain  lasts only from $t_{\rm 
min}$
to $t_{\rm max}$ or a time-window.  To improve the SPA estimate of a
{\it Newtonian} chirp, they compute
the next order contribution\footnote{
We shall give below the general result for any chirp.}
 (to the Fourier integral) by the method of steepest
descent, show that  it is of order $v_{{\cal M}}^{10}$ 
relative to the leading order SPA estimate and conclude that  it is small enough
to be justifiably neglected. [Here  $v_{{\cal M}}$ is 
an invariantly defined `velocity'\footnote{ 
Note that, following DIS, we shall use
$v\equiv (\pi m F)^{1/3}$, instead of $v_{{\cal M}}={\eta}^{1/5} v$,
in all our analysis. We also use units such that $G=c=1$.}
 related to the instantaneous gravitational wave frequency $F$ and
chirp mass ${\cal M}$ by 
 $ v_{{\cal M}}=(\pi {\cal M} F )^{1/3}$. The chirp mass is related
to the total mass $m=m_1+m_2$ and dimensionless mass ratio
$\eta=\frac{m_1m_2}{(m_1+m_2)^2}$ by ${\cal M}
=\eta^{3/5}m $.]
They  point out the importance of windowing,  
estimate the amplitude  and phase modulations
induced in the frequency-domain
 by the time-window and  conclude that {\it in all cases }  these modulations  have
 negligible effect on overlaps.
However, their analytic  expression for the effects of windowing is only valid for values 
of frequencies {\it well 
away from the boundaries} of the natural frequency-window induced by the 
time-window,
 denoted by $F_{\rm min}= F(t_{\rm min})$ 
and $F_{\rm max}=F(t_{\rm max})$ ---  the gravitational wave frequencies at  times 
$t_{\rm min}$ and
$t_{\rm max}$, respectively.
 In this paper we provide a formalism 
allowing one to compute analytic approximations to the Fourier-transform
of a time-windowed signal in the most crucial {\it edge-frequency-domains}
$f\sim F_{\rm min}$ and $f\sim F_{\rm max}$ (including $f<F_{\rm min}$
and $f>F_{\rm max}$).
As  first  noticed  in DIS and discussed in detail in this present work, the   
effect
of window oscillations on overlaps (claimed to be negligible in \cite{dkpo99}) 
 starts to be noticeable when the total mass $m\gtrsim 13 M_\odot$ and becomes
 very significant for  $ m \gtrsim 20\, M_{\odot}$. [Here we consider
equal mass systems $\eta=1/4$.] 
 Since the  difference between the statements in DIS  and Droz {\it et al}
 \cite{dkpo99} can
be disconcerting and a serious source of confusion to the potential user community,
we discuss this in further detail next.  

In DIS,  what was meant in Table II (the only place where it
was used) by ``stationary phase approximation'' was  the product
of the usual SPA by a simple Heaviside step function $\theta(F_{\rm max} -f)$
{\it i.e.} $\tilde{h}(f)$ was truncated above a
Fourier frequency $f = F_{\rm max}$ where $F_{\rm max}$ is the instantaneous
gravitational wave  frequency at which the  time-domain signal is  itself terminated,
assumed to be (in DIS and here) the frequency at the last stable orbit $F_{\rm LSO}$.
[In the following, we shall, for brevity, refer to this frequency Windowed
Usual Stationary Phase Approximation as the `uSPAw'.]
We were motivated to do this from the stationary phase result itself. The
SPA (to the Fourier transform of the chirp) says that the
dominant contribution to a certain Fourier amplitude
$\tilde{h}(f)$ comes from a neighbourhood of time (in the Fourier
integral) when the instantaneous
frequency $F(t)$ numerically reaches the corresponding Fourier frequency $f$. It
is therefore to be expected that the signal essentially terminates at $f =
F_{\rm LSO}$ {\it i.e.} that  there is no significant power in the Fourier transform
of the signal
beyond $F_{\rm LSO}$. 
This is indeed true in the first approximation, as is
evident from Fig.~\ref{fig:power.newt}  below, which shows that the power in the exact Fourier
transform of the time-windowed signal
[computed via a discrete Fourier transform (DFT)] 
falls off much faster than the SPA for $f> F_{\rm LSO}$. 
Moreover, as is discussed in detail below, in the relativistic case the usual SPA
breaks down   at $F_{\rm LSO}$ and cannot be meaningfully extended for 
 $f > F_{\rm LSO}$. Hence
the values quoted in Table II of DIS were obtained by computing the
overlap of the DFT of the truncated time-domain waves with the
truncated SPA representation of the wave.

On the other hand, a critical examination of  Droz et al \cite{dkpo99}  
reveals that their claim regarding the adequacy of the SPA 
in fact has  only a  restricted  domain of validity.
It is relevant to SPA  considered as a {\it mathematical} algorithm
to be applied to a generic smooth signal and low mass binaries ($m\lesssim 13 M_\odot$). 
As acknowledged by the authors, they do  not  address  physical issues
related to an eventual time-domain cut-off of the signal at $F_{\rm LSO}$. 
What they call ``Newtonian signals'' are  unphysical,
formally defined chirps whose instantaneous frequencies
are extended to $F_{\rm max} = F_{\rm Nyquist} \gg F_{\rm LSO}$, 
in fact,  better described as `analytically extended Newtonian signals'.
It is the SPA of  this formal, analytically extended signal which is
 shown to produce overlaps  with  exact FFTs   better
than $0.99$ even for massive binary systems of chirp mass ${\cal M}= 
10\,M_{\odot}$, corresponding to a total mass of
 $m\sim 23 M_\odot$ for an equal mass system
 ($\eta=1/4$). 
These large overlaps, in our view, are {\it not} a proof of the
validity of the SPA to compute, physically relevant, accurate frequency-domain inspiral
templates, as they do not address the important issue of inspiral-signal
termination at or near the  $F_{\rm LSO}$ when $F_{\rm LSO} \sim f_{\rm det}$, 
the frequency at which the broad-band noise of the detector is the least. 
It turns out that for binary
systems of total mass $m \gtrsim 28 M_{\odot}$  the power in the Fourier
domain beyond $f = F_{\rm LSO}$ for a relativistic signal, is
a significant fraction $(>10 \%)$ of the total power.
If the usual (frequency-windowed) SPA is used in
constructing frequency-domain inspiral waves we are risking
the loss of more than 30 \% of the events from binaries with
masses $\gtrsim 28 M_\odot$.
[This will be illustrated in Fig.~\ref{fig:zero} below.]
This is in addition to the losses induced by the inaccuracy
of the post-Newtonian waveforms and the discreteness of the bank of
templates used in data analysis.  

\subsection{Massive black hole binaries and first detections in LIGO/VIRGO}
\label{sec:ID}

Let us first establish our notation. We define the Fourier transform (FT)
$\tilde{h}(f)$ of a time-domain signal $h(t)$ by
\begin{equation}
h(t)=\int_{-\infty}^{\infty}\,df\,e^{-2\pi i f t}\,\tilde{h}(f)\,;\;\;
\tilde{h}(f)=\int_{-\infty}^{\infty}\,dt\,e^{2\pi i f t}\,h(t)\,.
\label{d1}
\end{equation}
We write the (suitably transformed) output of the detector as
\begin{equation}
h_{\rm out}= h(t) +n(t)\,,
\label{sha1}
\end{equation}
where $h(t)$ is the signal and $n(t)$ the noise.
The correlation function of the noise reads
\begin{equation}
\overline{n(t_1)n(t_2)}=C_n(t_1-t_2)=\int_{-\infty}^\infty\,df\,S_n(f)\,e^{2\pi i f(t_1-t_2)}\,,
\label{sha2}
\end{equation}
where $S_n(f)=S_n(-f)$ is the two-sided noise power spectral density.
In all the present work, we shall consider a noise curve of the type expected
for initial interferometers. For initial LIGO we take\cite{strain},
\begin{mathletters}
\begin{eqnarray}
S_n(f)&=&\frac{S_0}{2}\left[2 +2\left(\frac{f}{f_0}\right)^2 +\left(\frac{f}{f_0}\right)^{-4}\right],\,\,f\ge f_s\,,
\label{ligonoise}\\
&=&\infty,\,\,f<f_s\,.
\label{ligonoisea}
\end{eqnarray}
\end{mathletters}
 with  $f_s=40 $ Hz, $f_0=200$ Hz and 
$S_0=1.47\times 10^{-46}$ Hz$^{-1}$.
In the above we have included a factor of $1/2$
 [$S_n^{\rm one-sided} \equiv 2 S_n^{\rm two-sided}$]
 because Eq. (\ref{ligonoise})
gives the {\it two-sided} noise; the {\it one-sided} noise would be given
by the same formula without the factor of $1/2$. 
The minimum of $S_n(f)$ is at $f=f_0$ and is equal to $S_{\rm min}=2.5S_0$.
However, a physically more relevant quantity is the minimum of
the dimensionless quantity $h_n^2(f)\equiv fS_n(f)$ (effective GW noise,
see below). This is reached at  the {\em characteristic detection
frequency}  $f=f_{\rm det}=0.8347 f_0$,
and is equal to $(h_n^{\rm min})^2=2.2761 f_0 S_0$. 
The above numerical value for $f_0$ and $S_0$ leads to
$f_{\rm det}=167 $ Hz and  corresponds to 
$h_n^{\rm min}=2.5868\times 10^{-22}$.

\begin {figure}[t]
\caption{
This Figure compares the signal-to-noise ratio (SNR) for an inspiral signal
searched by means of two different filters, in two different interferometers
 [initial LIGO, Eq. (\ref{ligonoise}) or VIRGO, Eq.  (\ref{virgonoise})].
 The exact signal $h$ is assumed to be a time-truncated
Newtonian chirp (sufficiently well approximated by the inSPA). 
The solid lines represent the optimal SNR, obtained when the filter $k$ 
is identical with the exact signal $h$.
The dotted lines represent the sub-optimal SNR, obtained when the filter $k$ is the
frequency-windowed usual SPA (uSPAw). 
In the case of LIGO, for low mass binaries uSPAw 
extracts the full SNR,
 but at higher masses, which are the most crucial ones for
initial interferometers, it loses SNR up to a factor
of 1.5 leading to a loss in the number of  detectable events
up to 70\%. In the case of VIRGO, uSPAw performs quite well when the total mass $m < 50 M_\odot$, 
and requires inSPA for heavier binaries.
The loss in the number of events caused by uSPAw for heavier mass 
binaries in the range $ m> 50 M_\odot$  is unacceptable.}
\centerline {\epsfxsize 5 true in  \epsfbox {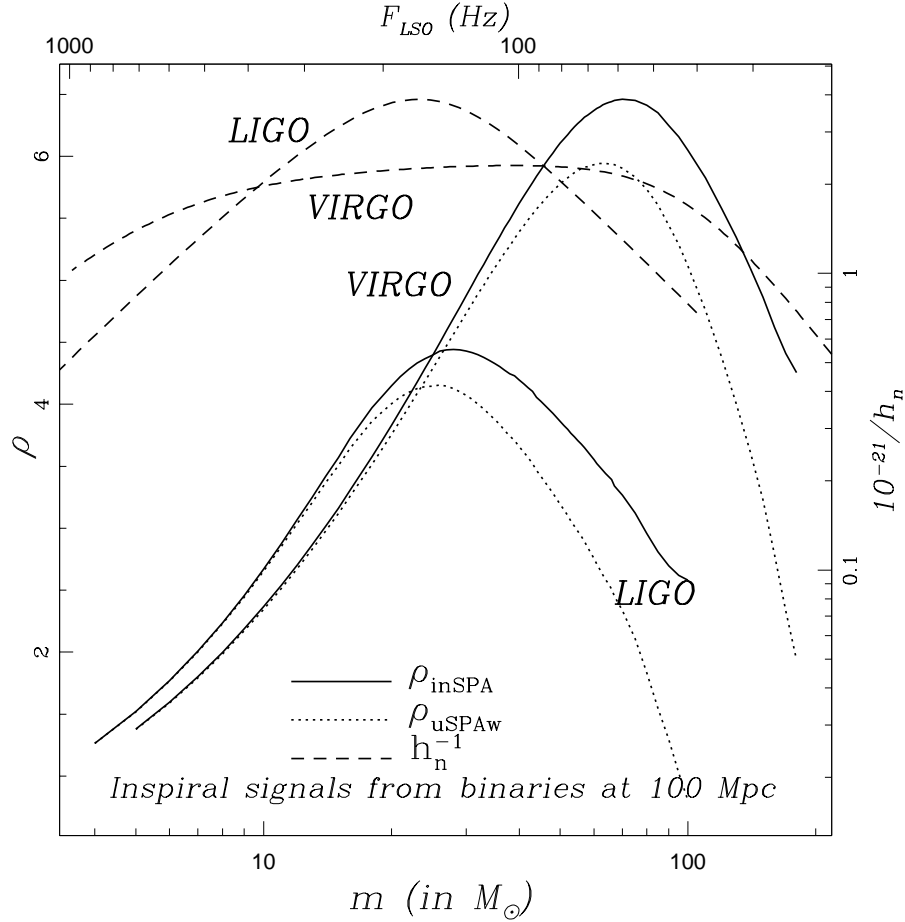} }
\label {fig:zero}
\end {figure}

For VIRGO on the other hand, the corresponding noise curve 
is given by
\begin{mathletters}
\begin{eqnarray}
S_n(f) &=&  \frac{S_0}{2}\left [ 10^3 \left(\frac{f_s}{f}\right)^5 + 
2 \left(\frac{f_0}{f}\right) + 1 + \left(\frac{f}{f_0}\right)^2\right ],\,\, 
f \ge f_s\,,
\label{virgonoise}\\
     &=&  \infty,\,\, f < f_s\,, 
\label{virgonoisea}
\end{eqnarray}
\end{mathletters}
In this case, $f_s=20$ Hz, $ f_0=500$ Hz  while $ S_0=3.24\times 10^{-46}$
Hz$^{-1}$\cite{bcc96}.
The minimum of $h_n(f) = \sqrt{ f S_n(f)}$ is reached at $f = 103 $ Hz   and
   is equal to $h_n^{\rm min} = 4.2902\times 10^{-22}$. 
It should be noted that the VIRGO noise curve is used
 only in this Section, while discussing Fig.~\ref{fig:zero}. 
In  the rest of the paper and 
all the   Figures and Tables, the scalar product is defined using
the LIGO noise curve.

 Anticipating on formulas to be discussed in \ref{sec:IIB}, 
the square of the signal to noise ratio (SNR)  is given by
\begin{equation}
\rho^2 = \left(\frac{S}{N}\right)^2 = \frac{\langle k,\,h\rangle^2}{\langle k,\,k\rangle},
\label{2.4r}
\end{equation}
 where the scalar product is defined by
\begin{equation}
 \langle k,\,h\rangle \equiv \int_ {-\infty}^ {+\infty}\, df\, 
\frac {\tilde{k}^*(f) \tilde{h}(f)} {S_n(f)}.
\label{2.4rr}
\end{equation}
 Here, $h$ denotes the exact signal, and $k$ the filter used in the data analysis.
 We assume in this paper that the signal $h$ is given by a time-truncated
adiabatic inspiral signal. 
[For simplicity, we consider in this subsection Newtonian waveforms,
 and we approximate the Fourier transform of a time-truncated Newtonian signal
by the very accurate {\it improved Newtonian stationary phase approximation} (inSPA)
to
 be constructed below.] 
In computing $\rho^2$ we average over all the angles 
(determining both the detector and the source	  orientations), 
and we place the source at a fiducial distance of 100 Mpc.
[Note that a coalescence rate of $10^{-5}$
 per galaxy and per year implies that in two years
 one event should happen within 100 Mpc.]

In most of the literature one uses as Fourier-domain filter $\tilde{k}(f)$ the
   frequency-windowed usual stationary phase approximation (uSPAw) to estimate
the
   SNR for an inspiral signal. We illustrate in Fig.~\ref{fig:zero} 
 the loss in signal strength extracted
   by using as filter the uSPAw in LIGO and VIRGO 
 [cf. Eqs. (\ref{ligonoise}) and (\ref{virgonoise})], 
instead of using the
   optimal filter $k = h$ 
(leading to the optimal SNR $\rho^2 = \langle h,h\rangle$). 
The plot also shows on the top horizontal axis
the last stable orbit frequency corresponding to the total mass in question.
The left vertical axis shows the SNR extracted and the right vertical axis
shows the sensitivity, $h_n^{-1} \equiv [f S_n(f)]^{-1}$, 
(both of which are dimensionless)
of LIGO-I and VIRGO instruments.  
While reading the sensitivity curve one should use the top and right axes 
and while reading the SNR curve one should use the bottom and left axes.
   The SNR values plotted in Fig.~\ref{fig:zero}  have been
computed numerically by inserting the relevant values of 
$\tilde{h}(f)$ and $\tilde{k}(f)$ in Eq. (\ref{2.4r}).
Though we did not use it, it might help the reader to see the {\it analytical}
expression of $\rho^2$ obtained in the simple
approximation where  $\tilde{k}(f) \simeq \tilde{h}(f) \simeq
\tilde{h}^{\rm uspaw}(f)$.
   Using the equations of Sec.~\ref{sec:USE}  below 
(and averaging over angles as explained
in Sec.~\ref{sec:IVA} below) leads to
\begin{equation}
\langle\rho^2\rangle\simeq \frac{\eta}{15 \pi}
\left(\frac{m}{d}\right)^2\int_{0}^{F_{\rm LSO}}\,\frac{df}{f} \;
\frac{1}{v(f)}\;\frac{1}{f S_n(f)}\,,
\label{2.6new}
\end{equation}
where $\langle..\rangle$ denotes the angular average,
$d$ the distance to the source,
$v(f)=(\pi m f)^{1/3}$ and $S_n(f)$ the
{\it two-sided} noise given by Eq. (\ref{ligonoise}).
We indicated no precise detection threshold in Fig.~\ref{fig:zero}
 because this depends on many parameters (like the number of detectors involved). 
The reader should, however,	have in mind that a reasonable
   detection threshold is, at least, $\rho_{\rm threshold} \sim 5$.

We note that the {\it effective }
sensitivity of LIGO-I peaks near a frequency of 167 Hz which
is the last stable orbit frequency for a binary of total mass
of about  $27 M_\odot$.
The effective sensitivity of VIRGO peaks at a much lower frequency of 103 Hz. 
This low-frequency sensitivity of VIRGO means two important 
things: Firstly, lighter binaries (i.e., $m\lesssim 30 M_\odot$) are 
integrated for a longer time in the low frequency regime and,
therefore, the corrections to the Fourier transform introduced 
in this paper are less important for such systems.  
This means that the uSPAw is quite good
in extracting the full signal power of such binaries as evidenced
in Fig.~\ref{fig:zero}. 
Notice that, for VIRGO,  the uSPAw curve follows the inSPA curve
for $m\lesssim 30 M_\odot.$ On the contrary, LIGO's  lower sensitivity to
lower frequencies makes it important to include the corrections to the
Fourier transform of LSO-truncated  signals from binaries of mass 
$m\gtrsim 15 M_\odot$. In LIGO's case uSPAw extracts only 75\% of the
full SNR implying a loss of more than 40 \% of all massive binary
coalescences.  On the other hand, the low-frequency peak of VIRGO
sensitivity means that we will have to employ the accurate Fourier
domain models discussed in this paper for more massive binaries, i.e.
$m\gtrsim 30 M_\odot$.  
[Note, however, that the low-frequency sensitivity of VIRGO means that, 
for low-mass and medium-mass binaries, it is even more
 crucial to use P-approximants (instead of the usually considered
 T-approximants) than for LIGO,
 in order to accurately keep track of the phasing of the many
cycles accumulated at low frequencies.]

It is fair to say that at present the most well-understood gravitational
waveform  is the inspiral one  and thus the only reliable templates 
correspond to  inspiral signals.  It is also generally believed that  
binary black holes are better candidates for gravitational wave
sources than binary neutron stars due to their larger masses 
(the average mass of observed black hole candidates is around $8 M_\odot$
\cite{ts97}).
Theoretical computations based on stellar evolution indicate 
that binary black holes  with  individual masses $\lesssim  15 M_\odot$
may be  the only  known sources that exist  (hopefully) in 
sufficient numbers \cite{lpp97,fh98,bct98,pp99}.
 When looking at Fig.~\ref{fig:zero}, one  clearly sees the importance of dealing 
with binary black holes  with  total masses in the range of $28-30 M_\odot$.
They lead to  signals with   the best  SNR. 
However, it is precisely  for such systems that the $F_{\rm LSO}$ 
is around the middle of the detection bandwidth  for initial LIGO
{\it i.e.}  $F_{\rm LSO}\sim f_{\rm det}$.
The most likely sources to detect are in the problematic region
discussed in this paper.
 This makes it imperative  to not lose SNR when dealing
 with such signals and provides the other major motivation for our work.
 If our assumption that the best models of inspiral waveforms must be abruptly
 shut off in the time domain holds, it is essential to use the improved SPA
 formulas discussed in this work in order to maximise our chances of detecting
 inpiralling binaries. 
The analysis presented in this paper provides insights and
 techniques  to deal with  binary black hole  signals in probably the
most  crucial mass range.

\subsection{Summary of the present paper and proposals for data analysis groups}
\label{sec:IE}

In this paper we propose analytical approximations to
the Fourier-transform of the LSO-truncated
time-domain inspiral waves that are very accurate 
even for the massive black hole binaries, the most likely
sources for LIGO and VIRGO
[overlaps  with FFT
$>0.99$ for $m \lesssim 40 M_{\odot}$] and are at the same time
{\it computationally inexpensive}.  We call our final new,
frequency-domain filters the SPP approximants because they
combine the computational convenience of stationary phase approximants with
the accuracy of the (time-domain) P-approximants. Our strategy is two-fold: On the one
hand we introduce a correction factor ${\cal C}(f)$ to the usual SPA for
$f <f_{\rm up}\le F_{\rm LSO}$ which improves the SPA by taking into account 
the ``edge'' oscillations present when $f\lesssim f_{\rm up}$.
($f_{\rm up}$ will be defined below. For Newtonian-like signals $f_{\rm up}=F_{\rm LSO}$,
while for relativistic signals $f_{\rm up} < F_{\rm LSO}$.)
On the other hand, we introduce a new
approximation to the Fourier transform for $f > f_{\rm up}$ which efficiently recovers
the signal power  around and 
beyond the frequency corresponding to the last stable orbit. These 
features are important new   steps forward as there was no formalism
until now that could compute (especially for {\it relativistic} signals)
 Fourier transforms analytically for 
$f\sim F_{\rm LSO}$, and in particular
$f>F_{\rm LSO}$.
These
new features now make it possible to generate templates {\it directly in
the Fourier-domain}, leading to a saving on the
computational cost of template generation by a factor of 10 or more.

Our concrete proposal to the interferometer data analysis groups that
are building the gravitational wave  search software and wish to
have Fourier-domain filters which are both accurate and fast-computed, is thus the following:
First, we confirm that for {\it accurate} post-Newtonian
template generation of binary systems of total mass $m \lesssim 40 M_{\odot}$  
one needs to use a frequency-domain version of the P-approximant (previously
defined only in the time-domain).
For $m<5 M_\odot$ a straightforward (uncorrected for edge-effects) SPA
of the P-approximants  is sufficient.
(They match with the exact DFT  of the same time-signal with overlaps $ > 0.999$.)  
On the other hand, in the total mass range  $5 M_{\odot} \lesssim  m \lesssim
40 M_{\odot}$, and {\it assuming} that one wishes an accurate 
frequency-domain (f-domain) representation of a
time-windowed signal
%\footnote
%{In the unlikely case where a f-windowing turns out to be a better model, 
%one still needs the formulas given in this paper to generate the f-domain SPA version of the P-approximants}
it is crucial to use  our new SPP approximants.
For $m \gtrsim 40 M_{\odot}$   a straightforward DFT is recommended
(but, anyway, the signal is not known with enough precision
in this high mass range, where the plunge and merger signals
become observationally important).

It is important to stress the position we assume in this paper:
Given the absence of any detailed and precise information about
the plunge signal today, we suggest that a time-truncated
chirp (time-windowed signal) is currently our best bet and the modified
SPA presented in this paper is the appropriate  Fourier-domain 
representation one must use.
However, this should not be taken to imply that  we are claiming to
have logically excluded the other possibility that
the f-window may turn out to be  the  better choice,   
when we have further  
details  about the transition from inspiral to  plunge 
and, about the plunge waveform.
Even so, we emphasize  that  a definitive contribution
of the present work is to provide explicitly  for  the first time   the frequency-domain
version of the time-domain P-approximants which were shown in DIS to bring   indispensible
improvements over the usually considered T-approximants. Consequently, even in the unlikely case where a 
straightforward frequency-window turns out to be a better model than the time-window assumed in most of this work, 
one will still require the formulas given in this paper [with the trivial change of replacing 
the correction factors ${\cal C}(\zeta)$ by a $\theta$ function $\theta(F_{\rm LSO}-f)$] to
generate sufficiently accurate f-domain filters.
Thus this work may  not be  the complete final answer but
only  a step ahead and 
 a partial contribution towards defining good f-domain filters.
Assumptions that seem the best we can accept,
require  special tools  for their  analysis   
and this paper provides them.

This paper is organised as follows:
In Sec.~\ref{sec:IIA}, \ref{sec:IIB}, \ref{sec:IIC}
 as a prelude to later  technical  material, 
we introduce several useful   physical notions and employ them
to give a preliminary discussion of the questions
raised by the detectability of massive-binary signals.
In Sec.~\ref{sec:TTC} we summarise the mathematical tools used
in the paper to estimate the time-truncated chirps.
In Sec.~\ref{sec:SPA} we consider time-windowed Newtonian-like
signals.
Sec.~\ref{sec:IIIA} provides a short summary of the well known SPA.
Sec.~\ref{sec:BEYOND}  sets up the basic equations to discuss
the Fourier transform of time-windowed signals.
Sec.~\ref{sec:PLUS} estimates the edge contribution to the 
Fourier transform coming from
the non-resonant integral.
This is followed by Sec.~\ref{sec:IIIB} and \ref{sec:IIIC} 
where we elaborate in detail the construction of
optimal analytic approximations to the Fourier transform
of the time-windowed gravitational wave chirp ({\it improved Newtonian} SPA).
In Sec.~\ref{sec:COMPARE} we compare and contrast in detail
the usual SPA with our improved SPA for Newtonian-like signals.
Sec.~\ref{sec:IVA} addresses the new  issues related to
the Fourier-transform of time-windowed {\it relativistic}
signals. 
In Sec~\ref{sec:IVB} we present a new  method to estimate the small
non-resonant contribution in the relativistic case.
 In Sec.~\ref{sec:IVC} we construct a new form of improved SPA for such signals
({\it improved relativistic} SPA). Combining this improved relativistic SPA
with the P-approximants of DIS leads to the construction of
the frequency-domain SPP approximants.
In Sec.~\ref{sec:FESPP} we use the SPP approximants constructed earlier and investigate
their {\it faithfulness} and {\it effectualness} in detail for the test mass case.
Based on this we  comment on the corresponding situation in the  comparable mass case. 
In Sec.~\ref{sec:COMPU} we compare  the
computing costs for  template generation using the time-domain FFT with corresponding
costs  for the frequency-domain SPA and improved SPA  both 
for Newtonian and relativistic cases.
Sec.~\ref{sec:SUM} contains our concluding remarks.

\section{ Preliminary Discussion}\label{sec:USE}

As a preface to the technical treatments of the following Sections in which we shall
construct optimal analytic approximations to the Fourier transform of the 
gravitational wave inspiral  signal $h(t)$
let us start by discussing some general issues which are central
to this paper.

\subsection{Wiener filters and time-truncated inspiral signals}
\label{sec:IIA}

We briefly recall the principle underlying the optimal linear
filter technique (Wiener filter). A (real) linear filter is a linear functional of
 the detector's output, $h_{\rm out}$, Eq.(\ref{sha1}), say
\begin{equation}
K[h_{\rm out}]=
\int_{-\infty}^{\infty} dt\, K(t)\, h_{\rm out}(t)=
\int_{-\infty}^{\infty} df\,\tilde{K}(-f) \tilde{h}_{\rm out}(f)=
\int_{-\infty}^{\infty} df\,\tilde{K}^*(f) \tilde{h}_{\rm out}(f)\,.
\label{d2}
\end{equation}
Let us associate to any $K(t)$ the time-domain function $k(t)$
such that its FT equals $\tilde{k}(f)\equiv S_n(f)\tilde{K}(f)$ and
let us introduce the Wiener scalar product
(defined on real time-domain functions)
\begin{equation}
\langle g,h\rangle\equiv \int_{-\infty}^\infty \frac{df}{S_n(f)}
\tilde{g}^*(f)\tilde{h}(f)=
\int_{-\infty}^{\infty}
\int_{-\infty}^{\infty}
\,dt_1\,dt_2\,g(t_1)\,w_1(t_1-t_2)\,h(t_2)\,,
\label{d3}
\end{equation}
\begin{equation}
{\rm where},\;\; w_1(\tau)=\int_{-\infty}^{\infty}
 \frac{df}{S_n(f)}\,e^{2 \pi i f\tau}=w_1(-\tau), 
\label{n1}
\end{equation}
is the convolution inverse of the noise correlation function $C_n(\tau)=C_n(-\tau)$
{\it i.e.} 
\begin{equation}
(w_1*C_n)(t)=\delta(t).
\label{n2}
\end{equation}
[Here $*$ denotes the convolution product:
$(g*h)(t)\equiv \int_{-\infty}^\infty d\tau\,g(\tau) h(t-\tau)$].
With this notation the action of the filter $K$ on
$h_{\rm out}$ reads
\begin{equation}
K[h_{\rm out}]=\langle k, h_{\rm out}\rangle\equiv S+N\,,
\label{d4}
\end{equation}
where $S$ is the filtered `signal' and $N$ the filtered `noise',
defined by
\begin{equation}
S\equiv K[h]=\langle k,h\rangle\,;\;
N\equiv K[n]=\langle k,n\rangle\,.
\label{d5}
\end{equation}
The definition of the symmetric Wiener scalar product Eq. (\ref{d3}) is
such that the statistical average of a product of filtered noises simplify:
$\overline{\langle k_1,n\rangle\langle k_2,n\rangle}= 
\overline{\langle k_1,n\rangle\langle n, k_2\rangle}= 
\langle k_1, k_2\rangle$. In particular, the variance of the
filtered noise $N$ reads 
$\overline{N^2}=\overline{\langle k, n\rangle^2}=\langle k,k\rangle$,
so that the square of the signal-to-noise ratio (SNR) for the filter defined by
any function $k(t)$ reads
\begin{equation}
\rho^2\equiv \frac{S^2}{\overline{N^2}}=\frac{\langle k, h\rangle^2}{\langle k,k\rangle}
=\vert {\cal O}(k,h)\vert^2\langle h,h\rangle\,,
\label{d6}
\end{equation}
where we have defined the `overlap', or normalized ambiguity function,
between $k$ and $h$ 
\begin{equation}
{\cal O}(k,h)\equiv \frac{\langle k,h\rangle}{\sqrt{\langle k,k\rangle
\langle h,h\rangle}}\,.
\label{d7}
\end{equation}
Schwarz's inequality guarantees that $\vert{\cal O}(k,h)\vert\le 1$,
the equality being reached only when $k(t)=\lambda h(t)$. (We work
here in the space of real signals.) For a given signal $h(t)$, the
choice of filter $K\leftrightarrow k$ which {\it maximizes}
the SNR $\rho$ is, in view of Eq. (\ref{d6}), $k(t)=\lambda h(t)$,
where the proportionality constant is unimportant and can be
taken to be
one (`Wiener Theorem').
This optimal linear filter theorem applies when the full time-development
of the signal $h(t)$ is known, the noise is stationary and
has a known spectral distribution $S_n(f)$.

It is important to note that, saying that the best `associated' filter
$k(t)$ is simply $k(t)=h(t)$, means that the best time-domain filter
$K(t)$, which must be directly correlated with the detector's output,
$K[h_{\rm out}]=\int _{-\infty}^\infty dt\,K(t)h_{\rm out}(t)$, 
is a non-local (in time) functional of $h(t)$.
Explicitly,
\begin{equation}
 K(t)=(w_1*h)(t)=\int_{-\infty}^\infty d\tau w_1(\tau) h(t-\tau).
\label{n3}
\end{equation}
This poses the question whether, for an off-line analysis of the data, one
would like to store a bank of these non-local time-domain filters
$K(t)$, which densely cover the expected parameter space.
In the present paper, we shall assume that one anyway computes (nearly) on-line
the Fast Fourier Transform (FFT) of the detector's output (which is anyway needed to factor out the
frequency dependent effect of the interferometer on the GW signal), and we shall
set ourselves the task of providing the best possible analytical representations
 of the Fourier transform of the expected signals $\tilde{h}(f)$.
Moreover, the availability of a fast Fourier algorithm makes the filtering
problem computationally less intensive in the Fourier-domain. It is well-known
that the computation of a discrete correlation  for all (discrete)
 time lags between the output and the filter [ie the discrete version
 of Eq. (\ref{d2}) requires  ${\cal O}[N^2]$ 
operations in the time-domain while it takes
only ${\cal O}[N \log_2(N)]$ operations in the Fourier domain 
 (because a time lag $\tau$ adds a factor $\exp( 2\pi i f \tau)$ in the
f-domain version of Eq. (\ref{d2}), 
which is equivalent to computing a certain inverse Fourier transform).
Discrete correlation
in the f-domain suffers from spurious correlations for non-zero time lags
but this is easily taken care of by padding the tail part of a template
with large number of zeroes (see, eg. Ref \cite{bs91} for details.)

\begin {figure}[t]
\caption{
The whitening  kernel, $w_{\frac{1}{2}}(t)$, Eq. (\ref{br12}),
 is shown [for the initial LIGO noise curve, Eq. (\ref{ligonoise})]  plotted as a function of time.
We see that it is quite sharply peaked at $t=0.$ The peak drops
quite rapidly as we move away from the origin thus indicating that
we have almost local-in-time filters. The curve also indicates the
importance of knowing the plunge signal over a time-scale of several
10's of milliseconds so that the inspiral signal can have a good
overlap.}
\centerline {\epsfxsize 6 true in  \epsfbox {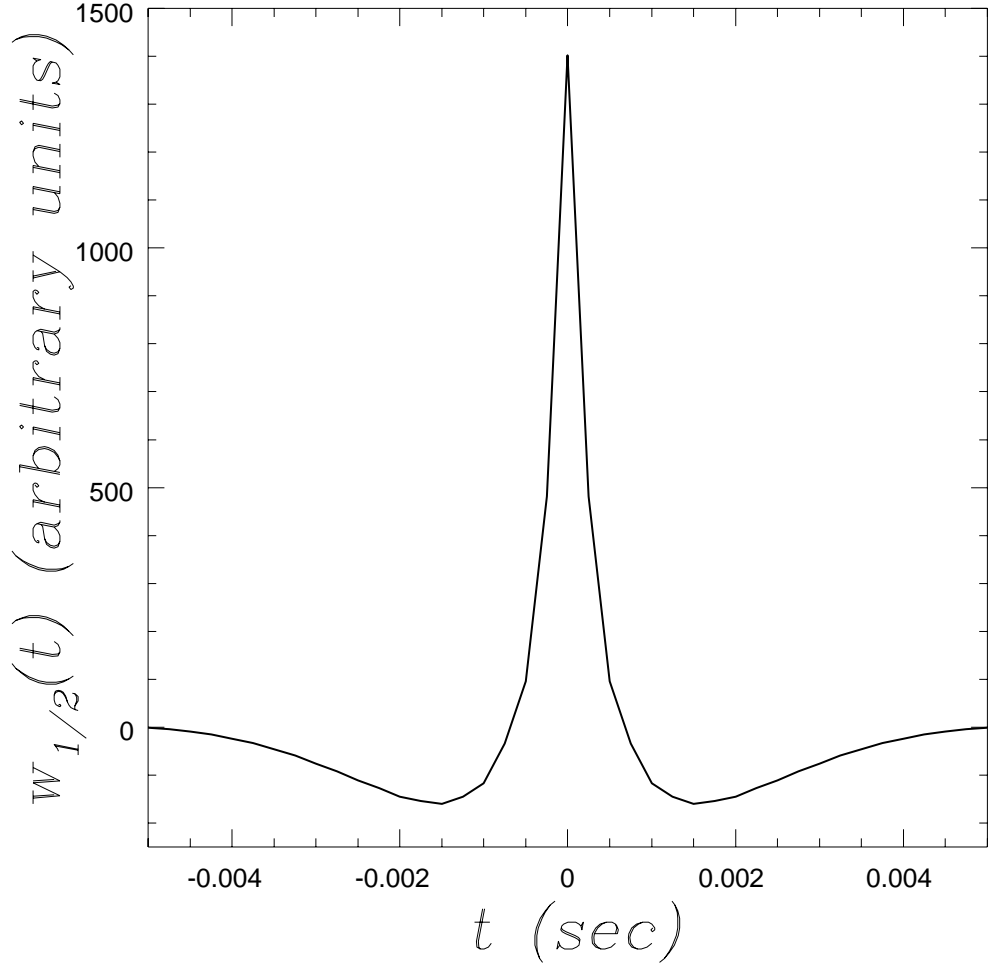} }
\label {fig:invft.snfm2}
\end {figure}

 As the problem of the  locality/non-locality in time
will be crucial to our discussion of the inspiral signals,
we wish to give an alternative discussion of Wiener's optimal
linear filter theorem.
Indeed, another proof of the theorem can be obtained by introducing a 
`whitening' transformation  say $w_{\frac{1}{2}}$, which simplifies the
properties of the noise. We define the `whitening-kernel'
$w_{\frac{1}{2}}(\tau)$ by
\begin{equation}
w_{\frac{1}{2}}(\tau)=\int_{-\infty}^{+\infty}
 \frac{df}{\sqrt{S_n(f)}}e^{2\pi i \tau f}\,;\;\; \tilde{w}_{\frac{1}{2}}(f)
\equiv\frac{1}{\sqrt{S_n(f)}}\,.
\label{br12}
\end{equation}
For any function $g(t)$, the action of the kernel $w_{\frac{1}{2}}$ on $g$ 
({\it i.e.} the `whitened' version of the function $g$) can be denoted by
\begin{equation}
g_{\frac{1}{2}}(t)\equiv (w_{\frac{1}{2}}*g)(t)=\int_{-\infty}^{\infty}
d\tau\, w_{\frac{1}{2}}(\tau)g(t-\tau).
\label{n4}
\end{equation}
The name `whitening kernel' comes from the fact that the transformed noise
$n_{\frac{1}{2}}(t)$ 
is `white' {\it i.e.} uncorrelated:
\begin{equation}
\overline{n_{\frac{1}{2}}(t_1) n_{\frac{1}{2}}(t_2)}= 
\int_{-\infty}^\infty df e^{2\pi i f(t_1-t_2)}=\delta(t_1-t_2)\,.
\label{br13}
\end{equation}
 Note that 
$w_{\frac{1}{2}}$ 
is simply the convolution square-root of the Wiener kernel
$w_1$ introduced above:
$w_{\frac{1}{2}}* w_{\frac{1}{2}}=w_1$.
The Wiener theorem states then that, after having whitened all the functions,
the optimal filter is simply the usual straightforward correlation
between the (whitened) output and the (whitened) signal {\it i.e.}
\begin{equation}
K_{\rm optimal}[h_{\rm out}]=
\int_{-\infty}^{\infty}\,dt\, 
h_{\frac{1}{2}}(t)\,
h_{\frac{1}{2}}^{\rm out}(t)\,,
\label{br14}
\end{equation}
where 
$h_{\frac{1}{2}}= w_{\frac{1}{2}}*h$, 
$h_{\frac{1}{2}}^{\rm out}= w_{\frac{1}{2}}*h^{\rm out}$.
In other words, we can think of the optimal filter as being local-in-time after the application, to all signals,
of the convolution kernel 
$w_{\frac{1}{2}}$.
When working with the transformed time-domain functions 
$h_{\frac{1}{2}}(t)$,
$h_{\frac{1}{2}}^{\rm out}(t)\cdots$, we shall say that we work in the `whitened time-domain'.
In this language, the best filter in the whitened time-domain is to simply correlate
(as when trying to visually superpose two time-functions) the output with a copy of the signal.
This `whitened time domain' is conceptually useful in the present context because it introduces
only a small non-locality (by small we mean here much smaller than the non-locality introduced
by the Fourier transformation).
Indeed, as the function $1/\sqrt{S_n(f)}$ has a rather flat maximum, its
Fourier transform 
$w_{\frac{1}{2}}(\tau)$
is nearly a delta function
as seen in Fig.~\ref{fig:invft.snfm2} wherein we have plotted the whitening kernel
for the initial LIGO interferometer.
More precisely, 
$w_{\frac{1}{2}}(\tau)$ is 
an even function
made of a positive spike around $\tau=0$, followed (on each side)
by a slightly negative wing which decays fast towards zero as $\vert\tau\vert\rightarrow +\infty$.
The half-width at half-maximum of the central spike is 0.18 ms.
The location of the wings is around $\tau=\pm 0.002$ s.
Therefore the non-locality contained in the whitening transformation is only 
between 0.2 ms and 2 ms (depending on the function on which it acts).
This non-locality together with the one and a  half cycle 
in $w_{\frac{1}{2}}$ 
is sufficient to efficiently damp both high and low frequencies
so that $h_{\frac{1}{2}}= w_{\frac{1}{2}}*h$ is 
a chirp whose amplitude is important only when the instantaneous frequency is around
$f_{\rm p}=165$ Hz (see below).
We shall use below this whitened time domain picture to discuss the important features of the 
expected chirp that we should try to model as well as possible.

In the present paper, we shall be primarily interested in massive compact 
binaries with total mass $m= m_1+m_2$ in
the approximate range $3 M_\odot\lesssim m\lesssim 40 M_\odot$.
We recall that the GW signal from a compact binary is made of
an inspiral signal, followed, after the last stable orbit is reached, 
by a plunge signal which leads to a final merger signal.
Thanks to the analytical work on the motion \cite{td83} of and GW emission
\cite{bdiww95,test} from general relativistic binary systems we have quite a good
analytical control of the inspiral signal.
In the present paper, we shall further argue that we have also a rather
good analytical control on the location of the last stable orbit (LSO),
{\it i.e.} on the transition
between the inspiral and plunge.
First, DIS \cite{dis98} introduced a new, more robust approach
to the determination of LSO based on the invariant function
$e(v)$.
More recently \cite{bd99}, a new approach to the dynamics of
binary systems has confirmed the result of DIS (saying
that the LSO was slightly more `inwards' for comparable
mass systems than as predicted by the test-mass limit) 
and predicts values for the important physical
quantities at the LSO (notably the  orbital frequency)
which are even nearer to the (Schwarzschild-like)
ones obtained in the test-mass limit.
We anticipate that further analytical
progress in the problem of motion
of binary systems \cite{3PN}, combined with the 
LSO-determination techniques given in
\cite{dis98} and \cite{bd99,djsprep} will  soon allow one to know with 
more certainty 
and more precision the gravitational wave frequency at the LSO.
Pending such a determination, we shall use 
as fiducial value for the GW frequency at the LSO 
--- when we need it for simple analytical estimates --- 
the usual `Schwarzschild-like' approximation
\begin{equation}
F_{\rm LSO}=4400\left(\frac{M_\odot}{m}\right)\;{\rm Hz}\,.
\label{d8}
\end{equation}
However, in our actual numerical calculations and plots we shall use the
$\eta$-dependent $F_{\rm LSO}$ corresponding to the approximation used for
the energy function $E_{\rm A}(v)$.
For instance, in the case of the 2PN P-approximant $P_4$ we have \cite{dis98}
\begin{equation}
F_{\rm LSO}^{P_{4}}=4397.2\left(\frac{1+\frac{1}{3}\eta}{1-\frac{35}{36}\eta}\right)^{\frac{3}{2}}
\left[2-\frac{1+\frac{1}{3}\eta}{\sqrt{1-\frac{9}{16}\eta +\frac{1}{36}\eta^2}}\right]^{\frac{3}{2}}\frac{M_\odot}{m}\;{\rm Hz}
\,.
\label{dis23}
\end{equation}
In the equal-mass case ($\eta = 1/4$) this yields $F_{\rm LSO} = 5719.4 M_\odot /m$ Hz. 
Note that the most recent determination of the LSO \cite{bd99} suggests
that when $\eta\neq 0$ the GW frequency at the LSO  lies between
Eq. (\ref{d8}) and Eq. (\ref{dis23}):
\begin{equation}
F_{\rm LSO}=4397.2 (1+0.3155\eta) \left(\frac{M_\odot}{m}\right)\;{\rm Hz}\,.
\label{d8a}
\end{equation}

Preliminary studies indicate 
that the plunge signal, emitted during the fast fall of the
two masses towards each other following the crossing of the LSO, 
will last (when $4\eta\sim 1$)  only for a fraction of an orbital period 
(see  \cite{bd99} and \cite{bdprep})\footnote{ 
We assume here that we are in the generic case where
the spins of the coalescing objects are smallish compared to their
maximal Kerr value.}.
As usual, one can also assume that 
the subsequent merger signal linked to the formation
of a black hole of total mass $\sim m$ contains significantly higher
frequencies than the inspiral ones.
Indeed, the characteristic frequency of the merger signal may be taken  
to be given by the real part of the most slowly damped quadrupolar normal mode of a
black  hole (which  when neglecting the black hole spin, 
has a complex circular frequency $\omega_1 m_{\rm bh}=
0.37367 -0.08896\, i$ \cite{cd75}), {\it i.e.} (with $m_{\rm bh}\sim m$)
\begin{equation}
f_{\rm merger}\sim f_{\rm bh}\simeq \frac{0.374}{2\pi m}\simeq 12000\,\frac{M_\odot}{m}\,
{\rm Hz}\,.
\label{d9}
\end{equation}
Eqs. (\ref{d8}) and (\ref{d9}) lead us to accept that there is a significant
frequency separation between the inspiral and plunge signals and the merger one:
$f_{\rm merger}/F_{\rm LSO}\sim 2.75$.
Therefore, if we restrict our attention to systems such that the characteristic
detection frequency $f_{\rm det}$ defined by the
noise curve, stays logarithmically nearer to $F_{\rm LSO}$ than
to $f_{\rm merger}$,  it seems plausible
that a good filter to use for GW detection can neglect the (ill-known)
merger signal,
but should try to model as accurately as possible the inspiral and plunge signal.
For the  initial LIGO noise curve, Eq. (\ref{ligonoise}),  
the characteristic detection frequency $f_{\rm det}$   is  167 Hz.
It is then for a total mass $m\le 43.5 M_\odot$ that 
$f_{\rm det}/F_{\rm LSO}\le f_{\rm bh}/f_{\rm det}$.

We shall see below that, just before reaching the LSO, the inspiral signal is still
significantly `quasi-periodic' (with $\gtrsim 6$ cycles before a significant change in 
instantaneous frequency). 
By contrast, though the plunge signal may not decay
monotonically and may be oscillating,
 it seems reasonable to assume that the plunge 
 lasts only for  a fraction of the orbital  period $T_{\rm LSO}=2 F_{\rm LSO}^{-1}$.
Thus, in the absence of a precise knowledge of
the plunge signal,  a good model of the time-domain signal consists 
in abruptly shutting off, by a step function $\theta(t_{\rm LSO}-t)$
the (adiabatic)
 inspiral signal beyond the time $t_{\rm LSO}$ when the last stable
orbit is reached.
We also tested formally the robustness of our approach by showing that our model above has a good overlap with a signal which decays smoothly on a time scale of a few (up to 3) $F_{\rm LSO}^{-1}$ beyond the LSO.
Because of the likely oscillatory behaviour of the plunge signal,
details of the oscillations are necessary for any further improvements and
we are currently working towards improving our understanding of the
transition between the inspiral and the plunge \cite{bdprep}.

These  considerations motivate us to propose that, in the absence of knowledge of
{\it the} optimal filter which should be $k_{\rm optimal}(t)=k_{\rm exact}(t)$,
our best bet is to use as (sub-optimal) filter 
the time-truncated inspiral signal 
$h_{\rm inspiral}(t)\theta(t_{\rm LSO}-t)$.
In other words, we think that the best strategy is to use all the
information available, in the time-domain, about the signal and to replace the
transient plunge and higher frequency merger signals by zero, 
as a measure of our current ignorance. But having settled on this tactic in the time-domain, the aim of this paper
is to provide the best possible frequency-domain description of such a time-windowed signal.
We shall see in detail below that the Fourier transform 
$\tilde{h}_{\rm tw}(f)={\rm FT}[h_{\rm tw}(t)]$ of the time-windowed signal
$h_{\rm tw}(t)\equiv h_{\rm inspiral}(t)\theta (t_{\rm LSO} -t)$
has a non-trivial structure
which is not captured by the usually considered frequency-windowed stationary
phase approximation $\tilde{h}_{\rm spaw}(f)$.
In particular, for massive systems, a significant fraction of the total power 
is contained in the `tail' of $\tilde{h}_{\rm tw}(f)$ beyond $f=F_{\rm LSO}$.
The general result Eq. (\ref{d6}) for SNR obtained with any filter
then says that the Fourier-domain filter $\tilde{h}_{\rm tw}(f)$,
though sub-optimal, should be  still significantly  {\it better} than $ \tilde{h}_{\rm spaw}(f)$ because
(under our assumptions about the plunge + merger signal) it has better overlaps
with the exact signal.
We shall return below to this important issue and give further arguments
(in the time-domain) to confirm the superiority of $\tilde{h}_{\rm tw}(f)$ over
$\tilde{h}_{\rm spaw}(f)$ (see Fig.~\ref{fig:visual.newt1} 
and Fig.~\ref{fig:visual.newt2} and text around it).

\subsection{The number of useful cycles}
\label{sec:IIB}

Often one mentions that, in the total time development of an inspiral signal,
the total number of gravitational wave cycles 
\begin{equation}
N_{\rm tot}=\frac{1}{2\pi}(\phi_{\rm end} - \phi_{\rm begin})=\int_{F_{\rm beg}}^{F_{\rm end}}\,dF\,\left(\frac{1}{2\pi}
\frac{d\phi}{dF}\right)\,,
\label{2.1}
\end{equation}
is very large. Here $\phi$ is the gravitational wave phase, $\phi_{\rm end}$ is the phase at the
end of the inspiral regime (defined by the last stable orbit for sufficiently
massive systems, {\it i.e.} for the black-hole-neutron-star and the
black-hole-black-hole systems), while $\phi_{\rm begin}$ is the 
phase when the signal enters the lower frequency (seismic) cutoff of the detector
bandwidth. We have also rewritten $N_{\rm tot}$ as an integral over the running instantaneous
gravitational wave frequency $F$. However, the large number $N_{\rm tot}$, Eq. (\ref{2.1}),
is not significant because the only really {\it useful} cycles are those which contribute
most to the signal-to-noise ratio (SNR). To have a clearer idea of what
one might want to mean by  
the notion of a {\it useful number of cycles}, 
let us first introduce the {\it instantaneous number of cycles} spent near some 
instantaneous frequency $F$. It is naturally defined by multiplying the integrand in 
 Eq. (\ref{2.1}) by $F$, considered as the length of an interval 
$\pm\Delta F=\pm \frac{F}{2}$ around
$F$ {\it i.e.}
\begin{equation}
N(F)\equiv\frac{F}{2\pi}\frac{d\phi}{dF}\equiv \frac{F^2}{dF/dt}\,,
\label{2.2}
\end{equation}
where we have used $d\phi/dt\equiv2\pi F(t)$.
Note that the instantaneous $N$ can be considered either as a function of the running frequency $F(t)$, or,
directly, of time.

The instantaneous number of cycles plays an important role both in defining the observability of a signal,
and in controlling partially the validity of the stationary phase approximation. 
The square of the {\it optimal} SNR reads
\begin{equation}
\rho^2\equiv\left(\frac{S}{N}\right)^2=\int_{-\infty}^{+\infty}\,df\,\frac{|\tilde{h}(f)|^2}{S_n(f)}\,.
\label{2.4}
\end{equation}
In the stationary phase approximation (discussed at length and improved below; but here we use standard results
for orientation) the modulus of the Fourier transform of the real signal 
$h(t)=2 a(t) \cos \phi(t)$, reads
$|\tilde{h}(f)|\simeq a(t_f)/\sqrt{\dot{F}(t_f)}$ where $t_f$ is the time when the
instantaneous frequency $F(t)$ reaches the value $f$.\footnote{
In the following, it will be necessary to distinguish
carefully between the instantaneous frequency $F$ and the  Fourier variable $f$.
In the present Section this distinction is not very important and we shall
freely change notation $f\leftrightarrow F$.
Similarly, 
the gravitational wave flux and factored   flux functions which 
following standard notation was denoted by
$F(v)$ and $f(v)$ in \cite{dis98} are denoted  by ${\cal F}(v)$ and $l(v)$
 in this paper
to avoid confusion with the instantaneous gravitational wave frequency
and Fourier variable respectively.}
Therefore, the squared modulus can be written as 
\begin{equation}
|\tilde{h}(f)|^2\simeq \frac{a^2(f)}{df/dt}\equiv \frac{1}{f^2}\,N(f) \,a^2(f)\,.
\label{2.5}
\end{equation}
Finally, the SNR can be rewritten as 
\begin{equation}
\rho^2\equiv\left(\frac{S}{N}\right)^2=\int_{-\infty}^{+\infty}\,\frac{df}{f} \;
\frac{N(f) a^2(f)}{f S_n(f)}
=\int_{-\infty}^{+\infty}\,\frac{df}{f}\,\frac{h_s^2(f)}{h_n^2(f)}\,,
\label{2.6}
\end{equation}
where
we have introduced the notation $h_s^2(f)\equiv N(f) a^2(f)$ and
 $h_n^2(f)\equiv f S_n(f)$.  Here,   $h_n^2(f)$ is  the usual
 squared amplitude of  the {\it effective gravitational wave 
noise} at the frequency $f$, {\it i.e.} the minimum gravitational wave
  amplitude detectable in a bandwidth $\pm f/2$
around frequency $f$. Eq. (\ref{2.6}) exhibits that the squared amplitude of the corresponding
 {\it effective  gravitational wave signal} is
$h_s^2(f)\equiv N(f) a^2(f)=f^2 \vert\tilde{h}(f)\vert^2$, {\it i.e.} that the ``bare'' amplitude 
$ a(f)\equiv a(t(f))$ is effectively
multiplied by $\sqrt{N(f)}$ \cite{kt87,bs91}.
Eq. (\ref{2.6}) also  exhibits  the relative weight with which each cycle counts for
detectability purposes. Per logarithmic frequency interval this weight is simply
\begin{equation}
w(f)\equiv a^2(f)/h_n^2(f)\,.
\label{2.7}
\end{equation}
Therefore it is natural to define the number of {\it useful} cycles as
\begin{equation}
N_{\rm useful}\equiv \left( \int_{F_{\rm min}}^{F_{\rm max}}\,\frac{df}{f}\,w(f)\, N(f)\right) 
\left(\int_{F_{\rm min}}^{F_{\rm max}}\,\frac{df}{f}\,w(f)\right)^{-1} \,,
\label{2.8}
\end{equation}
where $F_{\rm min}$ is the low-frequency seismic-cutoff below which $h_n^2(f)$ is
essentially infinite and where the upper-cutoff $F_{\rm max}$ is the frequency at which the signal
itself shuts off.
For illustration, we list in Table~\ref{table:cycles} the number of {\it useful} cycles
and the {\it total} number of cycles for representative systems and orders of approximation.
For Newtonian chirps the total number of cycles is
\begin{equation}
N_{\rm tot}^{\rm Newt}= \frac      
{(\pi m f_{\rm s})^{-5/3} - (\pi m f_{\rm max})^{-5/3}} 
{32 \pi \eta }\,,
\label{bs1}
\end{equation}
where $f_{\rm s}$ is the  seismic cutoff.
The total number of cycles for relativistic chirps is always smaller than
$N_{\rm tot}^{\rm Newt}$.
From  Table~\ref{table:cycles} it is clear that
the number $N_{\rm useful}$ is usually much smaller than $N_{\rm tot}$, Eq. (\ref{2.1}).
 Note that for massive systems $N_{\rm useful}$ becomes quite small.
The number of useful cycles given in Table \ref{table:cycles}  have been
computed
for the initial LIGO noise curve Eq. (\ref{ligonoise}). 
The corresponding numbers for the VIRGO noise curve Eq. (\ref{virgonoise})
 would be larger both because the VIRGO sensitivity curve peaks at
a lower frequency, and because it is flatter.

\begin {figure}[h]
\caption{ The Newtonian  instantaneous number of cycles $N(f)$, 
 the square of the amplitude $ a^2(f)$, 
their product $N(f) a^2(f)$, 
the reciprocal of the effective GW noise $h_n^2(f)= fS_n(f),$ 
and  $ d(SNR)^2/d(\log f)$
 are plotted as a function of $f$. 
The scale on the y-axis corresponds to $N(f)$ on the left and all other
functions are plotted on an arbitrary scale indicated on the right. 
The top panel is for a lighter mass binary $(m_1=1.4M_\odot,$ $m_2=10M_\odot)$ 
and the lower 
one for a heavier mass system ($m_1=m_2=10M_\odot$).}
\centerline {\epsfxsize 6 true in  \epsfbox {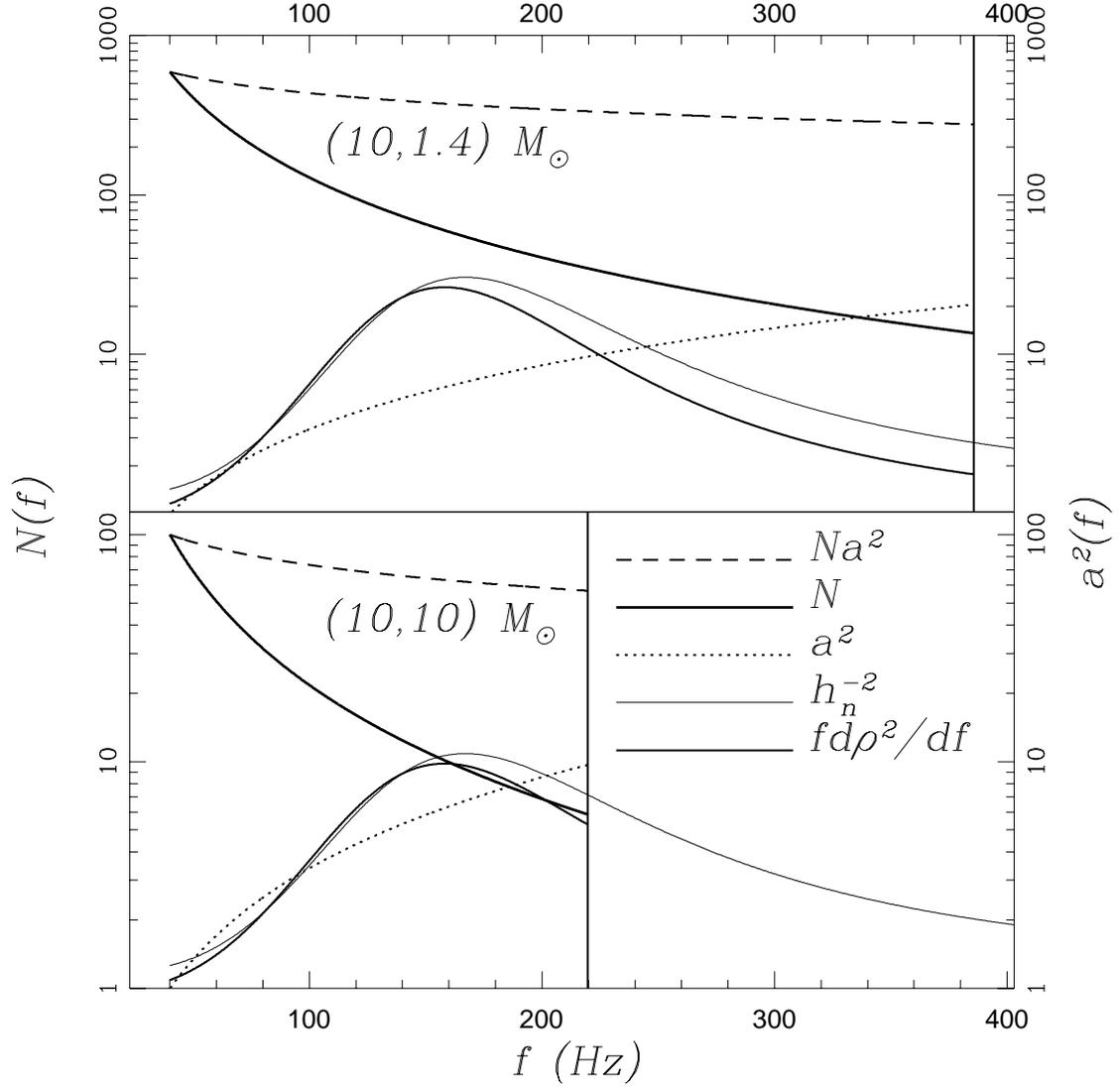} }
\label {fig:useful.cycles}
\end {figure}

To  explore in more detail the case  of  systems with  a small number of useful cycles 
we display in
Fig.~\ref{fig:useful.cycles} (on a linear-log plot)
 the various factors of the logarithmic integrand on the RHS of
Eq. (\ref{2.6}) for two different binary systems.
The instantaneous number of cycles $N(f)$ in the Newtonian approximation
is plotted,
together with 
square of the amplitude $ a^2(f)$,
their product the {\it effective} gravitational wave amplitude 
$h_s^2(f)=N(f)a^2(f)$, the reciprocal of 
effective noise $h_{\rm n}^2(f)=f S_n(f)$, 
(cut off after $F_{\rm max}=F_{\rm LSO}$) and the
power per log bin of the square of the SNR $d\rho^2/d(\log f)$. 
 On the left, the scale on the y-axis corresponds to $N(f)$. 
On the right it corresponds to the amplitude on an arbitrary scale.
Other quantities are on an arbitrary scale.
The top panel is for a lighter mass binary 
($m_1=1.4M_\odot$, $m_2=10M_\odot$)
and the bottom panel for a heavier one  
($m_1=m_2=10M_\odot$).

The last Figure exhibits two useful lessons which are well-known but are particularly important
to keep in mind when reading the present paper. First, because of the mass-scaling of the
 gravitational wave  frequency at the last stable orbit, 
given in the lowest (Schwarzschild-like) approximation  by Eq. (\ref{d8}),
it is only for systems with total mass $m=m_1+m_2\gtrsim 13 M_{\odot}$ 
that the {\it peak of the SNR logarithmic  frequency-distribution} $f_p$   becomes comparable,
within a factor of two, to 
$F_{\rm max} =F_{\rm LSO}$.
This statement critically depends on the characteristic frequency entering
the considered noise-curve. For instance,
in Fig.~\ref{fig:useful.cycles} we have used the initial LIGO curve Eq. (\ref{ligonoise})
for which $f_p=0.825 f_0=165$ Hz.
Note that the peak of the logarithmic SNR integrand, $f_p$, is very close to the 
minimum of the effective noise amplitude $h_n^2(f)=fS_n(f)$, which is located
(as mentioned above) at $f_{\rm det}=0.8347 f_0\simeq 167$ Hz.
This is because, the frequency dependence of the factor 
$N(f)\propto (\eta v^5)^{-1}\propto f^{-5/3}$
(which in the effective signal $h_s^2(f)=N(f)a^2(f)$,
favours lower frequencies) is nearly compensated by the frequency
dependence of the bare amplitude $a^2(f)\propto v^4\propto f^{4/3}$ 
(which favours higher frequencies).

\begin {figure}[h]
\caption{ Instantaneous number of cycles, Eq.(\ref{2.2}), near the 
LSO as a function of time for the Newtonian case, Eq.~(\ref{2.11}),
and the  relativistic cases, Eq.~(\ref{2.13}), defined by the 2PN approximant  $P_4$.
Also plotted is the  development of the wave-form $h_{P_4}(t)$ on an arbitrary scale.
The plots show how the number of useful cycles diminishes as one gets close
to the LSO. The Figure refers to $m_1=m_2=20 M_\odot$.}
\centerline {\epsfxsize 6 true in  \epsfbox {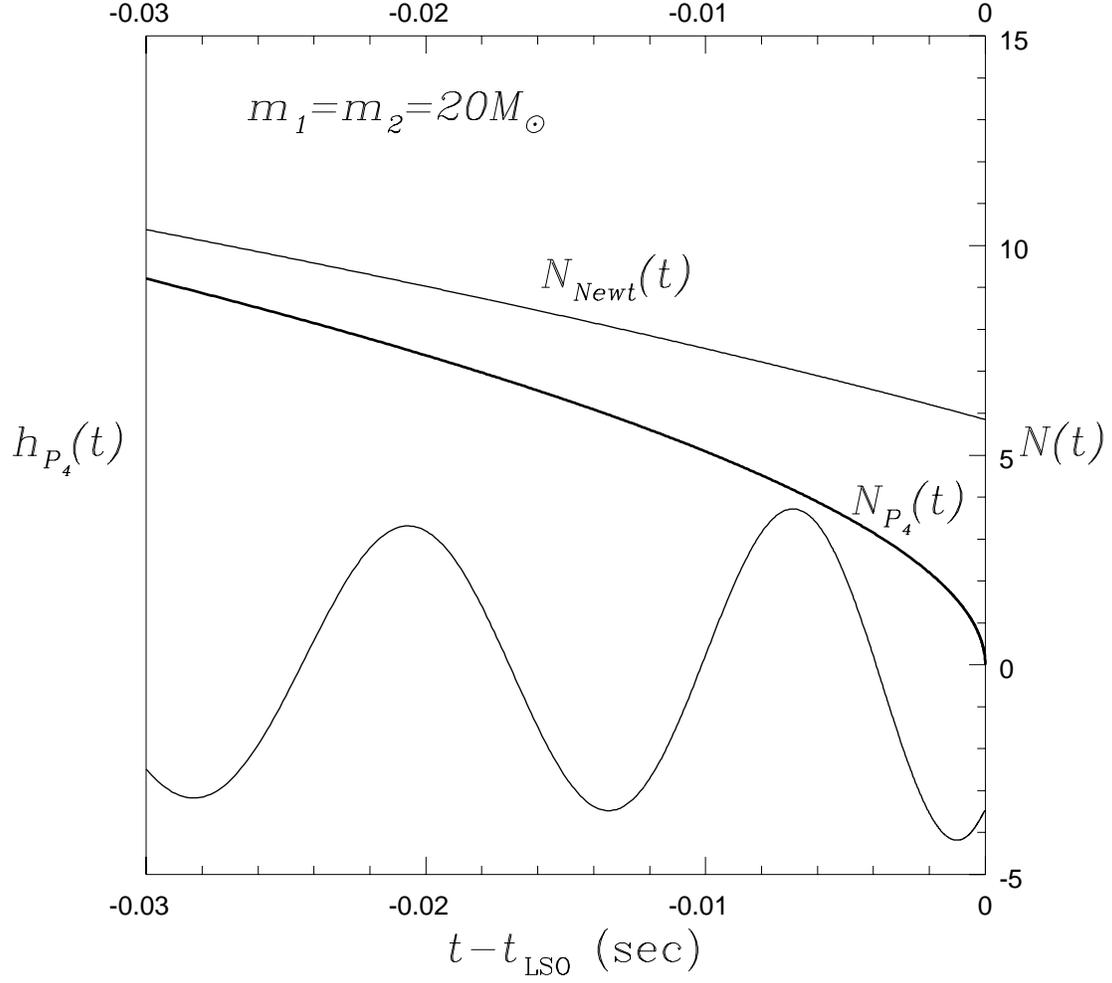} }
\label {fig:cycles}
\end {figure}

A second lesson, to be drawn from Fig.~\ref{fig:useful.cycles} is that the number
of useful cycles also becomes small in the same problematic case of
massive systems.
To see this more clearly, let us write down the
explicit expression for the instantaneous number of cycles.
In the Newtonian case (for which the basic formulas are recalled in Sec.~\ref{sec:IIIA} below), one has
\begin{equation}
N_{\rm Newtonian}(f)=\frac{5}{24 \pi} \frac{1}{4\eta} \frac{1}{v^5}\,,
\label{2.11}
\end{equation}
where $v=(\pi m f)^{1/3}$.
The lowest value of $N$ is physically that formally reached at the upper
frequency-cutoff $f=F_{\rm max}=F_{\rm LSO}$. For
$v_{\rm LSO}=1/\sqrt{6}$ (the ``Schwarzschild'' value), the above equation reads
\begin{equation}
N_{\rm Newtonian}(F_{\rm LSO})=\frac{5 }{24 \pi} 6^{5/2} \frac{1}{4\eta}\simeq \frac{5.8477}{4\eta}\,, 
\label{2.12}
\end{equation}
where we recall that $\eta\le 1/4$ and that the upper value $\eta_{\rm max}=1/4$ is reached
for equal mass systems $m_1=m_2$. Therefore for comparable-mass, massive systems, the Newtonian approximation suggests that
the useful number of cycles will be rather small 
$(\sim 6)$ and concentrated near the LSO.
As we shall see later, if we were interested in estimating the Fourier
transform (FT) of an
analytic Newtonian-like signal, even such a smallish number of cycles (and
even a smaller one, down to $N\sim 1)$ would be enough to ensure that
the leading correction to the stationary phase approximation is small.
However, the complication comes from the combination of two facts:
(i) the signal essentially terminates at the LSO crossing time $t_{\rm LSO}$,
and (ii) one crucially needs 
a relativistic description of the evolution near the LSO.
Using the formulas and the notation of  Sec.~\ref{sec:SPP} below, 
we find that the relativistic prediction for the instantaneous
number of cycles (in the adiabatic inspiral approximation) is
\begin{equation}
N_{\rm relat}(f)= - \frac{1}{3\pi} v^4 \frac{E'(v)}{{\cal F}(v)}\,.
\label{2.13}
\end{equation}
By definition of the LSO (see e.g. the discussion in DIS) the derivative $E'(v)$
vanishes at $v=v_{\rm LSO}$. Therefore,
the instantaneous number of
cycles is smaller in the
relativistic case than in the Newtonian one and actually tends to
{\it zero} near the LSO. We shall tackle  in Sec.~\ref{sec:SPP} below the problem that this
vanishing of $N(F_{\rm LSO})$ causes for the stationary phase approximation.
In this introductory Section let us only illustrate the problem by plotting 
the Newtonian and relativistic values of $N(F(t))$. In Fig.~\ref{fig:cycles}
we plot the instantaneous number of cycles for the Newtonian and second post-Newtonian
$P$- approximant waveforms in the last few cycles of the binary
inspiral for a 
$ (20 M_\odot$, $20 M_\odot)$ system.
We also show the development of the waveform in this interval on  an arbitrary scale.
These plots demonstrate how the number of useful cycles diminishes as one gets
close to the LSO and lead us to anticipate the subtleties in the
 detectability of signals whose
LSO is near the most sensitive part of the frequency response
of the detector.

\subsection{ Loss of SNR due to edge effects}
\label{sec:IIC}

As we already mentioned, in addition to the problem of a vanishing
instantaneous number of cycles near the LSO in the relativistic case,
the main problem with the acccuracy of the stationary phase approximation
comes from the fact that the Fourier transform of a time-windowed signal
$\tilde{h}_{\rm tw}(f)={\rm FT}[h_{\rm tw}(t)]$ differs from
the frequency-windowed SPA $\tilde{h}_{\rm spaw}(f)=\theta(F_{\rm max}-f)
\tilde{h}_{\rm spa}(f)$ because of some 
`edge effects'  in the frequency-domain, linked to the
abrupt termination of the signal in the time-domain.
These edge-effects comprise some additional oscillatory 
behaviour in $\tilde{h}(f)$ for $f<F_{\rm LSO}$,
as well as a decaying oscillatory tail in the usually
disregarded frequency-domain for $f>F_{\rm LSO}$.

Let us here anticipate on the results below and use a first-order approximation to discuss the main features 
of the corrections brought by the time-windowing. Roughly (see below) the exact Fourier
transform can be written as 
\begin{equation}
\tilde{h}_{X}(f)\simeq \tilde{h}^{\rm spa}_{\rm win}(f) + \varepsilon(f)\,,
\label{2.14}
\end{equation}
where $\tilde{h}^{\rm spa}_{\rm win}(f)= \tilde{h}^{\rm spa}(f) \theta(F_{\rm max}-f)$ 
is the usually considered {\it  frequency-windowed}
SPA.  The difference $\varepsilon(f)$ is approximately of the form
\begin{equation}
\varepsilon(f)\simeq {\cal D}(f) \tilde{h}^{\rm spa}(f)\,,
\label{2.15}
\end{equation}
\begin{equation}
{\rm where},\;\; {\cal D}(f)\equiv {\cal C}(f) -\theta(F_{\rm max} -f)\,,
\label{2.16}
\end{equation}
with the correction factor ${\cal C}(f)$  given by Eq. (\ref{3.20}) below (with any
choice of $\zeta(f)$ in the present approximation) and where $\tilde{h}^{\rm spa}(f)$
is some smooth continuation of $\tilde{h}^{\rm spa}_{\rm win}(f)$ from the domain $f<F_{\rm LSO}$
to the domain $f>F_{\rm LSO}$. (For the present purpose one can think that $\tilde{h}^{\rm spa}(f)$
is simply given by the Newtonian approximation).

\begin {figure}[t]
\caption{ The real and imaginary parts of the complex correction factor
${\cal C}(\zeta)$ --- in terms of which the new  frequency domain approximants are
represented in the Newtonian-like cases --- as a function of $\zeta$.
The real part is a `softened step' while the imaginary part an oscillatory
function vanishing at the origin and at large positive and negative values
of its argument.}
\centerline {\epsfxsize 6 true in  \epsfbox {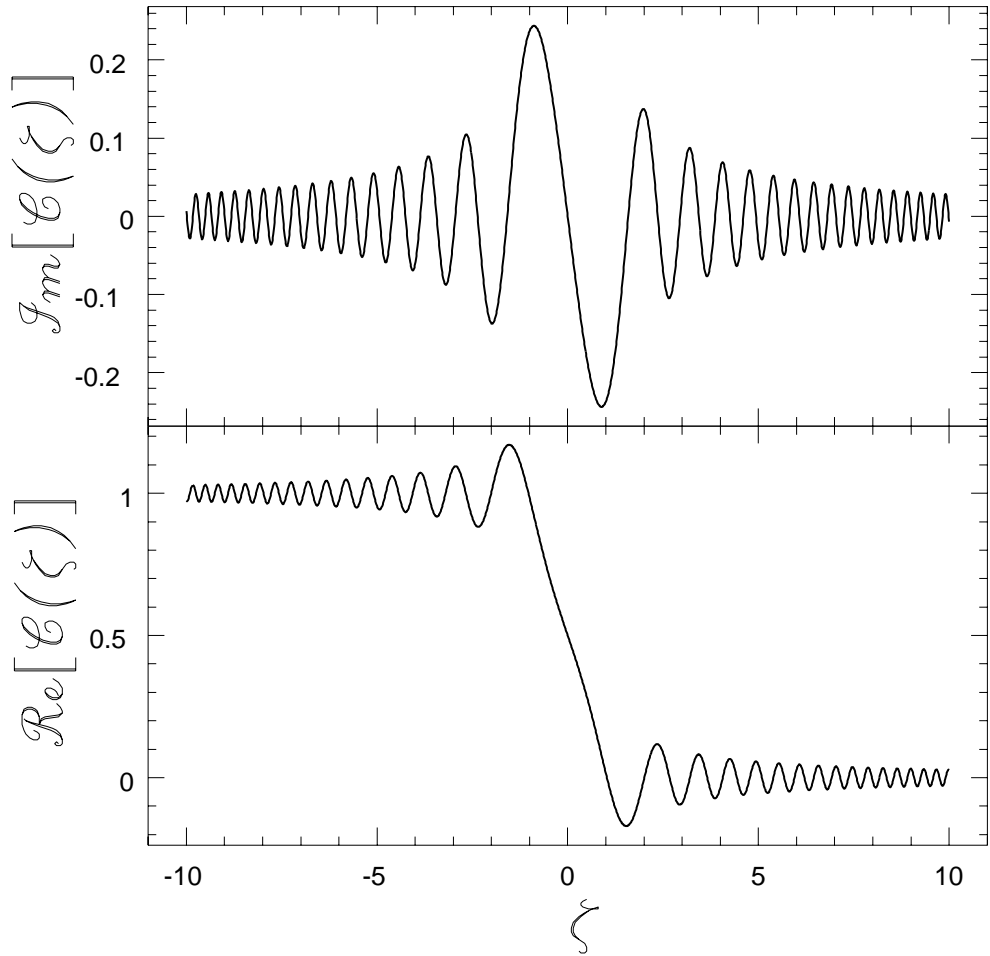} }
\label {fig:soft.step}
\end {figure}

Starting from Eq. (\ref{2.14}) one can compute the overlap
between the exact $\tilde{h}(f)$ and the usually considered 
frequency-windowed SPA  
$\;\tilde{h}^{\rm spa}_{\rm win}(f) $
\begin{equation}
{\cal O}=\frac{\langle h_X\,,\,h_{\rm win}^{\rm spa}\rangle}
{ \sqrt{\langle h_X\,,\,h_X\rangle\langle h_{\rm win}^{\rm spa}\,,\,h_{\rm win}^{\rm spa}}\rangle}\,.
\label{2.17}
\end{equation}
As recalled above [see Eq. (\ref{d6})] this overlap, if it is significantly
smaller than one, represents a loss in SNR.
To lowest order in $\varepsilon$ the overlap Eq. (\ref{2.17}) differs from $1$ by
\begin{equation}
1-{\cal O}\simeq\frac{1}{2\Vert h_X\Vert ^2}( \Vert\varepsilon\Vert^2 - 
\vert\langle\varepsilon\,,\,\hat{h}_X\rangle\vert^2)\,,
\label{2.19}
\end{equation}
where $\Vert\varepsilon\Vert^2\equiv \langle\varepsilon\,,\,\varepsilon\rangle$, and 
$\hat{h}_X\equiv h_X/\Vert h_X\Vert $. In inserting the explicit result 
Eqs.(\ref{2.15}) and (\ref{2.16})
for $\varepsilon$ one sees that the second term on the RHS of Eq. (\ref{2.19}) is much smaller
than the first (because the oscillations in ${\cal D}(f)$ are integrated against
the smooth variation of $\tilde{h}^{\rm spa}(f)$). Finally, if we define
the weight function
\begin{equation}
\sigma(f)\equiv\frac{f |\tilde{h}^{\rm spa}(f)|^2}{S_n(f)}\simeq \frac{N(f) a^2(f)}{h_n^2(f)}
=\frac{h_s^2(f)}{h_n^2(f)}\,,
\label{2.20}
\end{equation}
which is the full logarithmic weight function appearing in the squared SNR, 
Eq. (\ref{2.6}), we can
write 
\begin{equation}
1-{\cal O} \simeq \frac{1}{2}\;\; { \int_0^\infty\, \frac{df}{f}\, \sigma(f)\,|{\cal D}(f)|^2}
\left( {\int_0^{F_{\rm max}}\, \frac{df}{f}\, \sigma(f) }\right)^{-1}.
\label{2.21}
\end{equation}
As will be discussed later (see Fig.~\ref{fig:soft.step})
the function ${\cal D}(f)={\cal C}(f)-\theta(F_{\rm max}-f)$ is concentrated
in an interval of order $\sqrt{\dot{F}(t_{\rm max})}$ around $f\simeq F_{\rm max}$ and decays on
both sides [like $1/\zeta(f)\propto 1/(f-F_{\rm max})$] when $f$ gets away from $F_{\rm max}$.
The total integral of $|{\cal D}(f)|^2$ is finite and of order unity.
Thus, we see from Eq. (\ref{2.21}) that when the characteristic frequency $f_{\rm p}$ around which $\sigma(f)$
is concentrated satisfies $f_{\rm p}\ll F_{\rm max}$ we shall have a rough estimate
\begin{equation}
1-{\cal O}\sim \frac{\sigma(F_{\rm max})}{\sigma(f_{\rm p})} \frac{\sqrt{\dot{F}(t_{\rm max})}}{F_{\rm max}}
=\frac{ \sigma(F_{\rm max})}{\sigma(f_{\rm p})} \frac{1}{\sqrt{N(F_{\rm max})}}\,,
\label{2.22}
\end{equation}
while in the opposite limit where $f_{\rm p}\gg F_{\rm max}$, we get
the rough estimate
\begin{equation}
1-{\cal O}\sim \frac{\sqrt{\dot{F}(t_{\rm max})}}{F_{\rm max}}
=\frac{1}{\sqrt{N(F_{\rm max})}}\,.
\label{2.23}
\end{equation}
In the case where $f_{\rm p}\ll F_{\rm max}$ the factor $\sigma(F_{\rm max})/\sigma(f_{\rm p})$ in the RHS of
Eq. (\ref{2.22}) is very small.
Therefore, even if the number of cycles $N(F_{\rm max})$ is not very large, 
Eq. (\ref{2.22}) predicts
that a simple frequency-windowed SPA will have excellent overlap with the exact $\tilde{h}(f)$.
On the other hand, Eq. (\ref{2.23}) shows that in the reverse limit $f_{\rm p}\gg F_{\rm max}$ which means in fact when $f_{\rm p}
\gtrsim F_{\rm max}$, the overlap will become bad if $\sqrt{N(F_{\rm max})}$ is
not very large.
As we have seen that $\sqrt{N(F_{\rm max})}$ gets as low as $\sqrt{5.85}\simeq 2.4$ in the Newtonian case, 
and reaches smaller values in the relativistic case, we expect that the cases where the frequency-windowed SPA has a
bad overlap with $\tilde{h}(f)$ are those where $f_{\rm p}$ becomes comparable, 
say within a factor of two, to  $F_{\rm max}=F_{\rm LSO}$.
We recover the same conclusion as above, which was the conclusion already
pointed out in DIS: namely the signal from
{\em massive systems}
[$m\gtrsim  13 M_{\odot}$ 
if $f_{\rm p}=165$ Hz,  corresponding to $f_0=200$ Hz, 
more generally, $m\gtrsim  13 \left(165\, {\rm Hz}/f_{\rm p}\right)M_{\odot}$],
when treated (as they should be) {\em relativistically} will be badly represented by the usual
frequency-windowed SPA.
This conclusion obtained from an analytical approximation is
borne out by the numerical computations shown in Fig.~\ref{fig:zero}
above (see also Table~\ref{table:newtonian} below).
It is mainly for such systems that the work presented in the following Sections will be mandatory. But even for lower mass
systems, we shall construct here for the first time the Fourier-domain version of the (time-domain) P-approximants
introduced in DIS. 
Since P-approximants provide better
templates than the usually considered T-approximants 
\cite{dis98}, the work
of this paper will be useful for all types of systems,
even the less massive ones.

\subsection{Fourier transform of time-truncated chirps}
\label{sec:TTC}

To introduce the detailed analysis that we shall give in the following Sections,
let us start by delineating some general mathematical facts about the integrals
we have to deal with. We will be interested in evaluating the Fourier-transform 
$\tilde{h}(f)$
of a time-truncated chirp
$h(t)=2 a(t)\cos \phi(t) \theta(t_{\rm max}-t)$.
After decomposing the cosine into complex exponentials, the Fourier integral
leads to a sum of two integrals of the form 
$\int _{-\infty}^{t_{\rm max}}\,dt \,a(t)\,e^{i \psi_f^\pm (t)}$
with phases
$\psi_f^{\pm}=2\pi f t \pm \phi(t)$.
Let us then, for generality, discuss the properties of integrals of the type
\begin{equation}
I(\varepsilon)=\int_{t_a}^{t_b}\,dt\,a(t)\,e^{i\frac{\psi(t)}{\varepsilon}}\,.
\label{br15}
\end{equation}
Here, we have introduced a formal `small parameter'
$\varepsilon$ (set to unity at the end of the calculation) to formalize the
fast variation of the phase compared to that of the amplitude.

Let us first note that: 
(i) if the phase has no stationary point 
$\dot{\psi}(t)\neq 0$ for $t\in [t_a,t_b]$,
(ii) if the amplitude vanishes smoothly at the edge points $t_a$ and $t_b$,
which might be pushed to $\pm \infty$
and (iii) if the functions $a(t)$ and $\psi(t)$ are smooth $({\cal C}^\infty)$
within the interval $[t_a,t_b]$, the integral $I(\varepsilon)$ tends to zero
with $\varepsilon$, faster than any power.
This can be seen by integrating by parts.
To simplify the calculation we can [thanks to the assumption (i)] use
$\psi$ as integration variable. This yields 
\begin{equation}
I=\int_{\psi_a}^{\psi_b} \,d\psi\, A(\psi) e^{i\psi}\,,
\label{br7}
\end{equation}
where $A(\psi)=\left[a(t)/\dot{\psi}(t)\right]_{t(\psi)}$ where $t(\psi)$
denotes the (unique) solution in $t$ of $\psi=\psi(t)$ and where
$\psi_a=\psi(t_a),\,\psi_b=\psi(t_b)$.
Using $e^{{i\psi}/{\varepsilon}}=\frac{\varepsilon}{i}\frac{d}{d\psi}
\left(e^{{i\psi}/{\varepsilon}}\right)$,
integrating by parts, and using [thanks to assumption (ii)] the vanishing of $A(\psi)$ at
the edges, leads to
\begin{equation}
I(\varepsilon)=i\varepsilon\int_{\psi_a}^{\psi_b}\,d\psi A'(\psi)e^{\frac{i\psi}{\varepsilon}}\,.
\label{br16}
\end{equation}
Using [assumption (iii)] the vanishing of all the derivatives of  $A(\psi)$ at the edges, we
can iterate the result Eq. (\ref{br16}) to any order:
\begin{equation}
I(\varepsilon)=(i\varepsilon)^n\int_{\psi_a}^{\psi_b}\,d\psi A^{(n)}(\psi)e^{\frac{i\psi}{\varepsilon}}\,.
\label{br17}
\end{equation}
The result Eq. (\ref{br17}) means that, when $\varepsilon\rightarrow 0$,
$I(\varepsilon)={\cal O}(\varepsilon^n)$ for any integer $n$, {\it i.e.}
$I(\varepsilon)$ vanishes faster than any power of $\varepsilon$.
It does not mean that 
$I(\varepsilon)$ is zero for any finite (but small) value of $\varepsilon$.
For instance, under stronger assumptions about the existence and properties of 
an analytic continuation of the function $A(\psi)$ in the complex $\psi$ plane
it follows that $I(\varepsilon)\sim Ae^{-\frac{B}{\varepsilon}}$
for some constants $A$ and $B$.
For reasonably small values of $\varepsilon$ such exponentially small 
contributions are numerically negligible.
[We shall see later that the `small parameter' $\varepsilon$ (or better
$\varepsilon/B$) is typically of order $1/(2\pi N)$
where $N\equiv F^2/\dot{F}$, is the instantaneous number of cycles.]

We conclude, therefore that the integral $I$ will be (in most relevant cases)
numerically non-negligible only if the assumptions above are violated.
In other words, the value of $I(\varepsilon)$ will be dominated by the contributions
coming from either (i) from stationary-phase points, $\dot{\psi}(t_s)=0$,
or (ii) from the edge points $t_a$ and/or $t_b$.
Let us (for simplicity) assume that there is (at most) one stationary-phase
 point $t_s$, and that it is of the normal
parabolic type, {\it i.e.} $\psi(t)=\psi_s+\frac{1}{2!}\ddot{\psi}_s(t-t_s)^2+
{\cal O}((t-t_s)^3)$ with $\ddot{\psi}_s\neq 0$.
Let us also assume that $a(t_a)\neq 0$, $a(t_b)\neq 0$.
[We maintain here for the moment, the assumption (iii) above about the regularity of the
functions $a(t)$, $\psi(t)$ in the {\it closed} interval $[t_a,t_b]$.]
Then, assuming analyticity of the involved functions, the mathematically most
rigorous way of decomposing $I$ as the sum (modulo nonperturbative small 
contributions of the type discussed above) of a stationary-point contribution 
$I_{\rm stationary}$ and of edge contributions
$I_{\rm edge}= I_{\rm edge}^a+ I_{\rm edge}^b$
is to deform the original (real) contour of integration into the complex plane
\cite{BO,dkpo99}.
The deformed contour must be such that near $t_s$ it leads to a basic integral
of the type
$\int_{-\alpha}^{\beta}\,dx\,e^{-bx^2}\left[c_0+c_1x+c_2x^2+\cdots\right]$,
while near each end point it leads to integrals of the type
$\int_{0}^{\gamma}\,dy\,e^{-cy}\left[d_0+d_1y+d_2y^2+\cdots\right]$.
It is then easy to find the structure of the expansion of both
$I_{\rm stationary}$ and  $I_{\rm edge}$
in powers of $\varepsilon$, as  $\varepsilon\rightarrow 0$.  
For instance, it is convenient near $t_s$ to introduce a scaled variable:
$t-t_s=\varepsilon^{\frac{1}{2}}\tau$
(before rotating $\tau$ to complex values $\tau=e^{\pm \frac{i\pi}{4}}x$),
so that the phase scales as 
\begin{equation}
\frac{\psi}{\varepsilon}= \frac{\psi_s}{\varepsilon}
+\frac{1}{2!}\ddot{\psi}_s\tau^2
+\frac{\sqrt{\varepsilon}}{3!}\psi^{(3)}_s\tau^3+\cdots\,,
\label{sha3}
\end{equation}
where $\psi_s^{(3)}\equiv d^3\psi_s/dt^3$.
Expanding then the integrand of this $\tau$-integral in powers of 
$\varepsilon^{\frac{1}{2}}$
leads to an integral of the symbolic type
\begin{equation}
I_{\rm stationary}\sim \varepsilon^{\frac{1}{2}}
\int\,dx\,e^{-x^2}\left[1+
\varepsilon^{\frac{1}{2}} x^{\rm odd} 
+\varepsilon x^{\rm even} +
\varepsilon^{\frac{3}{2}} x^{\rm odd} +\cdots\right]\,,
\label{sha4}
\end{equation}
where $x^{\rm odd}(x^{\rm even})$ denotes a sum of terms $\sim
x^{2k+1}(x^{2k})$.
This yields (using the fact that the terms $\int _{-\alpha/\sqrt{\varepsilon}}
^{\beta/\sqrt{\varepsilon}}\,dx\,e^{-x^2}\,x^{\rm odd}$ are
exponentially small) an expansion of the type 
\begin{equation}
I_{\rm stationary}\sim 
\varepsilon^{\frac{1}{2}}\left[C_0+C_1\varepsilon+C_2\varepsilon^2+\cdots\right]\,.
\label{br18}
\end{equation}
The structure of the `edge' contribution $I_{\rm edge}$ can be obtained by a similar
technique.
Now the appropriate scaling is different, e.g.
$t-t_a=\varepsilon \tau$, and one ends up with integrals of the type
$\int_0^{\gamma/\varepsilon}\,dy\,e^{-y}y^n$. This yields,  
\begin{equation}
I_{\rm edge}\sim 
\varepsilon\left[D_0+D_1\varepsilon+D_2\varepsilon^2\cdots\right]\,.
\label{br19}
\end{equation}
The aim of this preliminary discussion was to point out the structures
Eqs. (\ref{br18}) and (\ref{br19}) of the two main contributions to a
generic oscillatory integral of the form Eq. (\ref{br15}).
Note that while the leading contribution is given by the lowest order
term in the stationary-point or saddle-point expansion Eq. (\ref{br18}),
the next-to-leading contribution comes from the lowest order edge-correction
Eq. (\ref{br19}).
One generally expects that $I_{\rm edge}$ will be only $\sqrt{\varepsilon}$,
{\it i.e.} $1/\sqrt{2\pi N}$, smaller than $I_{\rm stationary}$.
We shall give below the explicit expressions for the first two terms in both
expansions Eqs. (\ref{br18}) and (\ref{br19}).
We shall see that each coefficient $C_0,C_1,C_2\cdots\cdots D_0,D_1,\cdots$
in Eqs. (\ref{br18}) and (\ref{br19}) is a combination of derivatives 
(of increasing total order) of $a(t)$ and $\psi(t)$ evaluated at $t_s$
for Eq. (\ref{br18}) and at $t_a$ or $t_b$ for Eq. (\ref{br19}).
We note in advance that, for actual calculations, the simplest way to evaluate
the explicit forms of the expansions Eqs. (\ref{br18}) and (\ref{br19})
is not necessarily to follow the complex-contour route sketched above.
In the case of Eq. (\ref{br18}) one can deal directly with the original
stationary-point-expanded integral written as 
$\propto \varepsilon^{\frac{1}{2}}\int \,d\tau\,e^{i\ddot{\psi}_s\tau^2}\left[
\tau_0+c_1\tau+\cdots\right]$,
and in the case of Eq. (\ref{br19}) the simplest is to keep the boundary terms in the
integration-by-parts approach Eqs. (\ref{br16}) and (\ref{br17}) given above
for the simple case where they were neither stationary phase points, nor boundary
contributions.

To put in context the analysis that we shall perform below, let us finally mention
two serious limitations of the assumptions leading to Eqs. (\ref{br18}) and
(\ref{br19}). First, in the analysis above, based on the introduction of the
formal parameter $\varepsilon\rightarrow 0$, we have assumed that the stationary-phase point
$t_s$ was parametrically separated from the edges $t_a$ or $t_b$, {\it i.e.}
that $\vert t_s-t_a\vert$
and $\vert t_s-t_b\vert$ were much larger than the characteristic Gaussian width 
$\Delta t=\sqrt{\varepsilon/\ddot{\psi}_s}={\cal O}(\sqrt{\varepsilon})$
associated to the stationary point.
As we shall see, this limitation is unacceptable for the application we have in mind and we shall
have to introduce new tools to overcome it.
A second limitation (which compounds with the first and will lead to an
unavoidable complexity of our treatment) is the seemingly innocent assumption
(iii) above, namely the hypothesis that the functions $a(t)$ and $\psi(t)$
are infinitely differentiable within the {\it closed} interval $[t_a,t_b]$
({\it i.e.} including also the end points).
As we shall see, in the physically relevant case of a
{\it relativistic} (adiabatic) chirp the functions $a(t)$ and $\psi(t)$ will not  be
${\cal C}^\infty$ at the physically imposed upper cutoff
$t_b=t_{\rm LSO}$.
This will require us 
to introduce new types of expansions and new tools to deal 
with the relativistic edge contribution in addition to the 
 modification of the stationary-phase approximation needed 
in the case where $t_s$ is near the edge, in the sense that
$\vert t_s-t_b\vert={\cal O}(\sqrt{\varepsilon})$.

\section {Improved stationary phase approximation for time-windowed
Newtonian-like signals}\label {sec:SPA}

\subsection{The usual stationary phase approximation}\label{sec:IIIA}

Let us begin by a quick recall of the usual treatment of the stationary phase 
approximation to a chirp. Consider a signal,
\begin{mathletters}
\begin{eqnarray}
h(t)&=&2 a(t) \,
\cos \phi (t)= a(t) e^{-i \phi(t)} +a(t) e^{i \phi(t)}\,,
\label{3.1}\\
{\rm where},\;\; \frac{d\phi(t)}{dt}&\equiv &2 \pi F(t) >0\,.
\label{3.2}
\end{eqnarray}
\end{mathletters}

We shall say that a signal is ``Newtonian-like'' if the instantaneous
frequency $F(t)$, defined by Eq. (\ref{3.2}), increases without limit when $t$
runs over its full (mathematically allowed) range. (Note that we 
conventionally consider only positive instantaneous frequencies.)
For instance, at the Newtonian order, the explicit forms for the  chirp
 amplitude,
phase and frequency of the gravitational waves are respectively given by:
\begin{mathletters}
\begin{eqnarray}
a(t)&=&  {\cal C}_{{\cal M}} ( \pi {\cal M} F(t))^{2/3}\,,
\label{3.3}\\
\phi(t) &=& \phi_c - 2\, 
\biggl[ \frac{(t_c-t)}{5 {\cal M}} \biggr]^{5/8}, 
\label{3.4}\\
\pi {\cal M} F(t) &=& \biggl[ \frac{5 {\cal M}}{256(t_c-t)}
                    \biggr]^{3/8}, 
\label{3.5}
\end{eqnarray}
\end{mathletters}
where  ${\cal M}$
is the chirp mass given by ${\cal M}=\eta^{3/5} m$ in terms of the  total mass $m$ 
 and the symmetric mass ratio $\eta$;  
$\phi_c$ the gravitational wave phase at instant of coalescence $t_c$
and ${\cal C}_{{\cal M}}$ is  the product of  a function of different angles,
characterising the relative orientations of the binary
and the detector, with the ratio
$ {\cal M}/d$ where $d$ is the distance to the source (see below). 
Note that the function $F(t)$
increases without limit as $t$ tends to the formal coalescence time $t_c$.

Coming back to a general signal, the Fourier transform is defined by
Eq. (\ref{d1}).
Because the signal $h(t)$ is real, we have the identity 
$\tilde{h}(-f)\equiv (\tilde{h}(f))^{*}$. It therefore suffices to compute
the Fourier transform for positive values of the frequency $f$. [Note that
we use a lower case letter to distinguish the Fourier variable $f$ from
the instantaneous frequency $F(t)$.] The Fourier transform of a generic
signal of the form Eq. (\ref{3.1}) reads
\begin{mathletters}
\begin{eqnarray}
\tilde{h}(f)&=&  \tilde{h}_-(f)+  \tilde{h}_+(f) \,,\\ 
\label{3.7}
{\rm where},\;
\tilde{h}_-(f)&\equiv& \int _{-\infty}^{\infty} dt\; a(t) e^{ i(2\pi  f t -\phi(t))}\,, \\
\label{3.7a}
\tilde{h}_+(f)&\equiv& \int _{-\infty}^{\infty} dt\; a(t) e^{ i(2\pi  f t +\phi(t))} \,.
\label{3.7b}
\end{eqnarray}
\end{mathletters}
The integrands of $\tilde{h}_{\pm}(f)$ are violently oscillating and thus their
 dominant contributions come 
from
the vicinity of the stationary points of their phase (when such points exist).
 When $f> 0$ (which we shall henceforth assume), only the $\tilde{h}_-(f)$ term 
 has such a saddle-point. 
Therefore, we can write the approximation
\begin{mathletters}
\begin{eqnarray}
\tilde{h}(f)&\simeq&
\tilde{h}_-(f)\simeq
 \int _{-\infty}^{\infty} dt\;a(t)\; e^{ i\psi_f(t)}\,,
\label{3.8}\\
{\rm where}\,,\;\;\psi_f(t)&\equiv& 2 \pi f t -\phi(t)\,.
\label{3.9}
\end{eqnarray}
\end{mathletters}
The saddle-point of the phase $\psi_f(t)$ is the value, say $t_f$, of the time variable $t$
where $d\psi_f(t)/dt=0$, {\it i.e.} it is the solution of the equation 
$F(t_f)=f$. The dominant contribution to the integral Eq. (\ref{3.8}) now comes from a
time interval near $t=t_f$. When the second time derivative of the phase at the saddle-point
does not vanish, {\it i.e.} when $\dot{F}(t_f)\ne 0$, one can estimate Eq. (\ref{3.8}) by replacing
$\psi_f(t)$ [and $a(t)$] by truncated Taylor expansions near $t=t_f$, namely
\begin{mathletters}
\begin{eqnarray}
\psi_f(t)&\simeq& \psi_f(t_f) -\pi \dot{F}(t_f)(t-t_f)^2\,,
\label{3.10}\\
a(t)&\simeq& a(t_f)\,.
\label{3.11}
\end{eqnarray}
\end{mathletters}
(The zeroth-order term in Eq. (\ref{3.11}) is enough because the first order term $
\dot{a}(t_f)(t-t_f)$ vanishes after integration over $t$). This leads to a Gaussian integral
\begin{equation}
\tilde{h}(f)\simeq \int _{-\infty}^{\infty} dt\;a(t_f)\; e^{ i\psi_f(t_f) -i\pi \dot{F}(t_f)(t-t_f)^2}\,.
\label{3.12}
\end{equation}
Evaluating
this Gaussian integral, one finally obtains the well-known expression for the 
{\it usual } SPA (hereafter abbreviated as uSPA):
\begin{equation}
\tilde{h}^{\rm uspa}(f)= \frac{a(t_f)}{\sqrt{\dot{F}(t_f)}} e^{ i\left[ \psi_f(t_f) 
-\pi/4\right]}\,,
\label{3.13}
\end{equation}
where $\psi_f(t_f)$ is the value of $ \psi_f(t)$ at $t=t_f$.

The  conditions for the validity of the SPA are usually assumed to be
$\varepsilon_1\ll 1$, $\varepsilon_2\ll 1$, where 
\begin{equation}
\varepsilon_1\equiv \left\vert\frac{\dot{a}(t)}{a(t)\dot{\phi}(t)}\right\vert\,;\;
\varepsilon_2\equiv\left\vert\frac{\ddot{\phi}(t)}{\dot{\phi}^2(t)}\right\vert
= \left\vert\frac{1}{2\pi}\frac{\dot{F}(t)}{F^2(t)}\right\vert=\frac{1}{2\pi N}\,.
\label{d3.13a}
\end{equation} 
One can assess in a more precise  quantitative manner the accuracy
of the SPA by computing the leading correction to the integral
Eq. (\ref{3.12}).
This leading correction will be given by keeping more terms in
the Taylor expansions Eqs. (\ref{3.10}) and (\ref{3.11}).
To keep track of what one means by the `next order term' in the
SPA expansion it is convenient (as in Section~\ref{sec:TTC} above)  to formalize the fast variation of the
phases $\phi(t)$ and $\psi(t)$ by considering an integral
of the form $I=\int\,dt\,a(t) \exp(i\psi(t)/\varepsilon)$
with a `small' parameter $\varepsilon$ (set to unity at the end of the calculation).
It is then easy to see [e.g. after the introduction of a rescaled
time-variable: $t-t_f=\varepsilon^{1/2} \tau$
where $t_f$ denotes, as above, the saddle point of the phase $\psi_f(t)$]
that the leading correction to the result Eq. (\ref{3.13}) will be of fractional 
order $\varepsilon$ [as exhibited in Eq. (\ref{br18})]
and will come from keeping { \it two more terms} in both the
Taylor expansions Eqs. (\ref{3.10}) and (\ref{3.11}).
Expanding in powers of $\varepsilon^{1/2}$ leads to integrals
of the form $\int_{-\infty}^\infty d\tau\, \tau^n \exp(-i \pi \dot{F}(t_f)\,\tau^2)$
with $n\le 6$.
Finally, one finds that the sum of the usual SPA and of its leading correction
is equivalent to multiplying Eq. (\ref{3.13}) by the correcting phase
factor $e^{i\delta}$ where
($F^{(3)}$ denoting $d^3F/dt^3$)
\begin{equation}
\delta=\frac{1}{2\pi \dot{F}(t_f)}\left[
-\frac{1}{2}\frac{\ddot{a}}{a}+\frac{1}{2}\frac{\dot{a}}{a}\frac{\ddot{F}}{\dot{F}}
+\frac{1}{8}\frac{F^{(3)}}{\dot{F}}-\frac{5}{24}\left(\frac{\ddot{F}}{\dot{F}}\right)^2
\right]_{t=t_f}\,.
\label{d3.13b}
\end{equation}
Therefore, a quantitatively precise criterion for the {\it local}
validity of the SPA
is $\varepsilon_{\rm loc}\equiv\vert\delta\vert\ll 1$.
In the case of power-law chirps, $\varepsilon_{\rm loc}$
is equal to one-fifth of the criterion explicitly given in the
recent study \cite{cf98} of the validity of the SPA.
In the case of Newtonian chirps, Eq. (\ref{d3.13b}) yields 
\begin{equation}
\delta=\frac{23}{24}\left(\frac{1}{9\pi}\frac{\dot{F}}{F^2}\right)=
\frac{23}{24}\left(\frac{1}{9\pi N}\right)\,.
\label{d3.13c}
\end{equation}
Written in terms of $v=(\pi m F)^{1/3}$ this reads 
$\delta=(92/45)\eta v^5$ which agrees with the corresponding result in
\cite{dkpo99}.
It is interesting to note that Eq. (\ref{d3.13c}) formally predicts,
at the LSO, $\delta_{LSO}=(4\eta)\times 0.58\%$ which is quite small.
Alternatively, one can say that Eq. (\ref{d3.13c})
predicts that even if the instantaneous number of cycles were as small
as $N\sim 1$, the local correction to the SPA would be small
($\delta=0.0339/N$).
This result does {\it not} mean, however, that we can use the usual SPA
Eq. (\ref{3.13}) to estimate with sufficient accuracy the FT of a real
inspiral signal. Indeed, even if we were considering a Newtonian-like signal
(where $N$ stays away from zero at the LSO) the correcting phase factor
Eq. (\ref{d3.13b}) represents just the {\it local} correction to the
SPA, {\it i.e.} the correction due to higher-terms in the local
expansion near the saddle-point.
There are also {\it global} corrections to the SPA coming from the entire
integration domain, and, most importantly (as emphasized in \cite{cf98})
from the end-points of the time-integration.
In addition, there is also a correction coming from the neglected
contribution $\tilde{h}_+(f)$ in Eq. (\ref{3.7}).
Before considering them in detail, let us also note that Eq. (\ref{d3.13b})
indicates that $\delta$ blows up to infinity, at the LSO, in the case of a
{\it relativistic} GW chirp (because $F(t)\sim F_{\rm LSO} -a(t_{\rm LSO}-t)^{1/2}$
there; see below).
This shows again that, independently of the problems linked to the time-windowing,
relativistic signals will pose special difficulties.
But let us start by studying the simpler case of time-truncated
Newtonian-like signals, by which we mean that $N\equiv F^2/\dot{F}$
stays away from zero at the upper time cut-off. 

\subsection{Beyond the usual stationary phase approximation}\label{sec:BEYOND}

We therefore consider time-domain
signals of the form
\begin{equation}
h(t)=2 a(t) \,\cos \phi (t) \theta ( t_{\rm max} - t) \,,
\label{3.14}
\end{equation}
 where $\theta$ denotes the Heaviside step function, 
($\theta(x)=1$, for $x>0$ and $\theta(x)=0$, for $x<0$).
This time-windowing has three effects: (i) it induces oscillations in the usually
considered  frequency-domain $ f < F_{\rm max}$,  
(ii) it generates a tail
in the usually disregarded frequency-domain $ f > F_{\rm max}$,
and (iii) it renders non-negligible the `non-resonant' contribution $\tilde{h}_+(f)$. 
Here, $F_{\rm max}$ denotes
the instantaneous gravitational wave frequency reached at $t = t_{\rm max}$,
{\it i.e.} $F_{\rm max} \equiv F(t_{\rm max})$. The main purpose
of the present paper is to model and estimate, analytically, as accurately as
 possible all these effects.  The case where the saddle-point $t_f$
 is (below and) far away from the upper-cutoff $t_{\rm max}$ has been
 recently considered in \cite{dkpo99}. However, this case is not the
 physically relevant one. As pointed out in DIS, the case where the usual SPA
 becomes unacceptably inaccurate is the case of massive systems for which
 the signal emits not many cycles in the detector's bandwidth before
 crossing the last stable orbit. In this case the most important frequencies
 are located around the effective-cutoff frequency $F_{\rm max} \sim F_{\rm LSO}$
 (and as we shall see below, it is important to estimate 
 the Fourier transform accurately, both for $ f< F_{\rm max}$ and for  $f > F_{\rm max}$). 
 Here, we shall provide an approximate analytical treatment valid in
 this crucial range of frequencies. 

Let us first state clearly our notation.
We decompose the FT of the time-windowed signal Eq. (\ref{3.14}) as
\begin{mathletters}
\begin{eqnarray}
\tilde{h}(f)&=&\tilde{h}_-(f)+\tilde{h}_+(f)
\label{br1}
\,,\\
{\rm where},\; \tilde{h}_-(f)&=&\int_{-\infty}^{t_{\rm max}}\,dt\,a(t)\,e^{i\psi_f^-(t)}
\label{br2}
\,,\\
 \tilde{h}_+(f)&=&\int_{-\infty}^{t_{\rm max}}\,dt\,a(t)\,e^{i\psi_f^+(t)}
\label{br3}
\,,\\
\psi_f^{-}(t)\equiv \psi_f(t)&\equiv& 2\pi f t -\phi(t)
\label{br4}
\,,\\
\psi_f^{+}(t)&\equiv&  2\pi f t +\phi(t)
\label{br5}
\,.
\end{eqnarray}
\end{mathletters}
We shall refer to the contribution $\tilde{h}_-(f)$ as the `resonant'
contribution because the equation defining the saddle point it contains,
$F(t_f)=f$, corresponds to a resonance between the Fourier-frequency $f$ and the
instantaneous gravitational wave frequency $F(t)$.

\subsection{Edge contribution to the  non-resonant integral $\tilde{h}_+(f)$}\label{sec:PLUS}

Before dealing with the oscillatory and tail corrections to the  
resonant contribution $\tilde{h}_-(f)$ of $\tilde{h}(f)$, we shall first
deal with the non-resonant contribution $\tilde{h}_+(f)$.
When $f>0$, the phase $\psi_f^+(t)$ in contrast to $\psi_f^-(t)$ has
no stationary point.
Let us therefore consider the general problem of approximating an integral of the form 
\begin{equation}
I=\int_{t_a}^{t_b} \,dt\, a(t) e^{i\psi(t)}\,,
\label{br6}
\end{equation}
where $\psi(t)$ is always monotonically varying $(\dot{\psi}\neq 0)$.
We can then use $\psi$ as integration variable.
This yields, as above,
\begin{equation}
I=\int_{\psi_a}^{\psi_b} \,d\psi\, A(\psi) e^{i\psi}\,,
\label{bri7}
\end{equation}
where $A(\psi)=\left[a(t)/\dot{\psi}(t)\right]_{t(\psi)}$ where $t(\psi)$
denotes the (unique) solution in $t$ of $\psi=\psi(t)$ and where
$\psi_a=\psi(t_a),\,\psi_b=\psi(t_b)$.
One can    treat  $\psi$ as a fast varying phase,  
so that  by comparison, the amplitude $A(\psi)$ varies slowly when $\psi$ varies by  $2\pi$
(as above this could be formalized by the formal replacement $\psi \rightarrow \psi/\varepsilon$).
In other words, instead of a SPA, we are in a WKB like approximation where one can expand in 
the slowness of variation of $A(\psi)$, {\it i.e.} we can expand in successive derivatives $d^nA(\psi)/d\psi^n$.
This expansion is obtained by successively integrating Eq. (\ref{br7}) by parts [using
$e^{i\psi}\equiv \frac{d}{d\psi}(e^{i\psi}/i)$].
Contrary to Section~\ref{sec:TTC} above we now keep the edge contribution coming from
the boundaries.
For instance at second order this leads to
\begin{equation}
I=\left[\frac{e^{i\psi}}{i}\left[A(\psi)+i A'(\psi)\right]\right]^b_a +
i^2\int_{\psi_a}^{\psi_{b}}\,d\psi\, A''(\psi)\,e^{i\psi}\,.
\label{br8}
\end{equation}
It is easy to re-express this result in terms of the original
time variable $t$ by replacing $A(\psi)\equiv a(t)/\dot{\psi}(t)$ and
$d/d\psi=(\dot{\psi})^{-1}d/dt$. 
We deduce from Eq. (\ref{br8}) the full `edge' contribution to the
integral $I$ (as explained in Section~\ref{sec:TTC}, the `bulk' contribution is
exponentially small): 
\begin{equation}
I_{\rm edge}=\left[\frac{e^{i\psi}}{i}\left( A(\psi) +iA'(\psi) +\cdots+ i^n A^{(n)}(\psi)+
\cdots \right)\right]^b_a\,.
\label{br20}
\end{equation}
This is the explicit form of the parametric expansion sketched in Eq. (\ref{br19}).
It can be expressed in terms of the time derivatives of $a(t)$ and $\psi(t)$
by using the replacement rules just mentioned.
In particular, at the leading and next-to-leading order it reads explicitly
\begin{equation}
I_{\rm edge}=\left[\frac{a(t)}{i\dot{\psi}(t)} e^{i\psi(t)}
\left\{1+\frac{1}{i\dot{\psi}(t)}\left( \frac{\ddot{\psi}(t)}{\dot{\psi}(t)}-\frac{\dot{a}(t)}{a(t)}
\right)\right\}\right]^{t_b}_{t_a}\,.
\label{br9}
\end{equation}
If we apply this general result to $\tilde{h}_+(f)$ Eq. (\ref{br3}), we get the estimate 
\begin{mathletters}
\begin{eqnarray}
\tilde{h}_+(f)&\simeq& \tilde{h}_+^{\rm edge}(f) \simeq
\frac{a(t_{\rm max})}{ i\dot{\psi}_f^+(t_{\rm max})} e^{i\psi_f^+(t_{\rm max})}
\left[1+\frac{1}{i\dot{\psi}_f^+(t_{\rm max})}\left( 
\frac{\ddot{\psi}_f^+(t_{\rm max})}{\dot{\psi}_f^+(t_{\rm max})}-
\frac{\dot{a}(t_{\rm max})}{a(t_{\rm max})}
\right)\right]\,
\label{bra10}\\
&\simeq&\frac{a(t_{\rm max})}{2\pi i(f+F_{\rm max})} e^{i\psi_f^+(t_{\rm max})}
\left[1+\frac{1}{ 2 \pi i(f+F_{\rm max})}\left( \frac{\dot{F}_{\rm max}}{(f+F_{\rm max})}
-\frac{\dot{a}(t_{\rm max})}{a(t_{\rm max})}
\right)\right]\,.
\label{br10}
\end{eqnarray}
\end{mathletters}
Note that, in the approximation where the amplitude is taken as Newtonian-like, {\it i.e.},
$a(t)=Cv_F^2\propto (F(t))^{2/3}$,
the last term in Eq. (\ref{br10}) reads 
$\dot{a}(t_{\rm max})/a(t_{\rm max})=
\frac{2}{3} \dot{F}_{\rm max}/F_{\rm max}.$

\subsection{Improved stationary phase approximation when $f<F_{\rm max}$}\label{sec:IIIB}

Let us now consider the dominant contribution $\tilde{h}_-(f)$, Eq. (\ref{br2}).
We start by  considering the case where the Fourier variable $f$ [of the Fourier transform of the 
time-windowed signal
Eq. (\ref{3.14})] is smaller than $F_{\rm max}$, but near $F_{\rm max}$. As
will become clear from the formulas below, the interval around $F_{\rm max}$ where it is needed to
improve the usual SPA, is the range,
\begin{equation}
|f-F_{\rm max}|\lesssim \; {\rm ( few ) }\;\;\; \sqrt{\dot{F}(t_f)}.
\label{3.15}
\end{equation}
In the case where $f$ is in the range Eq. (\ref{3.15}) with $f<F_{\rm max}$,
there is a saddle-point $t_f$ in the first term of the exact integral 
Eq. (\ref{3.7}), and one can still use the parabolic approximation Eq. (\ref{3.10}) to the phase
$\psi_f(t)$, and the lowest approximation Eq. (\ref{3.11}) to the amplitude
$a(t)$. [Indeed, the work of the previous Section has shown that the {\it local}
corrections to the integral Eq. (\ref{3.8}), coming from the inclusion of
more terms in Eqs. (\ref{3.10}) and (\ref{3.11}), were quite small as long as $N\gtrsim 1$].
 Therefore, in this case, the resonant contribution to  the Fourier transform becomes:
\begin{mathletters}
\begin{eqnarray}
\tilde{h}_-(f)&\simeq&\int _{-\infty}^{ t_{\rm max}} dt \; a(t)\; e^{ i \psi_f( t)}\,,
\label{3.16}\\
&\simeq& a(t_f) e^{ i\psi_f(t_f)}\int_ {-\infty}^{ t_{\rm max}} dt\; 
e^{ -i \pi \dot{F}(t_f)(t-t_f)^2} 
\,.
\label{3.17}
\end{eqnarray}
\end{mathletters}
The crucial difference between the Eqs. (\ref{3.12}) and (\ref{3.17}) is that
the full Gaussian integral has  become a 
complex Fresnel integral   which may be evaluated in terms of the 
complementary error function.
Let us recall that the complementary error function ${\rm erfc}(z)\equiv 1-{\rm erf}(z)$
is defined by
\begin{equation}
{\rm erfc}(z)=\frac{2}{\sqrt{\pi}}\int_{z}^{+\infty}\;e^{-x^2}\; dx\,.
\label{3.18}
\end{equation}
It takes on the real axis the particular values 
${\rm erfc}(+\infty)=0$, 
${\rm erfc}(0)=1$, and  
${\rm erfc}(-\infty)=2$. 
By rotating the integration contour in the complex plane $(x=-e^{\frac{i\pi}{4}}\xi)$
and shifting the new integration variable $\xi$, we get the  following useful integration
formula:
\begin{equation}
\int_{-\infty}^{\tau_m}\, d\tau\, e^{-i( a\tau^2 +2 b \tau +c)} =
\frac{1}{2}\sqrt{\frac{\pi}{a}}e^{-\frac{ i\pi}{4}} \,e^{i\frac{b^2-ac}{a}} 
\,{\rm erfc}\left[-e^{\frac{i\pi}{4}}
\sqrt{a}( \tau_m +\frac{b}{a})\right]\,.
\label{3.19}
\end{equation}
The formula Eq. (\ref{3.19}) motivates us to define the following auxiliary function
\begin{equation}
{\cal C}(\zeta)\equiv \frac{1}{2} \rm erfc\left(e^{\frac{i\pi}{4}}\zeta\right)\,.
\label{3.20}
\end{equation}
It is useful to note that 
$ {\cal C}(+\infty)=0$,
$ {\cal C}(0)=1/2$, and
$ {\cal C}(-\infty)=1$.
Moreover, the leading terms in the asymptotic expansions of ${\cal C}(\zeta)$ as $\zeta\rightarrow
\pm\infty $ are 
\begin{mathletters}
\begin{eqnarray}
\zeta\rightarrow -\infty,\;\;\; {\cal C}(\zeta)&\sim&1+ \frac{e^{-\frac{i\pi}{4}}}{2\sqrt{\pi}}
\frac{e^{-i\zeta^2}}{\zeta}\,,
\label{3.21}\\
\zeta\rightarrow +\infty,\;\;\; {\cal C}(\zeta)&\sim&\frac{e^{-\frac{i\pi}{4}}}{2\sqrt{\pi}}
\frac{e^{-i\zeta^2}}{\zeta}\,.
\label{3.22}
\end{eqnarray}
\end{mathletters}

As the auxiliary function ${\cal C}(\zeta)$ plays an important role in our work we plot it in 
Fig.~\ref{fig:soft.step}.  From the Figure we note that the real part 
of the above complex  function
is a softened  version of a step function 
 $\theta(-\zeta)$  and the
imaginary part an oscillating function vanishing in the limits $\pm \infty$, as well as at
$\zeta=0$. 

Armed with this definition one finds that the right hand side of Eq. (\ref{3.17})
yields the approximation
\begin{mathletters}
\begin{eqnarray}
\tilde{h}_-(f)&\simeq& {\cal C}(\zeta_0(f))\tilde{h}^{\rm uspa}(f)\,,
\label{3.23}\\
{\rm where},\;\; \zeta_0(f)&\equiv& \sqrt{\pi \dot{F}(t_f)}(t_f -t_{\rm max})\,.
\label{3.24}
\end{eqnarray}
\end{mathletters}
In words, when $f<F_{\rm max}$ one can correct for the `edge effects' caused by the
cutoff at $t_{\rm max}$ by multiplying
the usual SPA 
 $\tilde{h}^{\rm uspa}(f)$ given in Eq. (\ref{3.13})
by a complex `correction factor' ${\cal C}(\zeta_0(f))$.
  The expression Eq. (\ref{3.23}) gives very good overlaps with the exact discrete
Fourier transform (DFT) of the time-windowed signal Eq. (\ref{3.14}).
However,  it is possible to do even better by a slight modification of the argument $\zeta_0(f)$,  
Eq. (\ref{3.24}).

To understand how a slight modification of the argument $\zeta(f)$ of the auxiliary function
${\cal C}(\zeta)$ can improve both the visual agreement (even quite far away from $F_{\rm max}$) and the overlap 
with the exact DFT  of the time-windowed signal Eq. (\ref{3.14}) we have to take into account the
asymptotic expansion Eq. (\ref{3.21}).
Indeed, on the one hand, when inserting the expansion Eq. (\ref{3.21}) into Eq. (\ref{3.23}), using Eq. (\ref{3.13})
for $\tilde{h}^{\rm uspa}(f)$ and allowing for a more general frequency dependent argument $\zeta(f)$,
we find that $\tilde{h}(f)$ differs from $\tilde{h}^{\rm uspa}(f)$ (in the domain 
$\zeta(f)\rightarrow -\infty$ {\it i.e.}
$f\ll F_{\rm max}$) by a correction term proportional to
$e^{-i\pi/2} e^{i[\psi_f(t_f)-\zeta^2]}/\zeta$.
On the other hand, a different way of estimating this edge
correction consists in writing Eq. (\ref{3.17}) as an integral between $-\infty$ and $+\infty$
minus a ``correcting'' integral between $t_{\rm max}$ and $+\infty$.
As was discussed in Section~\ref{sec:IIIB} the latter integral can be estimated by successive integration by parts.
This gives [see Eq. (\ref{br9})] a first order correction term proportional to 
$e^{+i\pi/2}\,a(t_{\rm max}) e^{i\psi_f(t_{\rm max})}/
\dot{\psi}_f(t_{\rm max})$.
The {\it phasing} of this edge correction  can be made to agree perfectly
with the phasing  predicted by the form Eq. (\ref{3.23}) written with a generalized
argument $\zeta(f)$ if 
$(\psi_f(t_f) -\zeta^2)=  \psi_f(t_{\rm max})$. 
This leads us to define, in the domain $\zeta<0$, {\it i.e.} $f<F_{\rm max}$, the new
argument
\begin{equation}
\zeta_{<}\equiv - \sqrt{\psi_f(t_f)-\psi_f(t_{\rm max})}\,.
\label{3.25}
\end{equation}
In the left part of the crucial region Eq. (\ref{3.15}) the argument $\zeta_<(f)$ Eq. (\ref{3.25}) is nearly
identical to the previous result $\zeta_0(f)$ Eq. (\ref{3.24}), as is seen from Eq. (\ref{3.10}).
However, we have checked that  the replacement of $\zeta_0$ by $\zeta_<$ 
improves both the visual agreement and the overlap with the
exact $\tilde{h}(f)$.
Let us also note in passing that an amplitude  proportional to $ \zeta^{-1}$ of the correction term
to $\tilde{h}^{\rm uspa}(f)$ derived from the asymptotic expansion of Eq. (\ref{3.21})
is consistent with the different analytical treatment used in \cite{dkpo99} which was valid only
for $f\ll F_{\rm max}$, {\it i.e.} large, negative $\zeta$. By contrast our approach
based on the function 
${\cal C}(\zeta)$ is adequate in the full range $ -\infty<\zeta\le 0$  without exhibiting
any fictitious blowup at $\zeta=0$ (remember ${\cal C}(0)=1/2$).
[Our approach is also valid in the region $\zeta >0$, {\it i.e.} $f>F_{\rm max}$, but
for an improved treatment of this domain, we shall find it convenient to modify
further the argument $\zeta (f)$  in the following Section.]

 Summarizing: we  propose as final result for the {\it resonant} part of 
our {\it improved } 
SPA for {\it Newtonian-like} signals  (or inSPA)
\begin{equation}
f\le F_{\rm max}:\;\; \tilde{h}_{-<}^{\rm inspa}={\cal C}(\zeta_{<}(f)) \frac{a(t_f)}{\sqrt{\dot{F}(t_f)}}
e^{i\left(\psi_f(t_f)-\pi/4\right)}\,.
\label{3.26}
\end{equation}
The corresponding  total improved approximation
 $\tilde{h}^{\rm intot}$ to the Fourier transform $\tilde{h}=\tilde{h}_-(f)+\tilde{h}_+(f)$
 is the sum of Eqs. (\ref{br10}) and (\ref{3.26}).
Note that the ratio $\tilde{h}_-/\tilde{h}_+$
is (when $t_f$ is near $t_{\rm max}$) of order $4\pi F_{\rm max}/\sqrt{\dot{F}_{\rm max}}=
4\pi \sqrt{N_{\rm max}}$ (this is consistent with the $\varepsilon$ scaling of
Eqs. (\ref{br18}) and (\ref{br19}), remembering that $\varepsilon \sim 1/2\pi N$).
The contribution of $\tilde{h}_+$ is expected to be non-negligible {\it only}
for signals which are really discontinuous in time.
As the real signal (whatever be the subsequent plunge signal) will
be continuous (and even smooth) it is clear that one should {\em not}
add any contribution from $\tilde{h}_+$ when applying our above treatment 
to real signals.
In fact, we shall see below that, even for discontinuous signals,
the addition of $h_+$ has only a minute effect on overlaps.

\subsection{Approximate Fourier transform when $f>F_{\rm max}$}\label{sec:IIIC}

Let us now consider the evaluation of $\tilde{h}_-(f)$ in the 
 case when the Fourier variable $f$ is larger than $F_{\rm max}$
[but near $F_{\rm max}$, in the sense of Eq. (\ref{3.15})].
In that case  the integral Eq. (\ref{br2})
giving $\tilde{h}_-(f)$ no longer has 
a saddle-point. 
However, it `nearly'  has a saddle-point and therefore we expect that
\begin{equation}
\tilde{h}_-(f)=\int _{-\infty}^{ t_{\rm max}} dt \; a(t)\; e^{ i \psi_f( t)}\,,
\label{3.27}
\end{equation}
will still dominate over $\tilde{h}_+(f)$.
One could think of two ways of analytically approximating the integral
Eq. (\ref{3.27}). A first way is to still use the fact that (for Newtonian-like 
signals where the mathematical function $F(t)$ continues to exist and 
increase beyond $t=t_{\rm max}$) though there is no saddle-point in
the domain of integration $[-\infty,t_{\rm max}]$, there exists a nearby saddle-point
of the analytically continued phase function $\psi_f(t)$.
More precisely, for Newtonian-like signals the mathematical equation
$F(t_f)=f$ still defines a unique value $t_f$ (with $t_f>t_{\rm max}$ when $f>F_{\rm max}$).
Capitalizing on the existence of this nearby saddle-point one can still try
to insert the expansions Eqs. (\ref{3.10}) and (\ref{3.11}).
This leads to a result of the form Eq. (\ref{3.26}) with the correction factor Eq. (\ref{3.24}),
{\it i.e.} now considered for positive values of the argument
$\zeta_0(f)\equiv\sqrt{\pi \dot{F}(t_f)}(t_f-t_{\rm max})$.
In other words, a simple uniform expansion to $\tilde{h}_-(f)$ on both sides of
$f\sim F_{\rm max}$ would seem to be simply
\begin{mathletters}
\begin{eqnarray}
\tilde{h}_{-}^{\rm cspa}(f) &=& 
{\cal C}(\zeta_0(f))
\frac{a(t_f)}{\sqrt{\dot{F}(t_f)}} e^{ i\left[ \psi_f(t_f) -\pi/4\right]} \,,
\label{3.28}\\
{\rm with}\;\;{\zeta_{0}}(f)&=&  \sqrt{\pi \dot{F}(t_f)}(t_f-t_{\rm max})\,.
\label{3.30}
\end{eqnarray}
\end{mathletters}
Here `cspa' means  (zeroth order) {\it corrected} SPA.
Note that when $f>F_{\rm max}$, $t_f$, and therefore all the quantities evaluated at $t_f$, are
 defined by using the
(supposedly existing) analytic continuation of the mathematical function $F(t)$ beyond $t=t_{\rm max}$.

In our first attempts at improving the SPA in presense of a time-windowing we came up
with the  simple proposal Eqs. (\ref{3.28})-(\ref{3.30}) and it gave excellent overlaps with the
exact Fourier transform. However, we realised later that we could further improve
on this simple proposal.
We already stated that for $f<F_{\rm max}$ our best proposal is to modify the
argument Eq. (\ref{3.30}) into Eq. (\ref{3.25}).
In the case where $f>F_{\rm max}$ our best proposal is neither to use the straightforward
argument Eq. (\ref{3.30}), nor the ``improved-phasing'' argument Eq. (\ref{3.25})
with a positive sign in front of the square-root
[which, however, still improves over the choice Eq. (\ref{3.30})] but to
follow a different tack which will turn out to be useful when considering 
the case of relativistic-like signals in the next Section.

To motivate our proposal in the case $f>F_{\rm max}$, let us  remark that 
the integral to be
approximated, {\it i.e.} Eq. (\ref{3.27}) having  no saddle-point in the domain
of integration, 
is formally of the general type Eq. (\ref{bri7}) with the  phase
 $\psi(t)=\psi_f(t)$ being a monotonically increasing function of $t$.
The important information we wish to deduce from Eqs. (\ref{br8}) and (\ref{br9}) is that
there exists an expansion (valid when $f\gg F_{\rm max}$, {\it i.e.} 
$\zeta_0(f)$ is large and positive) in which $\tilde{h}_-(f)$ is entirely
expressed in terms of the values of the functions $\psi_f(t)$ and $a(t)$, and
their derivatives, {\it evaluated at the edge point} $t=t_{\rm max}$.
This contrasts with the `corrected' result Eqs. (\ref{3.28})--(\ref{3.30})
which relied on the existence of the functions $\psi_f(t)$ and $a(t)$ in the ``unphysical''
region $t>t_{\rm max}$.
This motivates us to look for an approximation to Eq. (\ref{br7}) valid all over the
domain $\zeta_0(f)>0$ [and not only when $\zeta_0\gg 1$ which will be seen to be the
domain of validity of Eqs. (\ref{br8}) and (\ref{br9})] but expressed {\it entirely}
in terms of the edge values of $\psi_f(t)$ and $a(t)$.
We propose to define such an approximation by replacing 
the phase and amplitude in Eq. (\ref{3.27}) by
\begin{mathletters}
\begin{eqnarray}
\psi_f(t)&\simeq&\psi_f(t_{\rm max}) + 2 \pi (f- F_{\rm max})(t-t_{\rm max}) -\pi
\dot{F}(t_{\rm max})(t-t_{\rm max})^2\,,
\label{3.35}\\
a(t)&\simeq&a(t_{\rm max})\,.
\label{3.36}
\end{eqnarray}
\end{mathletters}
Thanks to the parabolic nature of the approximation Eq. (\ref{3.35}) this again leads to an
incomplete complex Gaussian integral ({\it i.e.} a Fresnel integral) which can be evaluated as before in
terms of the complementary error function. Using Eq. (\ref{3.19}), 
this leads to our final proposal for the (nearly) resonant part of $\tilde{h}_-(f)$
 for Newtonian-like signals
\begin{mathletters}
\begin{eqnarray}
f \geq F_{\max}:
\tilde{h}_{->}^{\rm inspa}(f)&=& {\cal C}\left(\zeta_{>}(f)\right)
\frac{a(t_{\rm max})}{\sqrt{\dot{F}(t_{\rm max})}}\;
\exp {i \left [\psi_f(t_{\rm max})+ \frac{\pi (f-F_{\rm max})^2}{\dot{F}(t_{\rm max})}-
\pi/4\right]} \,,
\label{3.37}\\
{\zeta_{>}}(f)&=& \frac{\sqrt{\pi}(f-F_{\rm max})}{\sqrt{\dot{F}(t_{\rm
max})}}\,.
\label{3.38}
\end{eqnarray}
\end{mathletters}
Note that, in the parabolic approximation where Eq. (\ref{3.10}) or
Eq. (\ref{3.35}) hold, the function $\zeta_>(f)$ is approximately
equal both to $\zeta_0(f)$, Eq. (\ref{3.30}), and to the analytic continuation of 
Eq. (\ref{3.25}), {\it i.e.} 
$ \zeta_{<}(f)\equiv {\rm sign}(f-F_{\rm max})
\sqrt{ \psi_f(t_f)-\psi_f(t_{\rm max})}$.
Note also that the phase factor in Eq. (\ref{3.37}) (which is explicitly expressed in terms of
edge quantities) is nearly equal to the analytic continuation of the usual SPA phase factor
appearing in Eq. (\ref{3.26}), {\it i.e.} $\exp{[i\psi_f(t_f)-i\pi/4]}$.
Finally, as required, the expressions, Eqs. (\ref{3.26}) and (\ref{3.37}) match continuously
at $f=F_{\rm max}$ with common value
\begin{equation}
\tilde{h}_{-<}^{\rm inspa}(F_{\rm max})=
\tilde{h}_{->}^{\rm inspa}(F_{\rm max})
= \frac{1}{2} \frac{a(t_{\rm max})}{\sqrt{\dot{F}(t_{\rm max})}}
e^{i  \left(\psi_f(t_{\rm max})- \pi/4\right)} \,.
\label{3.40}
\end{equation}
Summarizing: our best analytical estimate for the Fourier transform of discontinuous
 Newtonian-like signals is the sum 
\begin{equation}
\tilde{h}^{\rm intot}(f)=
\tilde{h}_-^{\rm inspa}(f)+
\tilde{h}_+^{\rm edge}(f)\,,
\label{sha5}
\end{equation}
where 
$\tilde{h}_+^{\rm edge}(f)$
is approximated by Eq. (\ref{br10}) and where 
$\tilde{h}_-^{\rm inspa}(f)$
is given, when 
$f\le F_{\rm max}$ by Eq. (\ref{3.26}) and for
$f\ge F_{\rm max}$ by Eq. (\ref{3.37}).
As stated earlier, 
we shall in fact recommend that the 
edge correction $h^{\rm edge}_+$ be not included  when applying our result to real signals 
(we shall
also see that it brings only a negligible improvement to the overlaps of time-windowed
signals).

\subsection{Comparison between the  improved SPA, the usual SPA and the `exact' SPA
 (numerical DFT)}\label{sec:COMPARE}

Before proceeding to  a quantitative comparison of the various
approximants in Table~\ref{table:newtonian}, 
it is important to remark that one needs to be more specific when using
the terminology uSPA. One could compute the uSPA truncated at
the $F_{\rm LSO}$ that we refer to as uSPAw ( where `w' stands for
`windowed') or the uSPA truncated
at the Nyquist frequency that we designate as uSPAn (where `n' stands for `Nyquist').
In Table \ref{table:newtonian} we have listed the overlaps,
as defined by Eq. (\ref{d7}), of a signal model
generated in the time-domain and then Fourier transformed using a numerical DFT 
algorithm\footnote{ This defines for us the `exact' Fourier representation of the signal,
after due care has been taken to use a smooth time-window below $f_s$, and a high
enough sampling rate.}
with the same signal model but directly generated  in the frequency-domain using 
the usual SPA (uSPA), the corrected SPA (cSPA) and the improved Newtonian SPA (inSPA)
discussed earlier. 
For simplicity, we consider only equal-mass systems $(\eta=1/4)$ and parametrize
them by the total mass $m=m_1+m_2$.
The total mass is the crucial parameter which measures the location of
the $F_{\rm LSO}$ with respect to the bandwidth of the detector.
The parameter $\eta$ is also important because it determines the number
of cycles near the LSO [Eq. (\ref{2.12}) shows that $N(F_{\rm LSO})$ scales as
$1/\eta$]. The worst case (for the sensibility to the shutting off of the signal
after the LSO) is $\eta=\eta_{\rm max}=1/4$, and this is why we focus on this case. 
[We are also motivated by the fact that 1/4 being the maximum value
of the function $\eta(m_1,m_2)$, the observed  values of $\eta$, 
corresponding to a random sample of of $m_1$ and $m_2$, 
are expected to have an accumulation point at $\eta_{\rm max} = 1/4$.]
As we have checked, if our filters exhibit good overlaps for $\eta=1/4$ they will
have even better overlaps for $\eta<1/4$ and the same value for $m$.

\begin {figure}[t]
\caption{ The power per logarithmic bin of the squared SNR
$d(\rho^2)/d(\log f)=f\vert\tilde{h}(f)\vert^2/S_n(f)$
 for an arbitrarily normalised Newtonian signal
 computed from its DFT and its various approximate representations
computed up to the Nyquist frequency:
uSPAn, cSPAn and inSPAn. 
In the most sensitive range of frequencies, our final proposal
inSPAn agrees with the FFT quite well.
The uSPAn 
  grossly overestimates
 the 
true signal power at frequencies $f>F_{\rm LSO}$.
 The last stable orbit frequency in this case ($m_1=m_2=10 M_\odot$) 
is at about 220 Hz (the vertical line). 
Observe that, therefore, even uSPAw would significantly overestimate
the signal power up to $F_{\rm LSO}$.}
\centerline {\epsfxsize 6 true in  \epsfbox {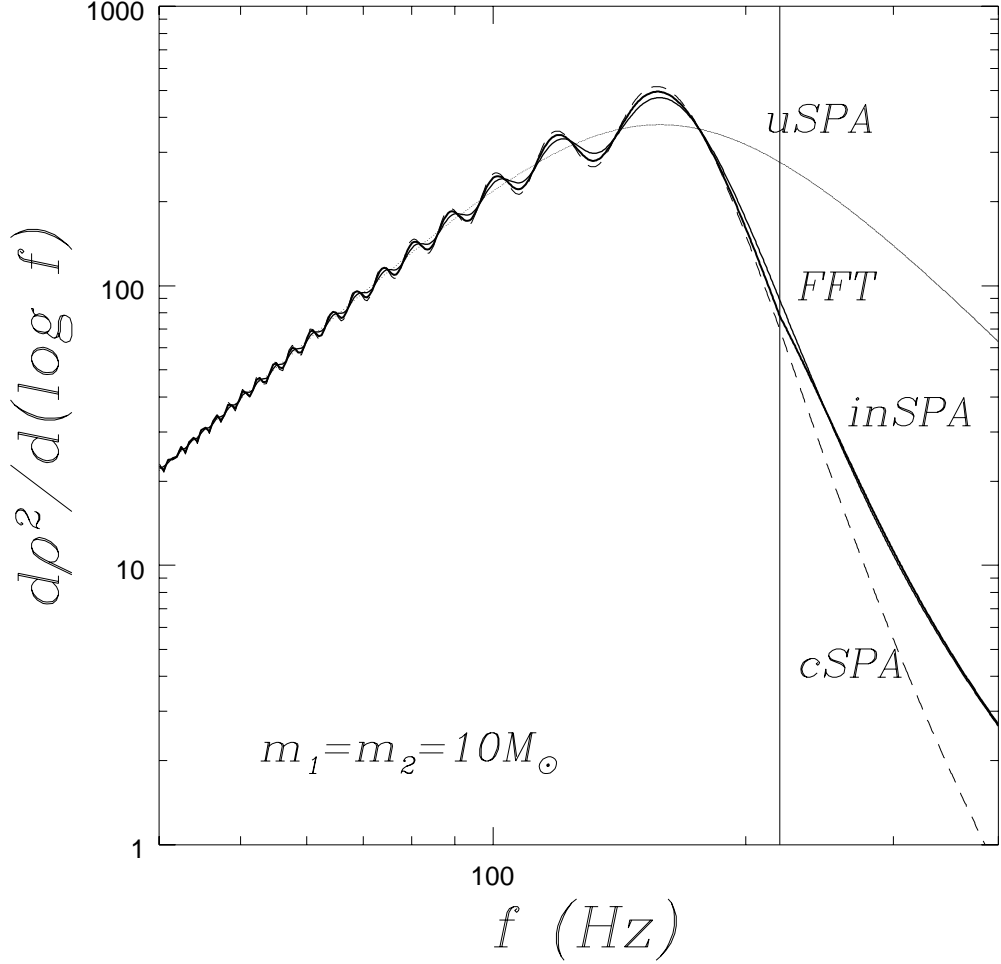} }
\label {fig:power.newt}
\end {figure}

The error function needed in computing
${\cal C}(f)$ is numerically computed using the NAG library S15DDF.
The overlaps are shown  for the usual SPA  with a frequency-windowing (uSPAw) 
together with the overlaps for the uSPAn,  cSPA and inSPA, 
computed up to $F_{\rm Nyquist}$.
Table \ref{table:newtonian} shows  that the improvements
on the SPA that we propose in this paper (both the simple cspa and our final
inspa) succeed very well in modelling the edge effects due to time-windowing.
The overlaps in the case of cSPA/inSPA are better than 0.99 for  equal-mass 
systems with total mass  $m< 40 M_\odot$. 
 For  a system of $m=40 M_\odot$ uSPAw  gives an overlap of
0.8589 resulting in a loss in the number of events by 37\%.
Although the overlaps of cSPA and inSPA seem to be always about the same,
we think that inSPA is a better representation of $\tilde{h}(f)$;
it has better overlaps in the case of the most massive
systems (see the first lines of Table \ref{table:newtonian}), and, as shown by 
Fig.~\ref{fig:power.newt}, it captures better the decay of $\tilde{h}(f)$ beyond $F_{\rm max}$.
In the Table for the usual SPA we have listed the overlaps
for uSPAw {\it i.e.}, uSPA  terminated at $f=F_{\rm LSO}=4400 (m_\odot)^{-1}$ Hz 
and  the overlap (uSPAn)   
up to the Nyquist frequency $f_{\rm Nyquist}=2~{\rm kHz}$.
As is very clear from these entries,
windowing of the SPA improves the overlaps for massive systems very much.
As remarked earlier, this is why in DIS, 
 while comparing the DFT to the SPA, the uSPAw was  used.
On the other hand, 
computing overlaps up to the Nyquist frequency, {\it i.e.} 
uSPAn, produces  much smaller  overlaps.

To understand this  further  we plot in Fig.~\ref{fig:power.newt} the power per
logarithmic bin of the squared SNR, $d\rho^2/d\log f=f\vert\tilde{h}(f)\vert^2/S_n(f)$,
which is the Fourier-domain quantity of most significance when discussing overlaps.
We compare this quantity for various approximations to the Fourier-transform of
an (arbitrarily-normalized) time-windowed signal:
 DFT,  uSPA, cSPA and inSPA. In the important range of
frequencies our best analytical approximant  inSPA agrees with the exact result (FFT)
quite well, the uSPA grossly overestimates and cSPA somewhat underestimates
the actual signal power.
This is why,  
though analytically continued up to $f_{\rm Nyquist}$, 
the uSPAn returns a smaller
overlap as compared to the windowed SPA (uSPAw) because it overestimates
the power in the signal beyond $F_{\rm LSO}$ 
\cite{effect.norm}.

In all the comparisons above, it is worth stressing that
 the  FFT calculation is delicate:
The `exact' time-domain chirp contains an infinite number of cycles
in the far past, with instantaneous frequencies tending to zero.
As what happens to frequencies  below the seismic cut-off $f_{\rm s}=40$ Hz
is not physically important, we wish to simplify the numerical calculation of
the FFT by essentially discarding the (infinite) part of the signal, having
instantaneous frequencies $F(t)< f_{\rm s}\equiv F(t_{\rm s})$.
We started doing that by simply time-windowing the signal for
$t<t_{\rm min}<t_{\rm s}$ by a sharp, lower time-window $\theta(t-t_{\rm min})$.
However, this method introduces
physically spurious oscillations (which are the lower-cutoff analogue
of the physically important upper-cutoff oscillations) present in both
$\tilde{h}_+(f)$ and $\tilde{h}_-(f)$.
One way to deal with this problem  is to subtract out from the FFT these
spurious edge oscillations by using the general formula Eq. (\ref{br9}),
which in the present context, can be applied both to $\tilde{h}_+^{\rm FFT}(f)$
and  $\tilde{h}_-^{\rm FFT}(f)$.
For instance, to lowest order the FFT corrected for these oscillations would read
\begin{mathletters}
\begin{eqnarray}
\tilde{h}_{\rm corrected}^{\rm FFT}&=& \tilde{h}^{\rm FFT} +\Delta^+_{\rm min}+\Delta^-_{\rm min}\,,\\
\label{low}
{\rm where},\; \Delta^{\pm}_{\rm min}&=& \frac{a(t_{\rm min})}{2\pi i (f\pm F_{\rm min})}
e^{i \psi_f^{\pm}(t_{\rm min})}\,,
\label{m1}
\end{eqnarray}
\end{mathletters}
where $F_{\rm min}=F(t_{\rm min})< f_{\rm s}$
[a better approximation can be derived from Eq. (\ref{br9})].
However, it seems better to use an alternative approach which does not require one to correct
by hand the FFT. The alternative approach we have actually used in our calculations consists
in imposing a {\it smooth} (rather than a sharp) lower time-window on the exact chirp,
acting below $t_{\rm s}$ and in a smooth-enough manner that it does
not introduce spurious edge-oscillations in the frequency-domain.
The smooth time-window that we used consists in multiplying the chirp
by the function 
\begin{equation}
\sigma(t,t_1,t_2)=\frac{1}{e^z+1}\,,\;\;z=\frac{t_2-t_1}{t-t_1} +
             \frac{t_2-t_1}{t-t_2} \,,
\label{br11}
\end{equation}
which smoothly interpolates between 0 when $t=t_1+0$ and 1 when $t=t_2-0$.
We used $t_1$ such that $F_1=F(t_1)=30$ Hz and 
 $t_2=t_{\rm s}$ {\it i.e.} $F(t_2)=f_{\rm s}$. 
Moreover, we need to be careful with sampling  and phase factors
to correctly reproduce the
edge  correction to $\tilde{h}_+^{\rm edge}$.             

In addition to the comparative evaluation of the various approximants,
Table~\ref{table:newtonian}  
also provides a numerical proof regarding the
effect of $h_+^{\rm edge}$ on the overlaps.
It is quite important to note that the inclusion of the non-resonant
edge term $h_+^{\rm edge}$ has only a very minute (but positive) effect
on overlaps.
This is good news for our formal time-windowing ansatz, because we
expect that this contribution will be (exponentially) negligible in the
case of real (continuous) signals. We interpret the fact that even for
our formal discontinuous model $h_+^{\rm edge}$ is negligible\footnote{
        Note that our statement here is only that $h_+^{\rm edge}(f)$
 can be effectively omitted without significantly worsening
 the overlaps. We are not claiming that
$h_+^{\rm edge}(f)$ is pointwise numerically negligible compared to $h_-(f)$. Indeed,
because the instantaneous number  of cycles is rather small near the LSO,
our analytical estimates above show that $h_+^{\rm edge}(f)$ is not very much
smaller than $h_-(f)$ near $f= F_{\rm LSO}$.}
as a confirmation that our improved SPA can adequately model not only
signals that vanish after the LSO, but also signals
that shut off rather quickly (on the $F_{\rm LSO}^{-1}$ time-scale) after the LSO.
It leads us also to propose, finally, to use as analytical representative
of the FT of real signals the $h_-^{\rm inspa}$ part of our formula above
(without the edge term). [To simplify the notation, we shall henceforth
drop the extra subscript minus on $h^{\rm inspa}$.]

In computing the above overlaps we have matched all
the parameters of the two waveforms, including the time of arrival and the
starting phase\footnote{
 The lag is set equal to zero in testing the accuracy of the Fourier
representation but chosen optimally when testing {\it faithfulness} of a family
of templates e.g. in Sec. \ref{sec:FESPP}.}. 
The overlaps in this Table as well as all other Tables
in this paper are found to be insensitive to the sampling rate at 
the level of a fraction of a percent, provided that it is large enough 
to obey Shannon's sampling theorem.

\section {Improved stationary phase approximation 
for relativistic  signals in the adiabatic approximation: 
The SPP-approximants}\label{sec:SPP}

Though one might a priori think that it is a simple matter to generalize
the improved SPA discussed above for Newtonian-like signals  to the
relativistic case, it does not turn out to be so.
What complicates matters is that there are serious qualitative, ``non-perturbative''
differences between the two cases: first, the value of $\dot{F}(t)$ formally
 tends to
$+\infty$ at the last stable orbit (LSO) which physically defines the
upper-cutoff $t_{\rm max}$ of the inspiral signal, and, second, the mathematical function
$F(t)$ does not admit a unique real analytic continuation beyond $t_{\rm max}=t_{\rm LSO}$.
(These two facts are evidently related; indeed we shall see that $F(t)$ behaves
in the non-analytic manner $F(t)\sim c_1+c_2(t_{\rm LSO}-t)^{\frac{1}{2}}$ when
$t\rightarrow t_{\rm LSO}^-$). Remembering the crucial role of a finite $\dot{F}(t)$
in the results Eqs. (\ref{3.26}) and (\ref{3.37}), it is clear that we need to tackle afresh the
problem of finding a good, analytic approximation to $\tilde{h}_-(f)$.
Similarly, in view of the appearance of $\dot{F}^{+1}(t_{\rm max})$
in the next-to-leading contribution to $\tilde{h}_+(f)$ Eq. (\ref{br10}),
we shall also need to revisit the calculation of $h_+(f)$ (though we
shall, again, find that it makes only a negligible contribution to the overlaps.)

\subsection{The phasing formula for relativistic signals in the adiabatic approximation}
\label{sec:IVA}

To extend the treatment of the previous Section and go beyond the Newtonian
approximation, let us begin
with  the phasing formulas for gravitational waves from compact binaries
written in a parametric form in terms of the  variable
$v_F\equiv (\pi m F)^{1/3}$ defined by the total mass $m=m_1+m_2$
and instantaneous  gravitational wave frequency $F$
\begin{equation}
t(v_F) = t_{\rm LSO} +m \int_{v_F}^{v_{\rm LSO}} dv \, \frac{{E}'(v)}{{\cal F}(v)} \, ,
\label{4.1}
\end{equation}
\begin{equation}
\phi (v_F) = \phi_{\rm LSO} + 2 \int_{v_F}^{v_{\rm LSO}} dv v^3 \, \frac{{E}'(v)}{{\cal F}
(v)} \, ,
\label {4.2}
\end{equation}
 where $E(v)$ is the dimensionless energy function related to the total
relativistic energy or Bondi mass
 by $ E_{\rm total}=m(1+ E)$, $\,{\cal F}(v)$  the flux function
denoting the
gravitational wave  luminosity
 of the system and  $t_{\rm LSO}$ is the time and $ \phi _{\rm LSO}$ is  the phase of the signal when $
v=v_{\rm LSO}$.
The parametric representaion  Eqs. (\ref{4.1} (\ref{4.2} of the phasing
formula $\phi = \phi(t)$ holds under the assumption of `adiabatic inspiral',
{\it i.e.}, that gravitational radiation damping can be treated as
 an adiabatic perturbation of a circular motion. See \cite{bdprep}
 for a treatment of radiation damping going beyond this approximation.

 In the restricted post-Newtonian approximation, one uses a Newtonian approximation for the
amplitude \cite{3mn}.
However, in order to extract an inspiral signal that may be buried in noisy data
by the method of matched filtering, 
we need to employ  post-Newtonian accurate representations  for the two
functions
${E}' (v)$ and ${\cal F}(v)$ that appear in the above phasing formulas.
To any approximant $E_A (v)$, ${\cal F}_A (v)$,
correspond [by replacing $E(v)\rightarrow E_A(v)$, 
${\cal F}(v)\rightarrow {\cal F}_A(v)$ in Eqs. (\ref{4.1}) and (\ref{4.2})]
some approximate parametric representation $t=t_A(v_F)$,
$\phi=\phi_A(v_F)$, and therefore a corresponding approximate
time-domain template 
\begin{equation}
h^A = h^A (t;{\cal C},t_{\rm LSO} ,\phi_{\rm LSO} ,m,\eta) \, , 
\label{4.3}
\end{equation}
obtained by replacing $v_F$, in the following $v_F$-parametric representation
of the waveform 
\begin{equation}
h^A (v_F) = {\cal C} \, v_F^2 \, \cos \, \phi_A (v_F) \, , 
\label{4.4}
\end{equation}
by the function of time $v_F=v_A(t)$ obtained by inverting
$t=t_A(v_F)$.

The standard approximants for $E(v)$ and ${\cal F}(v)$ are simply  their
successive Taylor approximants  $E_{T_n}$ and ${\cal F}_{T_n}$ respectively.
The DIS strategy
for constructing new approximants to $E(v)$ and ${\cal F}(v)$ 
 is two-pronged:
 Starting from the more basic energy-type and flux-type functions,  $ e(v)$
and $l(v)$ \cite{dis98} we construct Pad\'e-type approximants, say $e_{P_n}$,
$l_{P_n}$, of the ``basic'' functions $ e(v)$, $l(v)$\footnote{For explicit formulas representing $E(v)$ and ${\cal F}(v)$
see Eqs. (3.8),(4.2) and  (4.3) of DIS.
The associated $e(v)$ and $l(v)$ functions are given by Eqs. (3.7), (3.9) and
Eqs. (4.4)-(4.9) in DIS . See also Eqs. (3.5), (3.11) and 
Eqs. (3.18)-(3.23) there.}.
We then compute the required energy and flux functions entering the phasing formula.
The successive approximants
$ E[e_{P_n}]$ and  $ {\cal F}[e_{P_n} , l_{P_n}]$ 
have better convergence properties than  their Taylor counterparts 
$ E_{T_{n}}[e_{T_n}]$ and  $ {\cal F}_{T_{n}}[e_{T_n} , l_{T_n}]$. 
In DIS we were working directly with the time-domain signal $h(t)$.
As explained above  this necessarily  requires a numerical inversion of   the parametric
representation $t=t(v_F)$.
By contrast, if one wants to compute the usual stationary phase approximation
of $h(t)={\cal C} v_F^2(t)\cos\phi(v_F(t))$
there is no need to invert this parametric representation.
Indeed, from Eq. (\ref {3.13}), it is sufficient to know
the instantaneous amplitude and the phase at the time $t_f$ where 
$f=F(t_f)$.
This time is simply given by the same expression Eq. (\ref{4.1})
as above with the replacement of $v_F\equiv(\pi m F)^{1/3}$ by
$v_f\equiv (\pi m f)^{1/3}$, {\it i.e.} the stationary point $t_f$
is given by 
\begin{equation}
 t_f = t_{\rm LSO} + m \int_{v_f}^{v_{\rm LSO}} \frac{E'(v)}{{\cal F}(v)} dv \,.\label{s1}
\end{equation}
One then substitutes this value
of $t_f$ in Eq. (\ref{4.2}) to compute  
the phase $\psi_f(t_f)\equiv 2\pi f t_f -\phi(t_f)$ of the Fourier component:
\begin{equation}
 \psi_f(t_f) = 2 \pi f t_{\rm LSO} - \phi_{\rm LSO} + 2\int_{v_f}^{v_{\rm LSO}} (v_f^3 - v^3)
\frac{E'(v)}{{\cal F}(v)} dv \,.\label{s2}
\end{equation}
In terms of these quantities one has 
\begin{equation}
\tilde{h}^{\rm uspa}(f)=\frac{1}{2} {\cal C} \frac{v_f^2}{\sqrt{\dot{F}(t_f)}}
e^{i\left[\psi_f(t_f)-\frac{\pi}{4}\right]}
\label{d4.6a}
\end{equation}
The inclusion of relativistic effects in $\tilde{h}^{\rm uspa}(f)$ is
then simply accomplished by using relativistic accurate expressions for
$E'(v)$ and ${\cal F}(v)$ in the formulas giving $\psi_f(t_f)$
and $\dot{F}(t_f)$.
The coefficient ${\cal C}$, in Eq. (\ref{4.4}), determining the actual
amplitude of the waveform reads:
\begin{equation}
{\cal C}\left(r, i, \theta, \bar{\phi}, \bar{\psi}\right) = 
(4\eta) \left(\frac{m}{d}\right)
	 C\left(i, \theta, \bar{\phi}, \bar{\psi}\right)   ,
\label{kt1}
\end{equation}
 where $d$ is the distance to the source, and where
\begin{mathletters}
\begin{eqnarray}
C\left (i, \theta, \bar{\phi}, \bar{\psi}\right)& =& \sqrt{ A^2 + B^2},\\
\label{kt2}
 {\rm with},\;\;
 A&=& \frac{1}{2} \left( 1 + \cos^2 i\right) F_+;\;\; B = \cos i\;\, F_\times,
\label{kt3}
\end{eqnarray}
\end{mathletters}
 with the beam-pattern factors
\begin{mathletters}
\begin{eqnarray}
F_+\left(\theta, \bar{\phi}, \bar{\psi}\right) &=& 
\frac{1}{2} \left( 1 + \cos^2 \theta\right) 
\cos 2\bar{\phi} \cos 2\bar{\psi} -  \cos \theta \sin 2\bar{\phi} \sin 2\bar{\psi},\\
\label{kt4}
F_\times\left(\theta, \bar{\phi}, \bar{\psi}\right) &=& 
\frac{1}{2} \left( 1 + \cos^2 \theta\right) 
\cos 2\bar{\phi} \sin 2\bar{\psi} +  \cos \theta \sin 2\bar{\phi} \cos 2\bar{\psi}.\\
\label{kt5}
\end{eqnarray}
\end{mathletters}
 In these formulas the angle $i$ denotes the inclination of the orbit with respect
to the plane of the sky, and  the angles $\theta$, $\bar{\phi}$, 
and $\bar{\psi}$ parametrize both the propagation direction
 and the polarization of the gravitational wave with respect to the
detector (see \cite{kt87} for exact definitions; 
we added a bar over $\phi$ and $\psi$
 to distinguish them from the GW phase $\phi$ and Fourier phase $\psi$ respectively). 
Performing averages over the angles in the squared SNR
leads to : 
\begin{equation}
\langle F_+^2\rangle_{\theta,\bar{\phi},\bar{\psi}} = 
\langle F_\times^2\rangle_{\theta,\bar{\phi},\bar{\psi}}= \frac{1}{5},
\label{kt6}
\end{equation}
 and finally 
\begin{equation}
\langle C^2\rangle_{i,\theta,\bar{\phi},\bar{\psi}} = \frac{4}{25}.
\label{kt7}
\end{equation}
We are finally in a position to write down the rms and 
ideal SNRs. For a binary at a distance $d$ from the  earth 
consisting of stars of individual masses $m_1$ and $m_2$
(total mass $m\equiv m_1+m_2$ and symmetric mass 
ratio $\eta= m_1m_2/m^2)$ the rms and ideal SNRs,
obtained by using the rms and ideal values of $C,$ namely
$C=2/5$ and $C=1,$ respectively, 
when replacing Eq. (\ref{d4.6a}) in Eq. (\ref{2.4}), or
  equivalently, when replacing $a(f)=(1/2) {\cal C} v^2(f) =
  2 \eta m d^{-1} C v^2(f)$ in Eq. (\ref{2.6}) (with Eq. (\ref{2.11}) and a truncation
  at $F_{\rm LSO}$), are given by
\begin{equation}
\rho_{\rm rms} = \frac {m^{5/6}}{d\pi^{2/3}} 
\left ( \frac{ \eta}{15} \right)^{1/2}
\left [ \int_0^{F_{\rm LSO}} df 
\frac{f^{-7/3}}{S_n(f)} \right]^{1/2},\ \
\rho_{\rm ideal} = \frac{5}{2} \rho_{\rm rms}.
\label{final snr}
\end{equation}
Note that the SNR depends only on the combination 
${\cal M}=m\eta^{3/5}$ -- the chirp mass (see e.g. \cite{fc93}),
and that the first Eq. (\ref{final snr}) is equivalent to Eq. (\ref{2.6new}).

\begin {figure}[t]
\caption{ The instantaneous GW frequency $F(t)$  versus $t$ 
for the Newtonian and relativistic cases during the last 
few orbits before the plunge at LSO. Notice the rapid increase in the inspiral rate
close to the last stable orbit in the relativistic case.}
\centerline {\epsfxsize 6 true in  \epsfbox {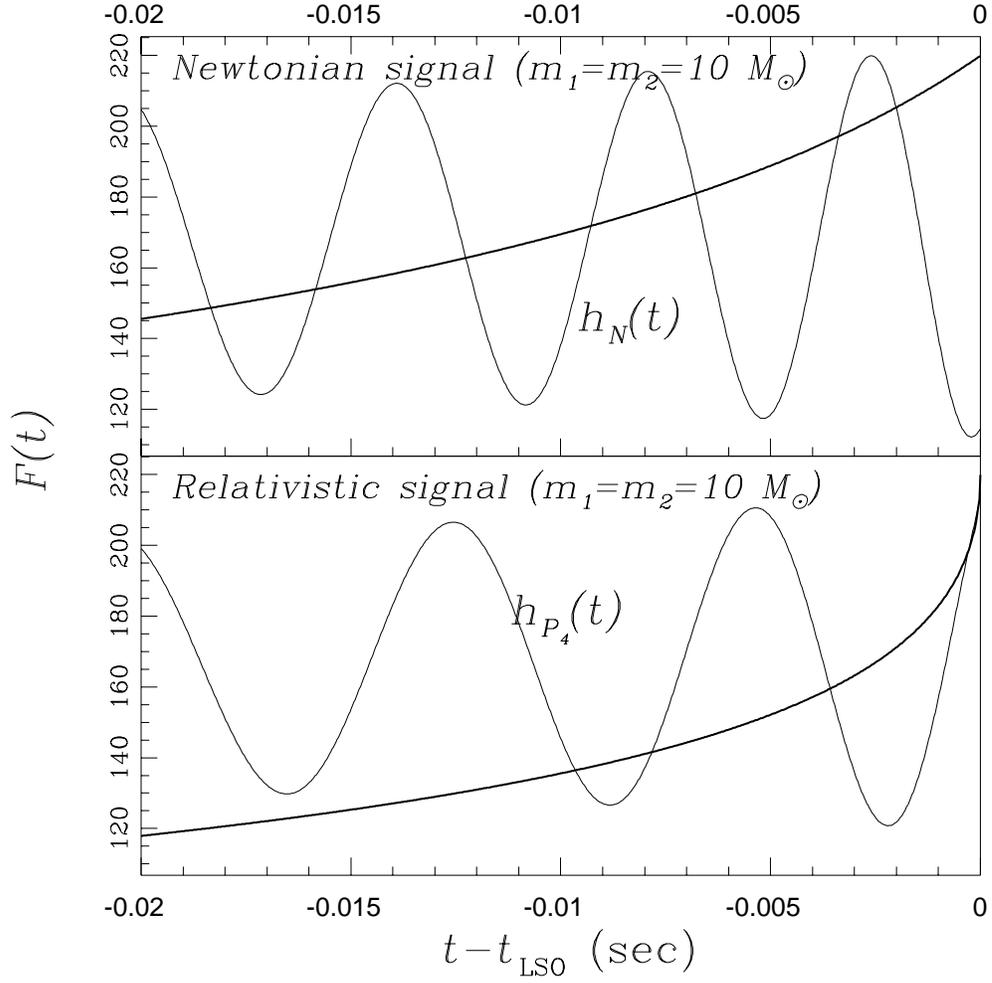} }
\label {fig:GWfreq}
\end {figure}

Let us  next delineate the qualitative differences between the relativistic and non-relativistic
cases by considering the function appearing as denominator in the uSPA,
Eq. (\ref{d4.6a})
\begin{equation}
\dot{F}(t)=\frac{1}{2\pi}\frac{d^2\phi}{dt^2}=-\frac{3v^2}{\pi m^2}  \frac{{\cal F}(v)}{E'(v)}.
\label{4.5}
\end{equation}
At the LSO, the gravitational wave flux ${\cal F}(v)$ is finite (it blows up
only later, when reaching the light ring \cite{dis98}) while, by definition, $E'(v)$ vanishes
linearly, $E'(v)\propto v-v_{\rm LSO}$. 
As we shall see below this means that $\dot{F}(t)$ blows up
as $(t_{\rm LSO} -t)^{-1/2}$.
A consequence of this blow up is that the last two terms in Eq. (\ref{d3.13b})
blow up like $(t_{\rm LSO}-t)^{-3/2}$ confirming the need for a special
treatment of the Fourier transform near the LSO.
We are here speaking of the exact behaviour of the functions $E(v)$ and ${\cal F}(v)$, as supposedly known
from combining the test-mass limit results \cite{test} with the best available results on the
physics underlying the existence of the LSO \cite{dis98}, and the emission of gravitational waves
in comparable mass systems \cite{bdiww95}. In DIS, we have incorporated this information
so that all the P-approximants $E_{P_n}\equiv E(e_{P_n}),\, {\cal F}_{P_n}\equiv {\cal F}[e_{P_n},l_{P_n}]$
that we define share, with the ``exact'' functions $E$ and ${\cal F}$ the crucial properties mentioned
above ({\it i.e.} finite ${\cal F}(v_{\rm LSO})$ and $E'(v)\propto v-v_{\rm LSO}$).
The (less-convergent) successive
T-approximants $E_{T_n}$
and ${\cal F}_{T_n}$ do not incorporate this information exactly, and only
as $n$ increases they tend to incorporate it.
In our opinion  the $T_n$ approximants   disqualify as `relativistic'
approximants since they  do not consistently incorporate 
the expectation (based on several different methods; see references in \cite{bd99})
that the frequency at the LSO is (for any $\eta\leq 1/4$)
numerically near the Schwarzschild-like prediction,
Eq. (\ref{d8}). Indeed, if we define the 2PN Taylor estimate of $F_{\rm LSO}$
by the value of $v=(\pi m F)^{1/3}$ where the straightforward
Taylor approximant  $E_{T_{4}}(v)=\sum _{k=0}^4 E_k(\eta) v^k$ 
reaches a minimum, we find, e.g. that
(i) when $m=40 M_\odot$ and $\eta=0$, $F_{T_4}=200$ Hz, very 
different from the exact value of 110 Hz, and
(ii) when $m=40 M_\odot$ and $\eta=1/4$, that $F_{\rm LSO}^{T_{4}}=221.4$ Hz,
very different from the other predictions $F^{P_4}_{\rm LSO}=143$ Hz and
$F^{{\rm Ref.}\cite{bd99}}_{\rm LSO}=118.6$ Hz.
We compare and contrast in Fig.~\ref{fig:GWfreq} 
the Newtonian and relativistic behaviours of the
wave amplitude and  instantaneous frequency $F(t)$ during the last couple of orbits before the LSO.
The blow up of $\dot{F}(t)$, {\it i.e.} the fact that the slope of $F(t)$ becomes vertical is
an effect which is localized in the last part of the last cycle before the LSO.
Note also in Fig.~\ref{fig:GWfreq} that a less localized consequence of this blow up
is that the average frequency a few cycles before the LSO is {\it smaller}
(for a given $F_{\rm LSO}$) in the relativistic case, than in the (unphysical) Newtonian one.
 Note that the physical origin of the blow up of $\dot{F}$ is that, just before the LSO
the `effective potential' for the radial motion becomes very flat
(before having an inflection point at the LSO).
In picturesque terms, the radial motion becomes ``groundless'' at the LSO.
Evidently, the blow up of $\dot{F}$ is due to our use of the `adiabatic'
approximation down to the LSO.
In reality, radiation reaction will cause a progressive transition between
the inspiral and plunge which will modify the evolution of $F(t)$ in the last cycle
before the LSO.
We shall discuss this issue in detail in a forthcoming paper \cite{bdprep} 
and subsequently its data analysis consequences.

\subsection{ Edge contribution to the non-resonant relativistic $\tilde{h}_+(f)$}
\label{sec:IVB}

As in the Newtonian-like case
we decompose $\tilde{h}(f)$ in two contributions, 
Eqs. (\ref{br2}) and (\ref{br3}).
The non-resonant contribution $\tilde{h}_+(f)$ will be dominated by the `edge' contribution 
to an integral of our usual type Eq. (\ref{br15}).
Though the problem is similar to the one we have generically solved 
in Section~\ref{sec:IIIB} we {\it cannot} apply the results Eqs. (\ref{br20}),
(\ref{br9}), (\ref{bra10}), (\ref{br10}),
because of the limiting hypothesis (iii) mentioned in our introductory discussion
Section~\ref{sec:TTC}. Indeed, the problem is that, in the (physically relevant) case of {\it relativistic}
signals the functions $a(t)$ and $\psi(t)$ are {\it not} smooth at the upper
edge $t=t_{\rm LSO}$. Let us see explicitly in what way they violate smoothness there.
 Let us first  define,
\begin{equation}
e_1(\eta)\equiv\left[\frac{d}{dv}\left(\frac{{E}'(v)}{{\cal
F}(v)}\right)\right]_{v_{\rm LSO}}\,,
\label{4.8}
\end{equation}
 so that near the LSO we may write:
\begin{equation}
\frac{{E}'(v)}{{\cal F}(v)}= e_1(v-v_{\rm LSO}) +{\cal{O}}[(v-v_{\rm LSO})^2]\,.
\label{4.9}
\end{equation}
If we were to use the test-mass approximation for the energy function $E'(v)$ and the 
Newtonian (quadrupole)
one for the flux function ${\cal F}(v)$ this would give
\begin{equation}
e_1^{P_0{\rm (tm)}}(\eta)\simeq 
\frac{15}{2}\,\frac{1}{4\eta}\frac{1}{v_{\rm LSO}^8(1-3v_{\rm LSO}^2)^{\frac{3}{2}}}
\simeq\frac{27492}{4\eta}\,.
\label{e1}
\end{equation}
 We have numerically estimated the function 
$\overline{e}_1^{P4}(\eta)\equiv 4 \eta e_1(\eta)$, when using the $P_4$-approximant
of \cite{dis98} in the definition of Eq. (\ref{4.8}).
We find that to a good approximation 
\begin{equation}
 4 \eta e_1^{P_{4}}(\eta)\equiv \overline{e}_1^{P4}(\eta)\simeq 
 26091.61194 \,\exp {(-4.474405683 \eta)}\,.
\label{sha6}
\end{equation}
 In terms of
\begin{equation}
\tau\equiv\frac{t_{\rm LSO}-t}{m}\;,\; \tau\ge 0\;\; {\rm for}\;\; v\le v_{\rm
LSO}\,,
\label{4.10}
\end{equation}
and using
\begin{equation}
\tau=-\int_v^{v_{\rm LSO}}\,dv\,\frac{E'(v)}{{\cal F}(v)}=\frac{1}{2}e_1(v-v_{\rm LSO})^2 +
{\cal O}[(v-v_{\rm LSO})^3]\,,
\label{4.11}
\end{equation}
and Eq. (\ref{4.2}) for $\phi(v)$, we find the following approximate representation 
(valid near the LSO) for the phase $\phi(t)$:
\begin{mathletters}
\begin{eqnarray}
t-t_{\rm LSO}&=& -m \tau\,,
\label{4.12}\\
\phi(t) -\phi(t_{\rm LSO})&\simeq& - 2 v_{\rm LSO}^3 \tau
+\frac{4\sqrt{2}}{\sqrt{e_1}} v_{\rm LSO}^2 \tau^{3/2}\,.
\label{4.13}
\end{eqnarray}
\end{mathletters}
Note also that Eq. (\ref{4.11}) gives the following representation for $v(t)$,
and therefore for the amplitude $a(t)={\cal C} v^2(t)$
\begin{mathletters}
\begin{eqnarray}
v&\simeq& v_{\rm LSO} -\frac{\sqrt{2}}{\sqrt{e_1}}\tau^{\frac{1}{2}}\,,
\label{br21}\\
a(t)&\simeq&a_{\rm LSO}\left[1-\frac{2\sqrt{2}}{\sqrt{e_1}}\,\frac{1}{v_{\rm LSO}}\,\tau^{\frac{1}{2}}\right]\,.
\label{br22}
\end{eqnarray}
\end{mathletters}
We are interested in evaluating the edge contribution to the integral
\begin{equation}
I=\tilde{h}_+(f)=\int_{-\infty}^{t_{\rm LSO}}
\, dt\, a(t) e^{i\psi_f^+}(t)=
m\int_0^{+\infty}\,d\tau\,a(\tau)\,e^{i\psi_f^+(\tau)}\,.
\label{br23}
\end{equation}
Near the LSO boundary {\it i.e.} near the edge $\tau=0$ in
the $\tau$-form of the integral, the amplitude behaves as
Eq. (\ref{br22}) while the appropriate phase $\psi_f^+(\tau)$ behaves, from Eqs. (\ref{4.12}) and (\ref{4.13})
as
\begin{equation}
\psi_f^+(\tau) \simeq \psi_{f{\rm LSO}}^+ - 2\pi m (F_{\rm LSO}+f)\tau +
\frac{4\sqrt{2}}{\sqrt{e_1}}v_{\rm LSO}^2 \tau^{3/2}\,,
\label{br24}
\end{equation}
where 
\begin{equation}
\psi_{f{\rm LSO}}^+ \equiv \psi_f^+(t_{\rm LSO})=2\pi f t_{\rm LSO}+\phi_{\rm LSO}\,.
\label{br25}
\end{equation}
The appearance of fractional powers of $\tau$ in the expansions 
Eqs. (\ref{br21}) and (\ref{br22}) show explicitly the violation
of the ${\cal C}^\infty$ property of $a(t)$ and $\psi(t)$ at the edge.
We cannot use the integration-by-parts method to evaluate the
expansion of $I_{\rm edge}$.
However, we can still use the general method sketched in Section~\ref{sec:TTC}.
Without rotating explicitly the the $\tau$-contour in the complex
plane the edge contribution to $I$ is obtained by inserting the expansions
Eqs. (\ref{br22}) and (\ref{br24}) in Eq. (\ref{br23}) 
and expanding everything out, except for the main phase,
$\psi_{f{\rm LSO}}^+-2\pi m (F_{\rm LSO}+f)\tau$ which must be
kept in the exponent. This yields
\begin{mathletters}
\begin{eqnarray}
I_{\rm edge}&=&ma_{\rm LSO}e^{i\psi_{f{\rm LSO}}^+}\int_0^\infty\,d\tau \,e^{-iy\tau}
\left(1-\frac{2\sqrt{2}}{\sqrt{e_1}}\,\frac{1}{v_{\rm LSO}}\,\tau^{\frac{1}{2}}+
\frac{i4 \sqrt{2}}{\sqrt{e_1}}v_{\rm LSO}^2\tau^{\frac{3}{2}}\right)\,,\\
\label{br26}
{\rm where},\; y&\equiv&2\pi m\left(F_{\rm LSO}+f\right)\,.
\label{br27}
\end{eqnarray}
\end{mathletters}
Note that, instead of rotating $\tau$ in the complex plane, we can (equivalently) consider that $y$
possesses a small negative imaginary contribution: $y\rightarrow y-i0$.
The integrals appearing in Eq. (\ref{br26}) are evaluated by the general formula
\begin{equation}
i_{\alpha}=\int_0^\infty\,d\tau e^{-iy\tau}\tau^\alpha=
\frac{e^{-i\frac{\pi}{2}(\alpha+1)}}{y^{\alpha+1}}\Gamma(\alpha+1)\,.
\label{br28}
\end{equation}
This yields finally
\begin{equation}
\tilde{h}_+^{\rm edge}(f)\simeq  \frac{m a_{\rm LSO}e^{ i[\psi^+_{f{\rm LSO}}]}}{iy}
         \left[ 1+ \left(\frac{3}{2}\frac{ F_{\rm LSO}}{F_{\rm LSO}+f}-1\right)
         e^{-i\pi/4} \frac{\sqrt{2\pi}}{\sqrt{e_1}}\frac{1}{v_{LSO}\sqrt{y}}\right]\,.
\label{br29}
\end{equation}
The leading contribution $(\propto (iy)^{-1})$ to the relativistic result Eq. (\ref{br29}) agrees with
the leading contribution in Eq. (\ref{br10}). Note that the next-to-leading contribution does not have the same dependence on
$f+F_{\rm LSO}$ as the corresponding term in the non-relativistic result in Eq. (\ref{br10}).
In spite of the breakdown of the formal expansion Eq. (\ref{br10}) the fractional correction
given by the last term in the bracket of Eq. (\ref{br29}) is checked to be numerically small. 
This check was the main motivation for us to compute $\tilde{h}_+^{\rm edge}$ to
next-to-leading order in the relativistic case. The lesson is that the
formal blow up of $\dot{F}$ near the LSO has only a small
numerical effect on $\tilde{h}_+^{\rm edge}$.
This is again a confirmation that our results are robust under a refinement
of our knowledge of the signal. We shall further check
below that, as in the Newtonian case, $\tilde{h}_+^{\rm edge}$ has only a negligible effect
on overlaps.

\subsection{Improved stationary phase approximation for relativistic signals}
\label{sec:IVC}
Let us now consider the resonant contribution 
$\tilde{h}_-(f)$, considered in the crucial domain where the stationary
point is near the edge $t_{\rm LSO}$.
As before also, 
 the optimal approximants
to $\tilde{h}_-(f)$ that we can construct are given by different analytical
expressions according to the value of $f$. However, we need now to introduce a new
definition of the two ranges of frequencies in which one must (minimally)
divide the $f$-axis. More precisely, we introduce a frequency $f_{\rm up}$,
near but below $F_{\rm max}$, and we shall construct a ``lower''
approximation $\tilde{h}_{-<}(f)$ in the range $f<f_{\rm up}$, and an ``upper''
one $\tilde{h}_{->}(f)$ in the range $f>f_{\rm up}$ (which includes
$f=F_{\rm max}$). The optimal value of $f_{\rm up}$ will be determined below.

In the lower range, $f<f_{\rm up}$, we can draw on the work of Sec.~\ref{sec:IIIB}. 
Indeed,
in that range there exists a saddle-point in the domain of integration.
However, as that saddle-point can become rather near $t_{\rm max}$ (because
$f_{\rm up}$ is near $F_{\rm max}$), we can significantly improve the usual SPA estimate by using
our previous result, {\it i.e.} by defining
\begin{mathletters}
\begin{eqnarray}
f\le f_{\rm up}\;:\;\;\tilde{h}_{-<}^{\rm irspa}(f)&=& 
{\cal C}\left(\zeta_{<}(f)\right)
\frac{a(t_f)}{\sqrt{\dot{F}(t_f)}}e^{i  \left[\psi_f(t_f)- \pi/4\right]} \,,
\label{4.6}\\
{\zeta_{<}}(f)&=& -\sqrt{\psi_f(t_f)-\psi_f(t_{\rm max})}\,.
\label{4.7}
\end{eqnarray}
\end{mathletters}
The label `irspa' in Eq. (\ref{4.6}) stands for {\it improved relativistic} SPA.

Let us finally explore the optimal analytic approximation to $\tilde{h}_-(f)$
in the upper range $f\ge f_{\rm up}$.
Proceeding as in Section~\ref{sec:IVB} in this case one has
\begin{equation}
\psi_f^-(t) \simeq \psi_{f{\rm LSO}}^-+ 2\pi m (F_{\rm LSO}-f)\tau -
\frac{4\sqrt{2}}{\sqrt{e_1}}v_{\rm LSO}^2 \tau^{3/2}\,,
\label{4.14}
\end{equation}
where 
\begin{equation}
\psi_{f{\rm LSO}}^-\equiv \psi_f^-(t_{\rm LSO})=2\pi f t_{\rm LSO}-\phi_{\rm LSO}\,.
\label{d4.16a}
\end{equation}
We shall use this expansion 
(which replaces the parabolic approximation Eq. (\ref{3.10}) used in the
Newtonian case) to evaluate the Fourier integral Eq. (\ref{3.27}). To this end we must introduce a new special
function  (characteristic of the relativistic phasing near the LSO) to replace the error function.
Let us define the function
\begin{equation}
g_{\frac{3}{2}}(x)\equiv\int_0^{\infty} d\hat{\tau} e^{i \left(3 x \hat{\tau}  - 2 \hat{\tau}^{3/2}\right)}\,,
\label{4.15}
\end{equation}
where the new variable $\hat{\tau}$ is related to $\tau$ by  $\tau=\alpha \hat{\tau}$
where 
\begin{equation}
\alpha=\frac{1}{2}v_{\rm LSO}^{-4/3}e_1^{1/3}\,.
\label{4.16}
\end{equation} 
In the test mass case corresponding to  Eq. (\ref{e1}) the value of
$\alpha$ is $49.83/(4\eta)^{1/3}$. 
The value defined by the $P_4$ approximant on the other hand is given by
combining Eqs. (\ref{dis23}) and (\ref{sha6}). In particular,
 $\alpha$ equals  30.055 (54.578) for $\eta= $ 0.25 (0.1) respectively. 
The index $\frac{3}{2}$ in $g_{\frac{3}{2}}(x)$
 alludes to the power $\hat{\tau}^{3/2}$ replacing the power $\hat{\tau}^2$
in the usual error function, and where the conventional coefficients $3$ and $2$ have been chosen
to simplify some formulas (although they complicate others!).
 The final result is
conveniently written in terms of a variable $x$ given by
\begin{equation}
x=\frac{2\pi}{3} \alpha m (F_{\rm LSO} -f)\,.
\label{ex}
\end{equation}
This {\it improved relativistic} SPA is  thus written as
\begin{equation}
 f\ge f_{\rm up}\,:\;
\tilde{h}_{->}^{\rm irspa}(f)=m \alpha e^{i \psi_f^{\rm LSO}} a(t_{\rm LSO})
g_{\frac{3}{2}}(x)\,.
\label{4.17}
\end{equation}
Note that $f<F_{\rm LSO}$ corresponds to $x>0$ (saddle-point domain), while $f>F_{\rm LSO}$
corresponds to $x<0$ (absence of a  saddle-point). Roughly speaking the variable $x(f)$
corresponds to $-\zeta(f)$ of the non-relativistic case, and $g_{\frac{3}{2}}(x)$ is the
relativistic analogue 
of the combination ${\cal C}(\zeta) e^{i\zeta^2}$ appearing in the previous
treatment (see, e.g., Eq. (\ref{3.37})).

\begin {figure}[t]
\caption{ The real and imaginary parts of the $g_{\frac{3}{2}}(x)$ function 
Eq.(\ref{4.22}) in terms of which the improved relativistic SPA (irSPA) is 
represented.  
The thick lines represent the asymptotic behaviour as $x\rightarrow\pm\infty$
given by Eqs.(\ref{4.20}) and (\ref{4.21}).}
\centerline {\epsfxsize 6 true in  \epsfbox {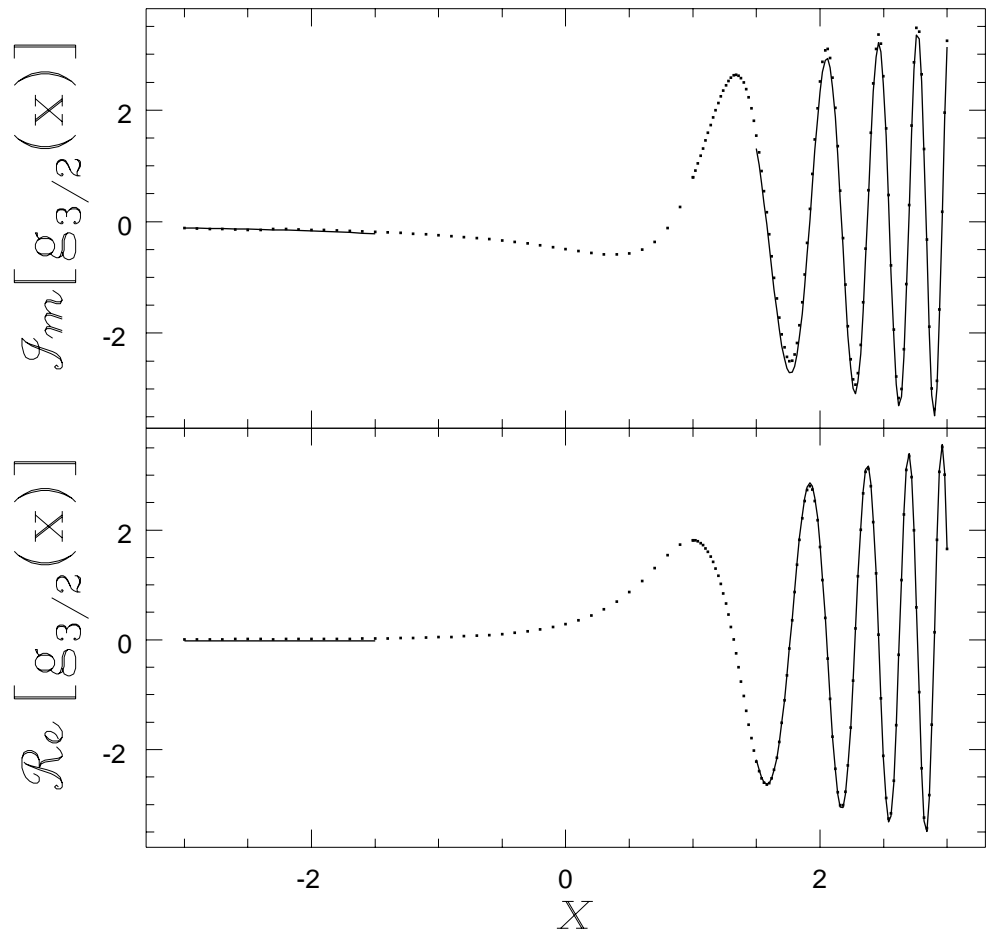} }
\label {fig:gee}
\end {figure}

It is useful to summarise some properties of the function $g_{\frac{3}{2}}(x)$:
\begin{mathletters}
\begin{eqnarray}
g_{\frac{3}{2}}(0)&=& \frac{1}{3}\frac{(1-i\sqrt{3})}{ 4^{1/3}}
\Gamma\left(\frac{2}{3}\right)=0.284347 - 0.492503\, i\,, 
\label{4.19}\\
g_{\frac{3}{2}}(x)&\sim&  \sqrt{\frac{4\pi x}{3}}e^{i (x^3- \pi/4)}\;,\;  x>0,\;x \gg 1\,, 
\label{4.20}\\
g_{\frac{3}{2}}(x)&\sim&\frac{i}{3x}\;;\;  x<0\,, \;-x\gg 1\,.
\label{4.21}
\end{eqnarray}
\end{mathletters}
By expanding the integrand of  $g_{\frac{3}{2}}(x)$ in powers of $x$, and integrating
term by term [using the properties of the Euler $\Gamma$-integral after having changed the variable
of integration: $\hat{\tau}=e^{-\frac{i\pi}{3}}(u/2)^{\frac{2}{3}}$], 
one proves that $g_{\frac{3}{2}}(x)$ is given by the following,
 everywhere convergent, Taylor-Maclaurin expansion:
\begin{equation}
 g_{\frac{3}{2}}(x)= \frac{2^{1/3}}{3} e^{-i \pi/3}
\sum_{n=0}^{n=\infty}\frac{\Gamma[\frac{2}{3}(n+1)]}{n!}
\left(\frac{3 x}{2^{2/3}}e^{\frac{i\pi}{6}}\right)^n\,.
\label{4.22}
\end{equation}
With about 300 terms the above series represents $g_{\frac{3}{2}}(x)$ accurately enough
for values of $x$ in the range $x\in [-2.3,\,2.3]$.
We used this series to generate the plot of $g_{\frac{3}{2}}(x)$ represented
in Fig.~\ref{fig:gee}.
  Though we do not use it in this paper,  note
that for $ -x\rightarrow \infty$,  the following (divergent)
asymptotic expansion  is also valid:
\begin{equation}
g_{\frac{3}{2}}(x) \sim - \frac{1}{3x}\sum_{n=0}^{n=\infty}
 \frac{\Gamma(\frac{3}{2} n +1)}{n!}
\left[\frac{-2}{(-3x)^{3/2}}\right]^{n }\, e^{-\frac{i\pi}{4}(n+2)}\,.
\label{4.23}
\end{equation}
In all our calculations of overlaps we shall define the frequency $f_{\rm up}$
separating the lower range  from the upper range by choosing 
$x_{\rm up}= 0.36$ as the right hand side of Eq. (\ref{ex}).
This value is chosen so that at $x_{\rm up}$  one has a smooth transition
from the lower to the upper approximation.
We have also  checked that the overlaps do not change very significantly for $x_{\rm up}$ between
0.2 and 0.4.

In summary  our  best analytic representation of time-windowed
 relativistic signals in the Fourier-domain would be  defined 
by combining the P-approximant construction of the functions $E'(v)$, ${\cal F}(v)$
\cite{dis98} with the total  {\it improved relativistic 
approximants} (irtot) defined as
\begin{equation}
\tilde{h}^{\rm irtot}(f)=\tilde{h}_-^{\rm irspa}+
\tilde{h}_+^{\rm edge}\,,
\label{sha7}
\end{equation}
where $\tilde{h}^{\rm edge}_+$ is defined in Eq. (\ref{br29}) and 
$\tilde{h}^{\rm irspa}_-$ is defined 
 for $f\le f_{\rm up}$ by
Eqs. (\ref{4.6}), (\ref{4.7}), and, for $f\ge f_{\rm up}$ 
 by Eq. (\ref{4.17}).
Actually, as in the case of Newtonian signals, we have found that the inclusion 
of $\tilde{h}^{\rm edge}_+$ has only a minimal (though favorable) effect on overlaps.
Moreover, such a contribution is absent in the case of real signals.
Therefore, our final {\it practical and best} proposal consists in using only
$\tilde{h}_-^{\rm irspa}$ (For simplicity we henceforth drop the subscript minus).
We shall henceforth refer to the improved frequency domain stationary phase
P-approximants based on the irSPA as the SPP approximants.

\subsection{Comparison between the usual SPA, 
the improved relativistic SPA and the `exact' SPA (numerical DFT)}\label{sec:COMPARE2}

\begin {figure}[t]
\caption{
Visual comparison of  a chirp generated directly in the time-domain
(solid line) with the inverse Fourier transforms of the usual SPA
terminated at LSO (uSPAw) (dashed line) and the improved relativistic SPA
(irSPA) extended up to the Nyquist frequency (dotted line), all during the
last few cycles before the last stable circular orbit frequency is reached.
The Figure corresponds to the relativistic 
 (second post-Newtonian P-approximant case $(P_4)$)  
 inspiral of  a binary of total mass $m=40M_\odot.$ }
 \centerline {\epsfxsize 6 true in  \epsfbox {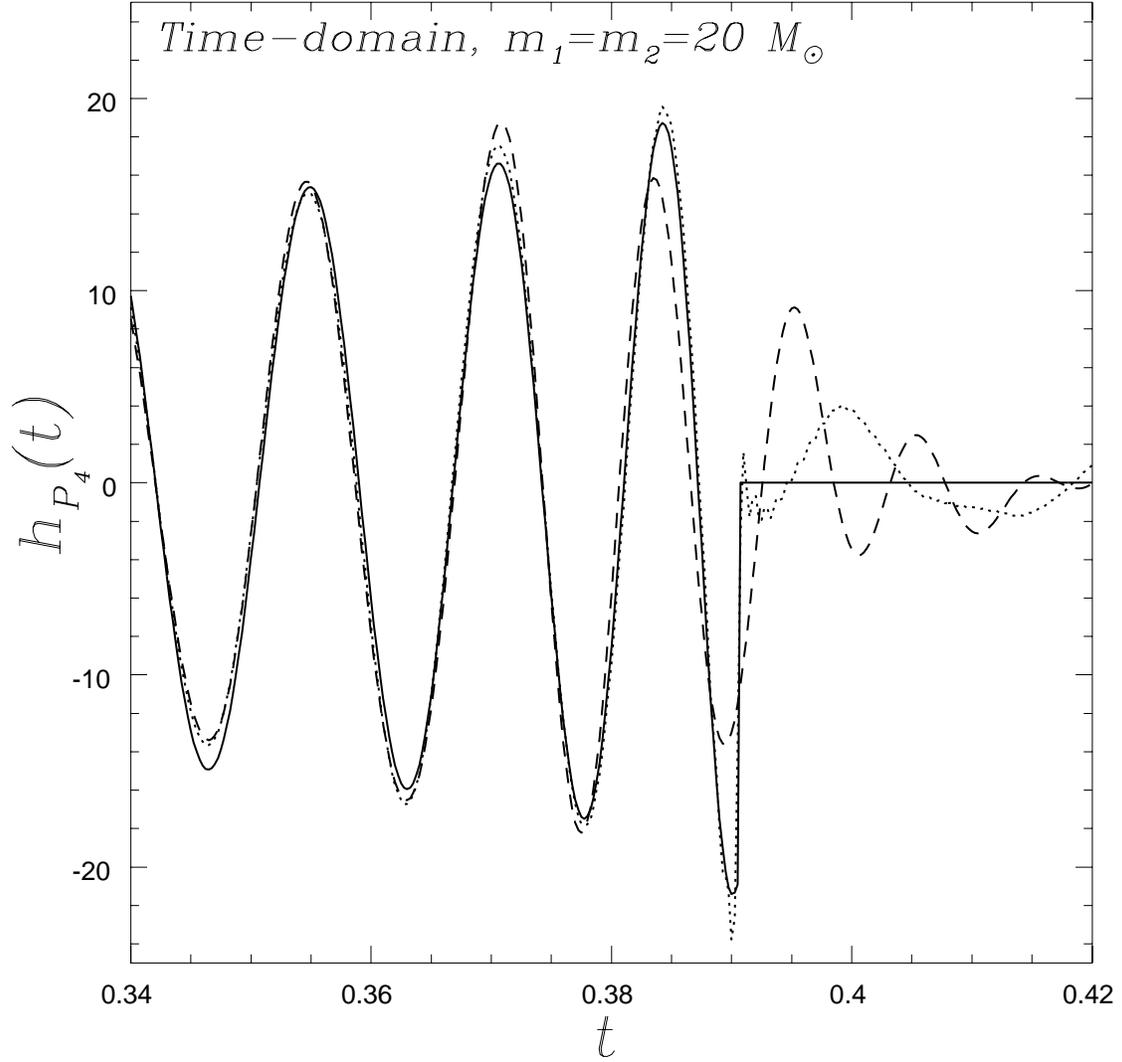} }
\label {fig:visual.newt1}
\end {figure}

In this Section, 
we test the accuracy of our analytical approximations in various ways. 
Fig.~\ref{fig:visual.newt1} compares an inspiral wave from a (20,20)~$M_\odot$ binary
generated by three different methods: (i) directly in the time-domain
and terminated when the instantaneous gravitational wave
frequency reaches the value at the LSO (solid line), (ii) in the
Fourier domain using the usual SPA but with a square window between
$f_{\rm min}$=40 Hz, $f_{\rm max}=F_{\rm LSO} $ (uSPAw) and then inverse
Fourier transformed (dashed line) and
(iii) again in the Fourier domain but using irSPA with Fourier
components computed up to Nyquist frequency and then inverse
Fourier transformed to obtain its time-domain representation
(dotted line). 
We only exhibit the comparison near the crucial LSO region [Much before
the LSO the uSPA is nearly equivalent to the irSPA
and they both do a good job in representing the actual signal].
We observe that the uSPAw begins to get out
of phase with the wave directly generated in the time-domain
during the last cycle and rings a few times beyond the
shut-off point. Our new proposal, irSPA, keeps in phase with the
time-domain signal until the last moment although it too has
a couple of low amplitude cycles beyond the  LSO.

\begin {figure}[t]
\caption{
This plot is the same as in Fig.~\ref{fig:visual.newt1}
 except that we compare whitened signals
to show the effect of the detector response function on the  time development.
The inset shows the entire time-domain signal starting from $F_{\rm GW}=$ 30 Hz
and terminating at $F_{\rm LSO}$.
}
\centerline {\epsfxsize 6 true in  \epsfbox {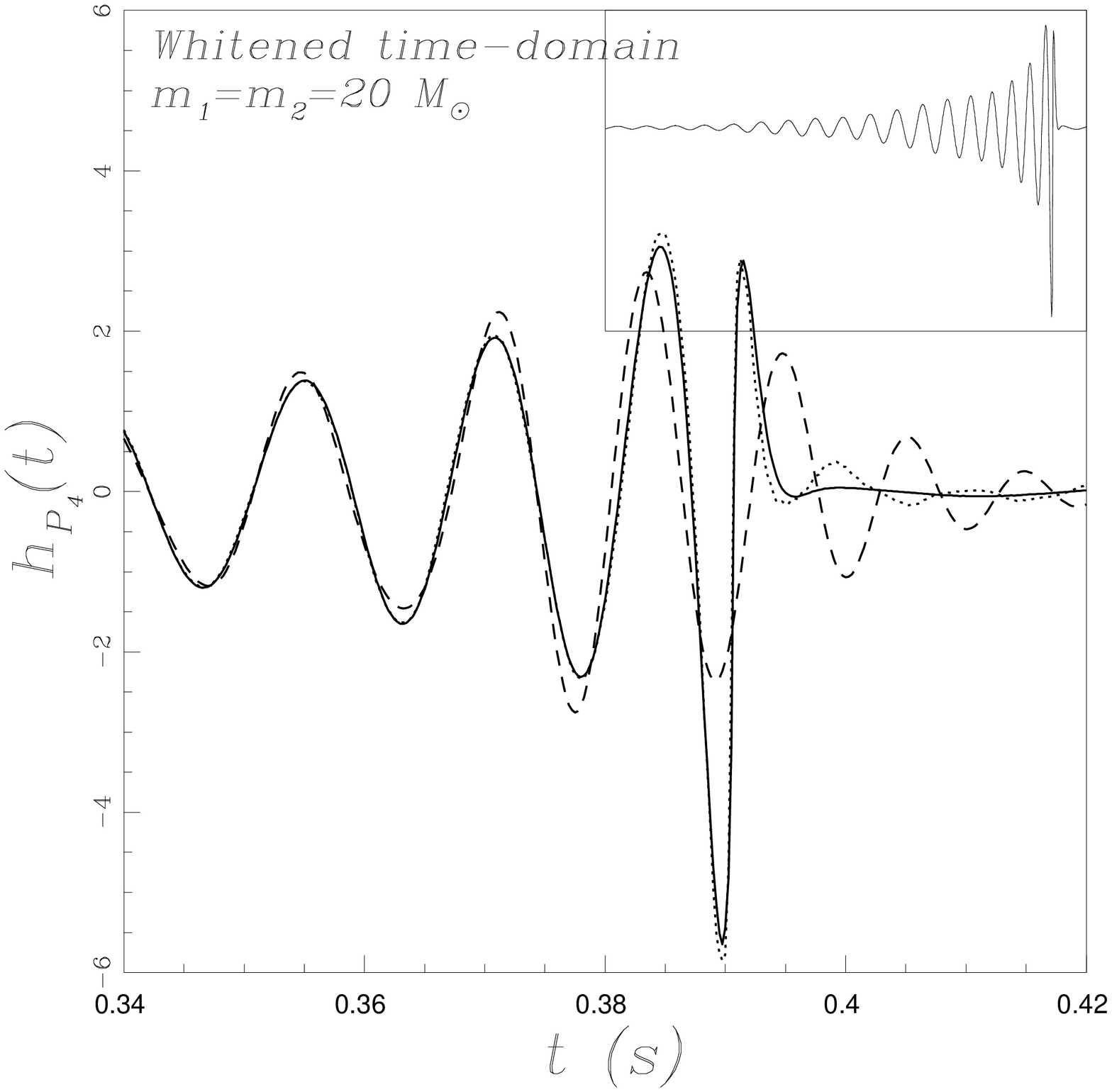} }
\label {fig:visual.newt2}
\end {figure}

Matched filtering involves not just the correlation of
two signals but rather their weighted correlation -- the weight
coming from the detector spectral noise density. 
To further compare and contrast our new $f$-domain approximants 
to the usual frequency-windowed SPA it is conceptually useful to compare
various approximations in the `whitened-time-domain' introduced in
Section~\ref{sec:IIA} above. As discussed above, in this picture
(and only in this picture) the optimal
filter consists of correlating the output of the detector with an exact
 copy of the expected signal.
The whitened [{\it i.e.}, convolved with the whitening kernel
$w_{\frac{1}{2}}$ Eq. (\ref{br12})] signals are plotted and compared in 
 Fig.~\ref{fig:visual.newt2} which is the same as 
Fig.~\ref{fig:visual.newt1} 
except that all
the waves here are whitened (i.e. divided by $\sqrt{S_n(f)}$
and then inverse Fourier transformed).
The inset in Fig.~\ref{fig:visual.newt2} 
shows the full whitened signal that was originally generated
in the time domain. Several observations are in order. First, we see
how low frequency components are suppressed relative to high
frequency components which occur in a more sensitive band of the
detector. Second, we can very clearly see the non-local
behaviour of the whitening kernel. It has the effect of softening
the window imposed on the wave that was  directly generated
in the time-domain and curbing the oscillations in the irSPA
beyond $F_{\rm LSO}.$  Finally, this same whitening is seen
to have worsened the mismatch of uSPAw with the whitened version
of the original time-truncated signal.
The conclusions drawn
from these visual comparisons are borne out by detailed numerical
experiments we performed.

To compare the approximants more quantitatively,
in Table \ref{table:relativistic} we list the overlaps of the exact Fourier
representation of a model waveform (i.e., a signal generated in the 
time-domain 
and then Fourier transformed using a DFT algorithm) with their approximate
Fourier representations analytically computed using one of the following:
the frequency-windowed  usual  SPA [{\it i.e.}, uSPAw, cf.Eq. (\ref{3.13})], 
the improved Newtonian SPA [inSPA is the same as irSPAw, cf. Eq. (\ref{4.7})] 
and the improved relativistic SPA [cf. Eq. (\ref{sha7})]. 
The uSPA and the inSPA used in
computing these overlaps are terminated at $f=F^{\rm A}_{\rm LSO},$ where 
$F^{\rm A}_{\rm LSO}$ is the last stable orbit frequency determined by the
condition $E_{\rm A}'(v)=0$ (hence the labels SPAw and inSPAw where `w'
stands for `windowed' --- in the frequency-domain).
This is because both uSPA and inSPA vanish at the LSO
(due to  the factor $1/\sqrt{\dot{F}_{\rm LSO}}$) and
 are  either not defined (in the case of the usual SPA)
or formally vanishing (according to the definition Eq. (\ref{3.37})
in the case of inSPA)  beyond  the LSO.
 Contrast this with the Newtonian case where it is 
possible to analytically extend the usual SPA beyond $F_{\rm LSO}$.

It is generally true, as stated in DIS,  that
the stationary phase approximation to the Fourier transform
worsens very significantly  as we consider  more
massive binaries. In this sense the uSPA poorly represents the
exact chirp. 
We conclude that, for massive systems with  total mass $m= m_1+m_2\lesssim 40 M_\odot$
the only uniformly acceptable analytic representation of
the Fourier transform is the irSPA.

\section{ Faithfulness and Effectualness  of SPP approximants}
\label {sec:FESPP}

So far we have concentrated on developing an accurate Fourier
representation of the inspiral waveform at various levels of
approximation from Newtonian to P-approximants. In order
to quantify the accuracy,  we used the overlap of the DFT of the waveform 
computed using a FFT of the time-domain signal with
an analytical approximation of the Fourier transform 
of the same time-domain signal  using the improved
SPA suggested in Sec.~\ref{sec:SPA} and \ref{sec:SPP}. 
However, an important question still remains: 
What is the total loss of accuracy due to combining the loss
of {\it precision} entailed by the use of an analytical
approximant to the FT (loss that we have shown how to minimize
by defining the irspa) with the loss of {\it accuracy}\footnote
{ We distinguish {\it precision} and {\it accuracy} in the same way that they
are distinguished in Metrology. }
 entailed by the
use of some finite-order in the post-Newtonian approximation
of the exact signal. In other words,  how accurate is the
approximate frequency-domain representation of a 
post-Newtonian  approximant 
in modelling the exact  FT of the exact general relativistic signal? 
More precisely, what fraction of
the SNR of a true signal is the Fourier-domain approximant likely to
extract? Additionally, one is also interested in knowing the biases
induced in the estimation of parameters when using the frequency-domain
approximants introduced in this work.

We shall follow DIS in saying that a representation of a signal is {\it faithful}
if it has a good overlap\footnote
{When discussing faithfulness and effectualness we always
assume, as in DIS, Eq 2.17 there,  that the overlap Eq. (\ref{d7}) 
is first maximized with respect to the
 relative time lag (and relative phase).}
 with the exact signal for the same  values of the (dynamical) parameters
 (or more precisely, if the overlap is maximized
for template parameters which have acceptably small biases with respect to the exact signal 
parameters).
As in DIS, we employ as necessary criterion for faithfulness the
requirement that the `diagonal' ambiguity function be larger than
0.965.
 On the other hand,
we shall say that a representation of a signal is {\it effectual} if the
overlap, maximized over the template parameters is very near one.
To use these definitions we follow DIS in introducing a 
{\it fiducial exact} general relativistic signal.
In Table~\ref{table:faithfulness.tm} 
and  Table~\ref{table:effectualness.tm}
we use as fiducial exact signal the
formal ``test-mass case'' for which the function $E(v)$ is known analytically
and ${\cal F}(v)$ numerically \cite{numflux}.
In Table~\ref{table:faithfulness.fm}
 we use as fiducial exact signal the
one defined in DIS for comparable masses [see  Eq. (4.11) there for the definition
of the exact new energy function, and Eqs. (7.1) (7.2) for the exact factored flux
function; we took the value $\kappa_0=47/39$
for the parameter defining formal higher PN-effects in Eq. (4.11) ]. 
As above we consider that the exact time-domain signal is shut off after
the LSO. [For each considered waveform, defined by some approximate energy and flux functions
$E_A(v)$ and ${\cal F}_A(v)$, we shut it off at the LSO defined by the
corresponding energy function $E_A(v)$.]

In Tables~\ref{table:faithfulness.tm} and \ref{table:faithfulness.fm}
we list the overlaps for different approximants for
the three `massive'  archetypal binaries
[$(1.4M_\odot,10M_\odot),\,
(10M_\odot,10M_\odot),\,$ and $ (20M_\odot,20M_\odot)$]
 that could be searched for in GEO/LIGO/VIRGO
data. These overlaps are computed using the expected LIGO noise Eq. (\ref{ligonoise})
by maximising over the lag parameter $\tau$ \cite{lag} and phase
$\phi_c$ but  without re-adjusting the intrinsic  parameters
{\it i.e.}, the masses of the two stars
in the approximants, to maximise the overlap. 
[This implies  that the value of $F_{\rm LSO}$ used in the
approximant is different from that in the `exact' signal.]
The overlaps  are therefore a
reflection of how accurate the various representations are in an
absolute sense.
In other words, they compare the {\it faithfulness} of the different
approximants.
Two independent aspects  of approximation  are  investigated in these Tables.  
Firstly, the comparison  between  the two alternatives in the frequency domain:
 the usual SPA (uspaw) and our improved relativistic SPA (irspa). 
And secondly, the post-Newtonian order to which the phasing is computed.
To investigate further the performance of these approximants
we summarise in 
Table~\ref{table:effectualness.tm}
the overlaps obtained by maximising over  all the parameters 
in the approximants including the intrinsic ones.
Thus in addition to  maximising over
the lag parameter $\tau$  and the  phase $\phi_c$  
one also extremises over the masses of the two stars 
$m_1$ and $m_2$.
In other words, we compare the {\it effectualness} of the various
approximants. We also compute the bias introduced in the total mass $m$.

From  Tables \ref{table:faithfulness.tm}--  
\ref{table:effectualness.tm} one can conclude the following:
 (i) The improved relativistic SPA is
 significantly more faithful and more effectual for massive systems with total
mass $m \gtrsim  20 M_\odot$, 
and {\it mandatory} for $m \gtrsim 26 M_\odot$, 
(ii) Comparing with DIS, we see that the frequency-domain irspa 
does as well as the time-domain waveform even
  for massive binaries up to $40M_\odot$;
  (iii) The 2.5PN SPP approximant is both a faithful and an effectual
approximant for a wide range of binary systems $(m\lesssim 40M_\odot)$.
  It only introduces a small bias. Note also that,
in regard to effectualness,
  the gain in going from 2PN to 2.5 PN accuracy is quite significant (mainly in
  decreasing the biases) and especially for low-mass systems (which have many
useful cycles), while the gain in going from 2.5PN to 3PN seems very slight.

To summarise: 
{\it   If one would like  to lose no more than a tenth
of the events that would be observable had one known the exact 
general-relativistic signal, then the  2.5PN SPP-approximants are a must. 
Furthermore, unbiased parameter estimation requires 2.5PN SPP-approximants
in all cases.}

\section{Why are Time Domain Relativistic Signals more expensive
to compute?}\label{sec:COMPU}

The main purpose of this work is to provide a set of tools to the
experimenters so that they can generate templates 
with a minimal computational cost.
We next, therefore, address the issue of
computational costs of various algorithms for template generation.

First, though the signal is initially given in the time-domain,
the time-domain version of the Wiener filter contains a double
time integration [see second form of  Eq. (\ref{d3})] which is (given the existence of FFT
algorithms) much more computationally expensive than the single frequency-domain version
of the Wiener filter  [see first form of Eq. (\ref{d3})].
Therefore, 
in the computation of the correlation of a template with the detector
output what is required is the Fourier transform of the matched filter. However, the
DIS proposal was to compute the templates in the time-domain and
compute their exact DFT using FFT algorithms. Admissibly, this procedure is
still highly computation-intensive. Let us reason out why
this is so.

To compute the time-domain signal we need a phasing formula $ \phi =
\phi(t)$. Since there is no explicit expression for the phasing of
inspiral waves as a function of time the standard approach is to use
the implicit formula,  Eqs. (\ref{4.1})-(\ref{4.2}).
The binding energy $E(v)$ and the gravitational wave flux ${\cal F}(v)$ have been computed,
e.g. using Pad\'e techniques, as explicit functions of  $v$ 
and these when used in Eq. (\ref{4.1})
and (\ref{4.2}) yield an implicit relation between $\phi$ and $t$.
However, the problem is that we need $\phi$ at equal intervals of
time (to enable us to use the standard FFT algorithms) and this makes
the computation of $\phi(t)$ expensive: every time-sample $ \phi_i
\equiv \phi(t_i)$ is computed by first solving Eq. (\ref{4.1})
iteratively for $v_i$, the lower limit in the integral for a given
$t_i$, and then using this $v_i$ as the lower limit in the integral of
Eq. (\ref{4.2}). Though the second step is the computation
of a single integral, the first step is a  rather slowly converging
($\sim 10$ iterations for every $t_i$) computation.

This problem could have  been circumvented if it had been  adequate to use the explicit
analytical expression 
$\phi(t)= b_0 (t_{\rm LSO} -t )^{5/8} +\sum _{k\ge 1} b_k (t_{\rm LSO} -t)^{(4-k)/8}$
(modulo logarithms) obtained by :
(i) expanding the quantity  $E'(v)/{\cal F}(v)$ in the integrands, in a 
straightforward  expansion in powers of $v$,
(ii) integrating term by term, and
(iii) inverting analytically by successive iterations (see e.g. \cite{bdiww95}).
However, this straightforward PN expansion of the phasing formula
defeats the very purpose of P-approximants
and loses all the benefits brought by the constructions given in
\cite{dis98}.
Consequently, DIS had to use the iterative procedure to compute the signal
phasing.
By contrast, using (any form) of SPA, {\it i.e.} an explicit analytical
f-domain expression, brings a 
 tremendous reduction in computational costs. On the one hand, as we
shall discuss  below there is no iterative procedure involved in
computing SPA. Secondly they are computed directly in the
frequency-domain and hence lead to a  further cost reduction, since
time-domain waveforms need to be Fourier transformed using FFTs
--- costing $N \log_2 N$ floating point operations --- in addition to 
floating point operations 
required to compute time-domain templates.

Let us recall that the usual SPA is given by Eq. (\ref{3.13}).
In this expression $t_f$ is the stationary point of the phase in the
integral of Eq. (\ref{3.9}). 
At a Fourier frequency $f=v^3_f/\pi m$ 
the stationary point $t_f$ is given by Eq. (\ref{s1}),
which is a non-iterative computation. 
One then substitutes this value
of $t_f$ in  Eq. (\ref{s2}) to compute $\psi_f$ --- the phase of the Fourier component.
Moreover, the derivative of the frequency which occurs in the
amplitude of the Fourier transform can be computed using Eq. (\ref{4.5})
while the factor $a(t_f) \propto f^{2/3}$ 
from Eq. (\ref{3.3}). 
Every quantity that appears
in the SPA is computed using a straighforward integral or a mere
algebraic expression. Hence, from the computational-cost
point-of-view, it is desirable to use some  SPA to generate templates.
Since the usual SPA has been shown to be inadequate for representing time-windowed signals from
massive binaries, we have proposed the
use of corrections to (for $f\le f_{\rm up} \le F_{\rm LSO}$) and analytic extensions of
(for $f \ge F_{\rm up})$ the usual SPA. In Table~\ref{table:computecost}  we compare for archetypal binaries,
the computational costs of templates that are generated in the
time-domain and Fourier transformed using an FFT algorithm with
the computational costs for the uSPA,  inSPA and irSPA.
This Table clearly shows that it is sensible to generate templates in
the Fourier domain.
{\it The SPA is up to a factor 100 times faster
and the irSPA is up to a factor 10 times faster than the
corresponding time-domain construction and Fourier transformation.}
 Table \ref{table:computecost} together with Table~\ref{table:faithfulness.fm} (of overlaps)
demonstrates that SPP approximants while more expensive to generate
 than the  usual SPA are nevertheless `affordable', and are anyway necessary for 
 efficient searches of inspiral signals in gravitational wave interferometer data.

\section { Concluding Remarks}\label {sec:SUM}

After nearly two decades of detector-technology development
long-baseline interferometric gravitational wave  antennas 
LIGO/VIRGO are
scheduled to become operational in about 2-4 years with target
sensitivities that are good enough to detect inspiral events from
massive ($m > 20 M_\odot$) binaries at an optimistic rate of a few
per year. Searches are planned to be carried out over a range of
$0.2$-$50 M_\odot$ by the method of matched filtering. 

An important issue in matched filtering is the number of cycles
accumulated in the correlation integral since the SNR grows as the
square-root of the number of  cycles. While this is strictly true, if the
noise power spectrum of the instrument is independent of frequency, in
practice one can only improve the SNR in proportion to the square-root
of a ``useful'' number of cycles $N_{\rm useful}$ which is determined by a
combination of the detector noise power spectrum and the
signal's power-spectrum. We have pointed out how the number of useful
cycles can be a lot smaller  than the actual number of cycles for massive
and relativistic systems: 
e.g. a $(10M_\odot,10M_\odot)$ [ $(20M_\odot,20M_\odot)$] binary system 
has only  7.6 [3.4] useful cycles in the detector's bandwidth
(see Table~\ref{table:cycles}).
A priori , it may seem that  the fewer number
of cycles should make it easier to model the  massive black hole  binaries   compared to   
the lighter  neutron star-neutron star ones with its corresponding  large number of cycles
to phase.
Tables \ref{table:faithfulness.tm}--\ref{table:effectualness.tm}
 show that there is some truth in this, but that
for very
massive black hole binaries, these fewer cycles are in fact more difficult 
to model than the neutron star-neutron star, 
or neutron star-black hole cases for two
reasons: (i) they are near the end of the inspiral, 
{\it i.e.} when the radiation reaction effects drives a faster drift of the 
frequency which has to be modelled accurately (this is why we need P-approximants introduced in DIS);
(ii) they might terminate due to the transition from inspiral to plunge while
in the detector's bandwidth, and  this poses the problem of accurately describing
the Fourier transform of a time-windowed signal (this requires the  correction factors introduced in this paper).
All this places stringent demands in modelling the waveform in the Fourier-domain  and due  attention
 needs to be paid to  delicate issues of detail.
This task is all the more important that the first detections
expected from LIGO/VIRGO are likely to concern massive systems with $m \sim 25
\pm 5  M_\odot$, for which the LSO frequency lies 
near the middle of the sensitivity curve [see Fig.~\ref{fig:zero}].

To this end, the present work makes two new robust ({\it i.e.}
assumption-independent)  contributions:
\begin{itemize}
\item the proposal of stationary phase P-approximants (SPP)  
which combine the excellent performance of our time-domain
 P-approximants \cite{dis98} with the
analytic convenience of the stationary phase approximation 
without serious loss of event-rate.
These Fourier-domain P-approximants  perform as well as their
time-domain counterparts in extracting the true general relativistic signal.
\item the definition of a universal Newtonian-like 
`edge-correction' factor ${\cal C}(\zeta(f))$,
as well as its relativistic complement $g_{\frac{3}{2}}(x(f))$
which take into account the frequency-domain effects, concentrated
around (and on both sides) of $F_{\rm max}=F(t_{\rm max})$, for signals
which are abruptly shut off, in the time-domain, after $t_{\rm max}$.
\end{itemize}
In addition to these new achievements, let us mention two other useful
contributions, of a more technical nature: 
(i) our recommendation to systematically use a smooth time window
at the lower frequency side to conveniently and efficiently suppress spurious
oscillations due to a numerical low frequency cutoff
and 
(ii) the emphasis on the comparison of the form of the signals in the
`whitened' time-domain.

Based on the detailed analysis presented in  this paper we find that 
for post-Newtonian template generation of binary systems of total mass $m \lesssim 5 M_{\odot}$  it suffices to use the usual SPA
(without correction factor) of the P-approximants defined in DIS.
On the other hand, 
in the total mass range  $5 M_{\odot} \lesssim m \lesssim
40 M_{\odot}$, it is crucial to use  our new SPP approximants
 to construct the frequency-domain templates.

In addition to the construction of the SPP approximants, 
the paper has examined in detail the 
Fourier-domain effects  entailed by a sharp time-domain windowing.
 As emphasized in the introduction, at our  present stage of knowledge, 
one cannot be sure that a template waveform terminated 
(in the time-domain) at the LSO is an accurate-enough representation of a real GW signal coming
from massive binaries (say with $m< 40 M_\odot$). 
We have given several plausibility arguments
towards justifying this assumption: brevity of the plunge,
and an expected frequency separation from the merger signal.
In the absence of knowledge of the transient plunge signal and of the
final merger signal, we have argued 
that it is best to use a template waveform which is
terminated at the LSO.
[Actually, we anticipate that the effectualness of the
template
waveform will be increased if we allow it to be	 terminated at a frequency
somewhat
larger than  $F_{\rm LSO}$ (thereby allowing it to approximately represent the plunge
waveform).]
Consequently,
this work has concentrated on signal models that are truncated in the
time-domain by a step-function and has  aimed at constructing
the best associated Fourier-domain analytical representation
for this possibility.

We have also pointed out that the opposite assumption of an abrupt termination
at $F_{\rm LSO}$ of the usual SPA in the frequency domain implies,
when viewed from  either the time-domain or 
the whitened time-domain, the existence of  
 some coherent oscillations `ringing' after the LSO crossing.
We have done another numerical experiment on this issue, 
by appending to the inspiral signal a smooth
decay taking place over less than $3 F_{\rm LSO}^{-1}$ time-scales.
We have found that our improved SPA was a reasonably good representation
of (the FT of) such a signal, and
definitely a better representation than the usual SPA one.
Let us finally  reiterate that, we do not claim to have  conclusively ruled
out the possibility that a frequency-windowed SPA may perform
better compared to the time-windowed SPA we propose here.
This important issue is not settled  though we conjecture that
this  is unlikely.
Anyway, this paper is the first one to explicitly construct the frequency domain
version of the time-domain P-approximants which were shown in DIS to bring   indispensible
improvements over the usually considered T-approximants. Therefore, even in the unlikely case where a 
straightforward frequency-window turns out to be a better model than the time-window assumed in most of this work, 
one will still require the formulas given in this paper (with the trivial change of replacing 
the correction factors ${\cal C}(\zeta)$ by a $\theta$ function $\theta(F_{\rm LSO}-f)$) to
generate sufficiently accurate f-domain filters.
In view of these comments,  we feel there is a urgent need to model   more precisely 
the transition from the inspiral signal  to the
plunge signal \cite{bdprep} close to the last stable orbit. 
We hope that the techniques (if not all the details of our
construction) used in this work to handle the blow up 
 of $\dot{F}(t)$ at the LSO will be useful
(maybe with some modifications) even if, on a later examination,
 this blow up turns 
out to be an artefact of an approximation which may drastically
alter with a better treatment of the transition to the plunge.
Only with this improved understanding 
and its implications for the  construction of templates 
can  one build even more  optimal templates for massive binaries and
 maximise our chance of detecting them.
Independently of issues such as windowing in time versus windowing in frequency
or the nature of the plunge we feel that in general
 P-approximants are much  better tools than the Taylor approximants.
We hope to come back to this question in a future work \cite{dis5}.

  Another aspect that needs to be looked into is the issue of whether
 whether  the interferometers  will work in the time-domain
 or the frequency-domain. 
  If indeed, they would  decide to  work in the time domain: 
{\it i.e.}, to store the raw output,
 and to transform it nearly online in the defiltered time-domain
 equivalent GW amplitude $h(t)$
the analysis of this paper would be irrelevant.
 In that case, one should store the Wiener transformed
 time-domain filter $ K(t) = w_1 * h(t)$.
However, with the presently available 
computational resources  it seems hopeless to filter in the
time-domain.
 We therefore anticipate that,
 though the raw detector output will be stored in the time-domain, all
filtering will be done in the Fourier domain.
In this event, the robust aspects of the present  analysis will be relevant
even if not the details.

The formalism developed in this paper can be applied not
only to initial interferometers but also to future 
generations of interferometers. We have refrained from
applying our formalism to the case of LIGO II since the
LIGO II design is at the moment in a state of flux and
any quantitative results we may quote will soon be irrelevant.
However, we should expect the results of this
work to be important for any detector that works with
a lower seismic cutoff and a broader bandwidth than LIGO I, 
since in such cases we will have to match the signal's 
phase for a larger number of effective cycles.

There are several notable and obvious improvements that need to be pursued.
The sensitivity to the value of $F_{\rm LSO}$ needs to be investigated [in particular,
our improved SPA will probably maximize their overlaps with the real signals
if we allow some flexibility in the choice of $F_{\rm LSO}$ (within some limits)].
Once the results of 3PN generation of gravitational waves are
available \cite{bij} and are combined with the 3PN results on
the dynamics \cite{3PN} they must be included in the construction of
templates.
In our discussion we have not  considered waveforms from binaries with
spinning compact objects. Nor have we included the effect of
eccentricity \cite{gi97} on the detectability \cite{mp99}. These are  unarguably
important physical effects that need to be incorporated in later data
analysis algorithms. Future research in this area should shed
light on these issues.

\acknowledgements

We thank J.Y. Vinet for informative communications concerning
the VIRGO noise curve.
BRI and BSS thank A. Gopakumar and B.J. Owen for discussions on
the validity of the stationary phase approximation.
BRI would like to thank AEI, Germany and  PPARC, U.K. for  visiting fellowships.

\appendix

\section{List of Symbols}

\begin{tabbing}
$a(t)$~~~~~~~~~~~~~~~~~~\= GW amplitude; $h(t)=2 a(t) \cos\phi(t)$\\
$\alpha$\>$=\frac{1}{2}v_{\rm LSO}^{-4/3}e_1^{1/3}$; Eq. (\ref{4.16})\\
cspa\> corrected SPA; Eqs. (\ref{3.28})-(\ref{3.30})\\
$C_n(t_1-t_2)$\> correlation function of noise\\
${\cal C}(\zeta)$\> $\frac{1}{2}{\rm erfc}(e^{i\pi/4}\zeta)$;
correction factor; softened step function\\
$\delta$\> leading phase correction to SPA; Eq. (\ref{d3.13b})\\
$\eta$\> symmetric mass ratio $\equiv m_1m_2/(m_1+m_2)^2$\\
${\rm erfc}(x)$\> complementary error function; Eq. (\ref{3.18})\\
$e_1(\eta)$\>$\equiv\left[\frac{d}{dv}\left(\frac{{E}'(v)}{{\cal
F}(v)}\right)\right]_{v_{\rm LSO}}$; Eq. (\ref{4.8})\\
$E(v)$\> dimensionless energy function\\
$\varepsilon_1$\>$\equiv \vert\frac{\dot{a}(t)}{a(t)\dot{\phi}(t)}\vert$; 
Eq. (\ref{d3.13a})\\
$\varepsilon_2$\>$\equiv\vert\frac{\ddot{\phi}(t)}{\dot{\phi}^2(t)}\vert
= \vert\frac{1}{2\pi}\frac{\dot{F}(t)}{F^2(t)}\vert=\frac{1}{2\pi N}$;
Eq. (\ref{d3.13a})\\
FFT\> Fast Fourier transform\\
f-domain\>frequency-domain\\
f-window\>frequency window\\
$F(t)$\>instantaneous GW frequency\\
${\cal F}(v)$\> flux function\\
$F_{\rm min}(F_{\rm max})$\> GW frequency at $t_{\rm min}(t_{\rm max})$\\
$F_{\rm LSO}$\> GW frequency at  LSO\\
$F_{\rm Nyquist}$\> Nyquist Frequency\\
$f$~~~~~~~~~~~~~~~~~~~ \>    Fourier frequency\\
%$f_0$\> ??BS\\
$f_{\rm det}$\> characteristic detection frequency;
 minima of effective GW noise $\sqrt{f S_n(f)}$\\
$f_{\rm p}$\>  frequency at which $d{\rm SNR}^2/d(\ln f)$ peaks  \\
$f_{\rm s}$\>   seismic frequency   \\
$f_{\rm up}$\> transition frequency between the low and high frequency approximations
for the   irSPA  \\
$g_{\frac{3}{2}}(x)$\>$\equiv\int_0^{\infty} d\hat{\tau} e^{i \left(3 x \hat{\tau}  - 2 \hat{\tau}^{3/2}\right)}$; Eq. (\ref{4.15})\\
GW \> Gravitational wave\\
$\Gamma$\>Gamma function\\
$h_s^2(f)$\>squared amplitude of effective GW signal; $\equiv N(f) a^2(f)$\\
$h_n^2(f)$\>squared amplitude of effective GW noise; $\equiv fS_n(f)$\\
$h(t)$\> time domain signal\\
$\tilde{h}(f)$\> Fourier transform of $h(t)$; $ 
\tilde{h}(f)\equiv\int_{-\infty}^{\infty}\,dt\,e^{2\pi i f t}\,h(t)$\\
$\tilde{h}_+(f)$\> Fourier transform of non-resonant part of  $h(t)$\\
$\tilde{h}_-(f)$\> Fourier transform of resonant part of  $h(t)$\\
$h^{\rm inspa}_-(f)$
\> improved Newtonian  SPA corresponding to $\tilde{h}_-(f)$; 
Eqs. (\ref{3.26}) (\ref{3.37})\\
$h^{\rm edge}_+(f)$
\> edge approximation to $\tilde{h}_+(f)$; Eq. (\ref{br10})\\
$h^{\rm intot}(f)$\>$=h^{\rm edge}_++h^{\rm inspa}_-$:
 total improved Newtonian   SPA of $h(t)$\\
$h^{\rm irspa}_-(f)$
\> improved relativistic SPA corresponding to $\tilde{h}_-(f)$;
Eqs. (\ref{4.6}), (\ref{4.7}), (\ref{4.17})\\
$h^{\rm irtot}(f)$\>$=h^{\rm edge}_++h^{\rm irspa}_-$:
 total relativistic   SPA of $h(t)$\\
inspa\>improved Newtonian SPA\\
irspa\>improved relativistic SPA\\
$l(v)$\> factored flux function\\
$m$\> total mass of the binary\\
${\cal M}$\> chirp mass $\equiv \eta^{3/5} m$\\
$n(t)$\> noise\\
$n_{\frac{1}{2}}(t)$\>whitened noise; $\equiv w_{\frac{1}{2}}(t)*n(t)$\\
$N_{\rm tot}$\> total number of cycles; Eq. (\ref{2.1})\\
$N(F)$\> instantaneous number of cycles; Eq. (\ref{2.2})\\
$N_{\rm new}(F)$\> instantaneous number of cycles in Newtonian case; Eq. (\ref{2.11}) \\
$N_{\rm rel}(F)$\> instantaneous number of cycles in relativistic case; 
Eq. (\ref{2.13}) \\
$N_{\rm useful}$\> useful number of cycles; Eq. (\ref{2.8}) \\
${\cal O}$\> overlap (Normalised Ambiguity Function); Eq. (\ref{d7})\\
$\phi(t)$\> GW phase\\
$P_n$\> P-approximant of order $v^n$\\
$\rho$\> signal to noise ratio\\
SPA\>stationary phase approximation\\
$S_n(f)$\> two-sided noise power spectral density\\
$\sigma(f)$\> weight function in $\rho^2$; Eq. (\ref{2.20})\\
$\sigma(t,z_1,z_2)$\> smoothing time window; Eq. (\ref{br11})\\
$t_{\rm min}$\> starting time of the signal\\
$t_{\rm max}$\> time at which  the signal terminates or is terminated\\
$T_n$\> Taylor approximant of order $v^n$\\
$\tau$\> $\frac{t_{\rm LSO}-t}{m}$\\
$\theta$\>Heaviside step function\\
uspa\>usual SPA\\
uspaw\>usual SPA frequency windowed \\
uspan\>usual SPA up to Nyquist frequency \\
$v$\> invariant velocity $(\pi m F)^{1/3}$\\
$v_{{\cal M}}$\> invariant velocity $(\pi {\cal M} F)^{1/3}$\\
$w(f)$\> weight factor $\frac{a^2(f)}{h_n^2(f)}$\\
$w_1(t)$\>correlation inverse of noise correlation function; Eq. (\ref{n1})\\
$w_{\frac{1}{2}}(\tau)$\>whitening kernel; Eq. (\ref{br12})\\
$x$\>$=\frac{2\pi}{3} \alpha m (F_{\rm LSO} -f)$; Eq. (\ref{ex})\\
$\zeta_0(f)$\>$\sqrt{\pi \dot{F}(t_f)}(t_f-t_{\rm max})$; Eq. (\ref{3.30})\\
$\zeta_<(f)$\>$=-\sqrt{\psi_f(t_f)-\psi_f(t_{\rm max})}$; Eq. (\ref{3.25})\\
$\zeta_>(f)$\>$=\frac{\sqrt{\pi}(f-F_{\rm max})}{\sqrt{\dot{F}(t_{\rm max})}}$;
Eq. (\ref{3.38}) \\
\end{tabbing}

\clearpage
%\documentstyle[prd,aps,eqsecnum,epsf]{revtex}
%\def \tablerule{\noalign {\vskip3truept\hrule\vskip3truept}}
%\begin{document}
\begin{table} 
\caption{Number of {\it useful} cycles for some representative systems
and approximants. For each of the cases the corresponding total number
of cycles is listed within rounded brackets in the line below.}
\begin{tabular}{ccc}
System &   Newtonian  &   Relativistic:$P_4$\\
\tablerule
 $1.4\,M_\odot-1.4\, M_\odot$  & 173      &      169   \\
                               & (1588)  &     (1586)\\
$1.4\,M_\odot-10\, M_\odot$    &  51      &      37    \\
                               & (348)    &      (320) \\
 $ 10\,M_\odot-10\,  M_\odot$  & 12       &      7.6   \\
                               & (57)     &      (47)  \\
 $ 20\,M_\odot-20\,  M_\odot$  & 9.2      &      3.4   \\
                               & (15)     &      (10)
\end{tabular}\label {table:cycles}
\end{table}

\begin{table} 
\caption{ Accuracy of the various approximations to the Fourier
transform of the chirp signal in the test mass limit for 
Newtonian waveforms. The accuracy is estimated via  the overlap, Eq. (\ref{d7}),
 of a  waveform  generated in the
time-domain and then Fourier transformed using an FFT algorithm,
with waveforms of exactly the same parameters generated directly in
the frequency domain via one of the following approximations:
The usual stationary phase approximation  
 truncated at $f_{\rm max}=F_{\rm Nyquist}$, 
(uSPAn, column 3), 
the usual stationary phase approximation  
 truncated at $f_{\rm max}=F_{\rm LSO}$ 
(uSPAw, column 4), 
the  corrected SPA (cSPA, column 5)
and the  improved SPA (inSPA, column 6), both extended up to $f_{\rm Nyquist},$
for different values of the total mass (column 1). The corresponding 
$F_{\rm LSO}$ is
listed in column 2.
[We deal only with equal mass binaries $\eta=1/4$.]
In the last column we also list the total overlap `intot' to exhibit
the relative importance of the `non-resonant' contribution.
In this table and all others the smooth time-window starts at 30 Hz;
the seismic cutoff for LIGO is at  $f_{\rm s}=40$ Hz.
}
\begin {tabular}{ccccccc}
$m/M_\odot$  & $F_{\rm LSO}$/Hz 
 &   $\left <\tilde h_{\rm N}^{\rm FFT}, \tilde h_{\rm N}^{\rm uSPAn} \right > $ 
 &   $\left <\tilde h_{\rm N}^{\rm FFT}, \tilde h_{\rm N}^{\rm uSPAw} \right > $ 
 &   $\left <\tilde h_{\rm N}^{\rm FFT}, \tilde h_{\rm N}^{\rm cSPA} \right > $ 
 &   $\left <\tilde h_{\rm N}^{\rm FFT}, \tilde h_{\rm N}^{\rm inSPA} \right > $ 
 &   $\left <\tilde h_{\rm N}^{\rm FFT}, \tilde h_{\rm N}^{\rm intot} \right > $ \\
\tablerule
  70.0 &    63& 0.1536& 0.6361& 0.9231& 0.9763& 0.9944\\
  60.0 &    73& 0.2294& 0.7302& 0.9489& 0.9721& 0.9873\\
  50.0 &    88& 0.3336& 0.8062& 0.9724& 0.9824& 0.9934\\
  40.0 &   110& 0.4708& 0.8589& 0.9862& 0.9952& 0.9993\\
  30.0 &   147& 0.6682& 0.9214& 0.9964& 0.9987& 0.9977\\
  20.0 &   220& 0.8811& 0.9681& 0.9968& 0.9974& 0.9986\\
  15.0 &   293& 0.9431& 0.9838& 0.9999& 0.9998& 0.9997\\
  14.0 &   314& 0.9528& 0.9868& 0.9995& 0.9997& 0.9993\\
  13.0 &   338& 0.9641& 0.9900& 0.9986& 0.9989& 0.9993\\
  12.0 &   366& 0.9700& 0.9912& 0.9996& 0.9997& 0.9995\\
  10.0 &   440& 0.9839& 0.9953& 0.9989& 0.9990& 0.9994\\
   5.0 &   880& 0.9983& 0.9988& 0.9991& 0.9991& 0.9992\\
   3.0 &  1466& 0.9987& 0.9987& 0.9985& 0.9985& 0.9984

\end{tabular}\label {table:newtonian}
\end{table}

\begin{table} 
\caption{Accuracy of the various approximations to the Fourier transform
of a chirp signal computed in the test mass limit  
for P-approximants at  different post-Newtonian orders $v^n$
(column 1) is measured by their overlaps with a waveform of exactly the
same parameters but computed in the time-domain and then Fourier transformed
using an FFT algorithm. The approximations considered are the usual
SPAw (column 2), inSPAw (column 3) and irSPA (column 4) 
(with the choice $x_{\rm up}=0.36$).
Both uSPA and inSPA are windowed in frequency beyond $F_{\rm LSO}$.
Only the irSPA has a spectrum extending beyond $F_{\rm LSO}$ though
 we put a numerical cutoff at  $x_{\rm cutoff}=-20.$
$x_{\rm up}=0.36$ and $x_{\rm cutoff}=-20$ in all subsequent tables.}

\begin{tabular}{cccc}
$n$  & uSPAw
     & inSPAw
     & irSPA\\
\tablerule
&   $\left <\tilde h_{{\rm P}_n}^{\rm FFT}, \tilde h_{{\rm P}_n}^{\rm uSPAw} \right > $ 
&   $\left <\tilde h_{{\rm P}_n}^{\rm FFT}, \tilde h_{{\rm P}_n}^{\rm inSPA} \right > $ 
&   $\left <\tilde h_{{\rm P}_n}^{\rm FFT}, \tilde h_{{\rm P}_n}^{\rm irSPA} \right > $ \\
\tablerule
&   \multicolumn{3}{c}{$m_1=1.4M_\odot,\ m_2=10M_\odot$}\\
\tablerule
4 & 0.9967 & 0.9986 & 0.9994\\
5 & 0.9965 & 0.9990 & 0.9997\\
6 & 0.9965 & 0.9986 & 0.9993\\
\tablerule
&   \multicolumn{3}{c}{$m_1=m_2=10M_\odot$}\\
\tablerule
4 & 0.9764 & 0.9762 & 0.9951 \\
5 & 0.9771 & 0.9775 & 0.9955 \\
6 & 0.9751 & 0.9786 & 0.9953 \\
\tablerule
&   \multicolumn{3}{c}{$m_1=m_2=20M_\odot$}\\
\tablerule
 4 & 0.8613  & 0.8613  & 0.9891\\
 5 & 0.8680  & 0.8773  & 0.9819\\
 6 & 0.8897  & 0.9038  & 0.9829\\
\end{tabular}\label {table:relativistic}
\end{table}

\begin{table}
\caption{{\it Faithfulness} of the Fourier domain 
$P$-approximants in the formal  {\it test mass} case. 
Values quoted 
 are the {\em minimax} overlap (see DIS)  of an  approximant
waveform generated directly in the frequency-domain with
the exact waveform generated in the time-domain and Fourier
transformed using the FFT algorithm.
 The approximations considered are the usual
uSPAw (column 2) and irSPA (column 3) 
As mentioned earlier $x_{\rm up}=0.36$, $x_{\rm cutoff}=-20$. 
 For  compactness of  presentation, in this Table and Tables
\protect{~\ref{table:faithfulness.fm}} and 
\protect{ \ref{table:effectualness.tm}} 
we use for the column headings
the abbreviation uSPAw to denote
$\left <\tilde h_{\rm X}^{\rm FFT}, \tilde h_{{\rm P}_n}^{\rm uSPAw} \right > $ 
 irSPA to denote    $\left <\tilde h_{\rm X}^{\rm FFT}, \tilde h_{{\rm P}_n}^{\rm irSPA} \right > $.} 
\begin{tabular}{ccc|cc|cc|cc}
 n & \multicolumn{2}{c|}{(1.4,10)} & \multicolumn{2}{c|}{(10,10)} 
 & \multicolumn{2}{c|}{(13,13)} 
 & \multicolumn{2}{c}{(20,20)} \\
\tablerule
& uSPAw   
&  irSPA 
& uSPAw   
&  irSPA 
& uSPAw   
&  irSPA
& uSPAw   
&  irSPA \\
\tablerule
 4   &  0.8360    &  0.8316 
    &  0.9596    &  0.9707 & 0.9513 & 0.9965 
    &  0.8248    &  0.9785\\ 
 5   &  0.9755    &  0.9729 
    &  0.9727    &  0.9914  & 0.9473 & 0.9973 
    &  0.8283    &  0.9855 \\
 6   &  0.9921    &  0.9903 
    &  0.9739    &  0.9938 & 0.9479 & 0.9972 
    &  0.8285    &  0.9856 
\end{tabular}\label {table:faithfulness.tm}
\end{table}

\begin{table}
\caption { {\it Faithfulness}, as in Table \protect{\ref{table:faithfulness.tm}} but for
waveforms constructed from {\it post-Newtonian} formulas for a binary
with stars of comparable masses. The value of the parameter 
$\kappa_0=47/39$.}
\begin{tabular}{ccc|cc|cc|cc}
 n & \multicolumn{2}{c|}{(1.4,10)} & \multicolumn{2}{c|}{(10,10)} 
 & \multicolumn{2}{c|}{(13,13)} 
 & \multicolumn{2}{c}{(20,20)} \\
\tablerule
& uSPAw   
&  irSPA 
& uSPAw   
&  irSPA 
& uSPAw   
&  irSPA
& uSPAw   
&  irSPA \\
\tablerule
 4   &  0.7919    &  0.7898 
   &  0.9596    &  0.9584 &0.9485&0.9665
    &  0.8764    &  0.9790 \\
5& 0.9765&0.9736&0.9820&0.9924&0.9657&0.9921
&0.8831&0.9815\\
 6   &  0.9965    &  0.9958 
    &  0.9835    &  0.9947 &0.9669&0.9921
   &  0.8860    &  0.9868 
\end {tabular}\label {table:faithfulness.fm}
\end {table}

\begin{table}
\caption{{\it Effectualness} of the Fourier domain 
$P$-approximants in the formal {\it  test mass} case. 
We list {\it minimax} overlaps of  the 
approximate waveforms generated directly in the frequency-domain
with the exact waveform generated in the time-domain and Fourier
transformed using an FFT algorithm. However, 
in addition to maximisation over the lag parameter $\tau$ and the initial phase of the approximate waveform we also maximize over the two masses $m_1$ and
$m_2$. The percentage bias in the  estimation of the  total mass $100(1-m_A/m)$
are listed in parenthesis below the overlaps.}
\begin{tabular}{ccc|cc|cc|cc}
 n & \multicolumn{2}{c|}{(1.4,10)} & \multicolumn{2}{c|}{(10,10)} 
 & \multicolumn{2}{c|}{(13,13)} 
 & \multicolumn{2}{c}{(20,20)} \\
\tablerule
& uSPAw   
&  irSPA 
& uSPAw   
&  irSPA 
& uSPAw   
&  irSPA
& uSPAw   
&  irSPA \\
\tablerule

 4     &0.9942        &0.9963   &  0.9798    &  0.9984  &  0.9651    &  0.9979 
    &  0.9166    &  0.9966 \\
    & ($-2.300$)  & ($-2.744$) & ($-0.664$) & ($-0.858$) & ($-0.962$)  & ($-1.081$)
    & ($-2.500$)  & ($-1.250$)\\
 5     &0.9882        &0.9987     &  0.9795  &    0.9979  &  0.9653    &  0.9968 
    &  0.9174    &  0.9965 \\
    & ($+0.480$) &  ($-1.301$) & ($-0.626$)  & ($-0.467$) & ($-0.631$)  & ($-0.481$)
 &   ($-1.875$) & ($-0.340$)\\
 6     &0.9970        &0.9982    &  0.9805    &  0.9964  &  0.9649    &  0.9963 
    &  0.9174    &  0.9966 \\
    & ($-0.311$)  & ($-1.166$) & ($0.000$)  & ($-0.626$) & ($-0.662$)  & ($0.000$)
   & ($-1.250$)  & ($-0.340$)\\
\end{tabular}
\label{table:effectualness.tm}
\end{table}

\begin {table}
\caption {In this Table we list times required to generate templates first in the 
time-domain and then Fourier transforming using an FFT (column 2) and compare
them with times required to construct the same templates directly in the Fourier 
domain using one of the three approximation schemes: uSPA (column 3), 
inSPA (column 4) and irSPA (5). For each post-Newtonian family (column 1) 
the time $t^{\rm FFT}_{{\rm P}_n}$ required to compute time-domain 
P-approximant templates is the highest and we normalise all times by this
value.}

\begin {tabular}{ccccc}
$n$  & {FFT}
     & {uSPA}
     & {inSPA}
     & {irSPA}\\
\tablerule
     & ${t_{{\rm P}_n}^{\rm FFT}}/{t_{{\rm P}_n}^{\rm FFT}}$ 
     & ${t_{{\rm P}_n}^{\rm uSPA}}/{t_{{\rm P}_n}^{\rm FFT}}$ 
     & ${t_{{\rm P}_n}^{\rm inSPA}}/{t_{{\rm P}_n}^{\rm FFT}}$
     & ${t_{{\rm P}_n}^{\rm irSPA}}/{t_{{\rm P}_n}^{\rm FFT}}$\\
\tablerule
  &   \multicolumn{4}{c}{$m_1=1.4M_\odot,\ m_2=10M_\odot$}\\
\tablerule
4    & 1   & 0.013  & 0.021  & 0.089\\
6    & 1   & 0.014  & 0.021  & 0.090\\
\tablerule
  &   \multicolumn{4}{c}{$m_1=m_2=10M_\odot$}\\
\tablerule
4   & 1   & 0.009  & 0.015  & 0.11 \\
6   & 1   & 0.009  & 0.016  & 0.11 \\
\end {tabular}\label {table:computecost}
\end {table}

%\end{document}

\clearpage

%\documentstyle[prd,aps,eqsecnum,epsf]{revtex}
%\def \tablerule{\noalign {\vskip3truept\hrule\vskip3truept}}
%\begin{document}

\clearpage

\clearpage

\clearpage

\clearpage

\clearpage

\clearpage

\clearpage

\clearpage

\clearpage
%\end{document}


\begin{references}
\bibitem{warn} We  stress the need to be careful
 when comparing  qualitative statements in 
publications by different authors regarding the PN order adequate for detection 
or the   necessity or otherwise of different approximants to the waveform.
 For the user, as always, the   fine print regarding the 
choice of various parameters  is  equally important: 
e.g. some papers \cite{dp97} consider 90\% overlap acceptable
(though it corresponds to a loss of 27\% in potential events), others
\cite{dis98} insist on considering 96.5\% (only 10\% loss); some papers  actually refer to Advanced LIGO which may be
irrelevant for initial LIGO; claims may refer only to lighter binaries rather
than  be  a comprehensive one including massive binaries.
\bibitem {effect.norm} Consider two signals $q_1$ and $q_2$ with supports
$S_1=[f_{\rm min}, f_{\rm max}]$ and $S_2=[f_{\rm min}, f_{\rm Ny}]$ with
$f_{\rm Ny} > f_{\rm max}.$ Even when $q_1$ and $q_2$ perfectly match with
each other in $S_1$ (both in amplitude and phase) their overlap will in
general be less than 1. Indeed, 
$\langle q_1, q_2 \rangle = \Vert q_1 \Vert/\Vert q_2 \Vert,$
where $\Vert q_i \Vert \equiv \langle q_i, q_i \rangle ^{1/2},$ $i=1,2.$
Since $S_1 \subset S_2,$ $\Vert q_1 \Vert \leq \Vert q_2 \Vert. $ This explains
why uSPAn returns less overlaps than uSPAw.
\bibitem{lag} In the case where one introduces a time lag $\tau$
between two signals the Wiener scalar product reads 
$\left < h,g\right >(\tau ;\lambda_k,\mu_k)
\equiv \int_{-\infty}^{\infty} \frac {df e^{2\pi i f \tau}} {S_n(f)}
\tilde h^*(f; \lambda_k) \tilde g(f; \mu_k)$,
where $^*$ denotes complex conjugation,  
$\lambda_k,$ $k=1,\ldots,n_\lambda,$ and
$\mu_k,$ $k=1,\ldots,n_\mu,$ are the parameters of the signals and
$n_\lambda$ and $n_\mu$ are the corresponding number of parameters
\cite{dis98}.
\end{references}
\end{document}